\documentclass[twocolumn,resetfootnote,trackchanges]{aastex701}

\usepackage{fontawesome}
\usepackage{hyperref}

\newcommand{\githubicon}{{\color{black}\faGithub}}

\graphicspath{{./}{figures/}}

\newcommand{\prot}{$P_{\mathrm{rot}}$}
\newcommand{\tpen}{$T_\mathrm{pen}$}
\newcommand{\tumb}{$T_\mathrm{umb}$}

\begin{document}

\title{KRONOS II: Solar-like Umbra and Penumbra Properties on the Young Sun V1298~Tau}
\shorttitle{Starspots on the Young Sun V1298~Tau}
\shortauthors{Murphy et al. 2026}
\received{June 15, 2026}
\revised{September 01, 2026 and October 21, 2026}
\accepted{November 09, 2026}

\author[orcid=0000-0002-8517-8857]{Matthew M. Murphy}
\affiliation{Department of Physics and Astronomy, Michigan State University, East Lansing, MI, USA}
\email[show]{mmmurphy@msu.edu}  

\author[orcid=0000-0002-9464-8101]{Adina D. Feinstein}
\affiliation{Department of Physics and Astronomy, Michigan State University, East Lansing, MI, USA}
\email[]{adina@msu.edu}

\author[0009-0007-9766-2324]{Meir E. Schochet}
\affiliation{Department of Physics and Astronomy, Michigan State University, East Lansing, MI, USA}
\email[]{meir@msu.edu}

\author[orcid=0000-0002-3627-1676]{Benjamin V. Rackham}
\affiliation{Department of Earth, Atmospheric and Planetary Sciences, Massachusetts Institute of Technology, 77 Massachusetts Avenue, Cambridge, MA 02139, USA}
\affiliation{Kavli Institute for Astrophysics and Space Research, Massachusetts Institute of Technology, Cambridge, MA 02139, USA}
\email[]{brackham@mit.edu}

\author[orcid=0000-0003-0156-4564]{Luis Welbanks}
\affiliation{School of Earth and Space Exploration, Arizona State University, Tempe, AZ, USA}
\email[]{luis.welbanks@asu.edu}

\author[0000-0002-0726-6480]{Darryl Z. Seligman}
\affiliation{Department of Physics and Astronomy, Michigan State University, East Lansing, MI, USA}
\email[]{dzs@msu.edu}

\author[orcid=0000-0002-7119-2543]{Girish M. Duvvuri}
\affiliation{Department of Physics and Astronomy, Vanderbilt University, Nashville, TN 37235, USA}
\email[]{girish.duvvuri@gmail.com}

\author[0000-0003-0973-8426]{Eva-Maria Ahrer}
\affiliation{Max Planck Institute for Astronomy (MPIA), K\"{o}nigstuhl 17, 69117 Heidelberg, Germany}
\email[]{ahrer@mpia.de}

\author[orcid=0000-0002-4881-3620]{John H. Livingston}
\affiliation{Astrobiology Center, 2-21-1 Osawa, Mitaka, Tokyo 181-8588, Japan}
\affiliation{National Astronomical Observatory of Japan, 2-21-1 Osawa, Mitaka, Tokyo 181-8588, Japan}
\affiliation{Astronomical Science Program, The Graduate University for Advanced Studies (SOKENDAI), 2-21-1 Osawa, Mitaka, Tokyo 181-8588, Japan}
\email[]{john.livingston@nao.ac.jp}

\author[orcid=0009-0000-6113-0157]{Saugata Barat}
\affiliation{Kavli Institute for Astrophysics and Space Research, Massachusetts Institute of Technology, Cambridge, MA 02139, USA}
\email[]{saugata@mit.edu}

\author[0000-0002-0875-8401]{Jean-Michel D\'{e}sert}
\affiliation{Leibniz Institute for Astrophysics Potsdam, An der Sternwarte 16, 14482 Potsdam, Germany}
\affiliation{DESY, Platanenallee 6, D-15738 Zeuthen, Germany}
\email[]{jmdesert@aip.de}

\author[0000-0002-4671-2957]{Rafael Luque}
\affiliation{Instituto de Astrof\'isica de Andaluc\'ia (IAA-CSIC), Glorieta de la Astronom\'ia s/n, 18008 Granada, Spain}
\email[]{rluque@iaa.es}

\author[orcid=0000-0001-8504-5862]{Catriona Murray}
\affiliation{Department of Astrophysical and Planetary Sciences, University of Colorado Boulder, Boulder, CO 80309, USA}
\email[]{catriona.murray@colorado.edu}

\author[orcid=0000-0001-8236-5553]{Matthew C. Nixon}
\altaffiliation{51~Pegasi~b Fellow}
\affiliation{School of Earth and Space Exploration, Arizona State University, Tempe, AZ, USA}
\email[]{matthewnixon@asu.edu}

\author[0009-0008-8016-6591]{Sydney Petz}
\affiliation{Department of Physics and Astronomy, Michigan State University, East Lansing, MI, USA}
\email[]{petzsydn@msu.edu}

\author[orcid=0000-0002-3328-1203]{Michael Radica}
\altaffiliation{NSERC Postdoctoral Fellow}
\affiliation{Department of Astronomy \& Astrophysics, University of Chicago, 5640 South Ellis Avenue, Chicago, IL 60637, USA}
\email[]{radicamc@uchicago.edu}

\author[0000-0001-9289-0570]{Hinna Shivkumar}
\affiliation{Anton Pannekoek Institute for Astronomy, University of Amsterdam, Science Park 904, 1098 XH, Amsterdam, The Netherlands}
\email[]{h.shivkumar@uva.nl}

\begin{abstract}
Transiting exoplanets provide a unique laboratory for studying stellar surface heterogeneities via starspot or facular occultations. When observed at multiple wavelengths, this configuration enables spectroscopic characterization of spot thermal contrasts, distributions, and morphology. In this work, we leverage JWST NIRISS/SOSS transit observations of the 20--30~Myr planets V1298~Tau~bcd to study the surface properties of their solar analog host star V1298~Tau. We identify 14 starspot crossing events across two visits. We derive $0.8-2.8\mu$m starspot contrast spectra and demonstrate the contrasts are best explained when accounting for the umbral and penumbral components of the starspots, robust to which stellar model grid is assumed. The spot temperatures are broadly consistent between visits, suggesting that V1298~Tau ($T_\mathrm{phot}=4866\pm33$\,K) has starspots with $T_{\mathrm{umbra}}$ = 3200--3400\,K umbrae and $T_{\mathrm{penumbra}}$ = 4400--4600\,K penumbrae, and are $\sim$25--30\% umbrae by area. The differences between these spot components and the stellar photosphere are consistent with sunspots. Additionally, the relation between the spot contrast and the ratio of umbral to penumbral area is similar to that of the Sun. Combining these JWST observations with long baseline multi-band photometry from the Las Cumbres Observatory, we also estimated the global unocculted spot distribution, revealing at least 5 additional large unocculted active regions assuming no facular contribution. Altogether, these measurements suggest that while the total spot coverage evolves in time, the relative temperatures of surface heterogeneities on Sun-like stars may be consistent throughout their lifetimes. Furthermore, these results demonstrate that JWST exoplanet transit observations can be useful for starspot substructure characterization.
\end{abstract}

\section{Introduction} \label{sec:intro}

Young stars are known to be more magnetically active than their main sequence counterparts. This activity manifests in a variety of ways, including strong global magnetic fields on the order of $\sim$kG \citep{hackman2016, folsom2016, kochukov2020}, enhanced stellar flare energies and rates \citep{ilin19, Feinstein2020_youngstarflares, ilin21, ealy24, feinstein24, Namekata2025_youngstarCMEs, mamonova25, tran26}, and photometric variability on the order of $\sim$1--5\% that traces the rapid rotational period of the star \citep{bouvier1993, boyle26}. This photometric variability primarily traces the presence of starspots rotating into and out of our line of sight, and has been the dominant method by which starspots are studied \citep[e.g.,][]{Morris2020_spotsVSage, luger21}. 

Starspots arise due to the suppression of convection in the presence of a strong magnetic field, allowing local plasma at the surface to cool and appear relatively dark compared to the surrounding photosphere. Resolved observations of sunspots predate the invention of the telescope \citep[e.g., see the historical discussions of][]{wittmann1987, yau1988, eddy1989}, but for stars other than the Sun, starspots can only be spatially resolved on the largest, most nearby stars for which interferometry is possible \citep[e.g.,][]{Roettenbacher2016_spotimaging, roettenbacher2026}. Therefore, for the vast majority of stars, we are unable to constrain the total number, sizes, locations, and contrasts of spots on the stellar photosphere. Techniques which invert the rotational light curves of these stars to infer the presence and general positions of starspots do exist. However, they are highly degenerate, as a large number of surface conditions can be generated to model observed variability \citep{luger21}. Multi-band photometry lessens this degeneracy by enabling the measurement of chromatic contrasts, allowing spot temperatures to be estimated independently of the geometric properties, but still lacks the necessary data to constrain the remainder of starspot properties \citep{biagini24, mori24, waalkes24}. Alternatively, with high-resolution spectroscopy, the position and size of spots can be inferred through their deformation of the stellar cross-correlation function \citep{maio2024}.

Transiting exoplanets offer a unique additional solution to this problem. A planet occulting a starspot produces a temporary increase in total system flux because it preferentially blocks the darker, cooler component of the photosphere. Starspot crossing events (SCEs) are readily observed as Gaussian-like features in the light curve, and are distinguishable from other sources of activity (e.g., flares) based on their temporal profile and lack of response from any stellar absorption or emission lines (e.g., H$\alpha$). SCEs have been observed in numerous transiting systems over the last two decades \citep[e.g.,][]{rabus2009, silva2010, tregloan2013, sanchis2013, mohler2013, mocnik2016, mocnik2017_wasp107, mocnik2017_qatar2, Fu2022, murray2026, triantafillides2026}. In fact, the first exoplanet transit observations of HD~209458~b potentially exhibited SCEs \citep{charbonneau2000_hd209458, silva2003}. 

The precision of JWST enables precise monitoring of the duration of SCEs and their chromatic amplitudes over a wide wavelength range  \citep{FournierTondreau2025, murray2026}. These allow for constraining starspot properties in significantly richer detail than possible with previous observatories. Combining JWST transits with simultaneous long-baseline photometry offers further means to break degeneracies in inferred heterogeneity properties \citep{Sagynbayeva25, Sagynbayeva26}.

Here, we present a starspot analysis of the young solar analog V1298~Tau. Initial estimates dated this system to be $\sim$23\,Myr \citep{david19_b}, though recent estimates suggest it may be slightly older \citep[$\sim$29\,Myr;][]{luhman2023, luhman2024}. The star exhibits heightened magnetic activity due to its age \citep{finociety_2023}, and hosts four known large ($R_p>5R_\oplus$) transiting planets \citep{david2019_v1298tau, david19_b}. This combination therefore offers a unique opportunity to characterize starspots through exoplanet transits. Due to the increased wavelength coverage and precision of JWST, we are able to resolve multiple starspot crossing events along the transit chords of three planets in this system, and study their thermal contrast spectra from $\sim$0.8--2.8\,$\mu$m. We demonstrate that these data can infer the properties of the starspot umbral and penumbral regions from their combined thermal contrast. These results make V1298~Tau only the second star other than the Sun for which its spot substructures have been characterized \citep{Jarvinen2018}, and are the first time this has been done using JWST exoplanet transit observations. We combine these observations with ground-based multi-wavelength photometry from the Las Cumbres Observatory (LCO) over multiple stellar rotation periods to model the photospheric surface of the star.

The remainder of the manuscript is presented as follows. In Section~\ref{sec:obs}, we detail our observations and data reduction. In Section~\ref{sec:stellarspectrum}, we present and model the out-of-transit stellar spectrum of V1298~Tau before moving on to our transit light curve fits in Section~\ref{sec:broadband_fits}. In Section~\ref{sec:broadband_fits}, we also present our best-fit occulted starspot map. Then, we present and model our thermal contrast spectra in Section~\ref{sec:contrast}. We contextualize these results using long-baseline photometry in Section~\ref{sec:lcophotometry}. Finally, we discuss the implications of our findings, including comparison to both the Sun and the young solar analog EK~Draconis, in Section~\ref{sec:discussion} before concluding in Section~\ref{sec:summary}. This manuscript aims to improve reproducibility and transparency of scientific research. All of the data presented are hosted on GitHub. Every figure caption includes a GitHub icon (\githubicon), which links to an interactive Jupyter notebook used to create that figure. We will also create a Zenodo archive upon final acceptance.

\section{Observations} \label{sec:obs}

We observed one transit of V1298~Tau~c \citep[$P=8.24$~days;][]{livingston2026_v1298tau} and one simultaneous transit of V1298~Tau~d ($P=12.40$~days) and V1298~Tau~b ($P=24.14$~days) with JWST's Near Infrared Imager and Slitless Spectrograph \citep[NIRISS; ][]{niriss1_overview} as part of the KRONOS program (JWST GO 5959). The details of the transit of V1298~Tau~c, taken on UT 2025 September 6 from 02:05:25--11:41:30, are presented in \cite{Murphy2026}. The simultaneous transits of V1298~Tau~d and b were taken on UT 2025 September 10--11, just four days later, from 16:15:45--03:15:39.  The rotation period of V1298~Tau is $P_\mathrm{rot}=2.97^{+0.03}_{-0.04}$\,days \citep{feinstein2022_v1298tauTESS}, so any heterogeneities present on the star should be similar, but the visible disk may be different. For the remainder of this work, we refer to the transit of V1298~Tau~c as ``Visit 1'' and the simultaneous transits of V1298~Tau~d and b as ``Visit 2.'' 

\subsection{JWST Data Reduction} \label{subsec:obs_JWSTdatareduction}

The observing setups between Visits 1 and 2 were largely identical. The exact details of the observation of Visit 1 are presented in \cite{Murphy2026}. For Visit 2, we obtained time series spectroscopy using the \texttt{SUBSTRIP96} subarray with \texttt{NISRAPID} readout mode, limited to SOSS' Order 1 (0.85--2.83\,$\mu$m). We took 2976 integrations, each of 5 groups and $\sim$13.3\,s, for a total exposure of 11\,hr. This setup was driven by the brightness of our target ($V$=10.115, $J$=8.687), and successfully avoided saturation across the detector. The pre-transit baseline was $\sim$1.6\,hr, but there was no clear post-transit baseline. 

We reduced the Visit 2 data using version 2.3.1 of the \texttt{exoTEDRF} pipeline \citep{exotedrf, radica2023_wasp96, feinstein_early_2023}. We follow the standard Stage 1--3 steps of \texttt{exoTEDRF} and the JWST calibration pipeline. Since several pipeline steps can be optimized by inputting outputs from later steps, we performed an iterative reduction. Namely, after the initial reduction, we used informed bad pixel masks, estimates of the out-of-transit integrations, and an estimate of the normalized spectroscopic light curve in our second reduction, which is the final output used in this work. In Stage 1, in the \texttt{DQInit} step, we pass the hot pixel map generated from the initial reduction pass. We perform background subtraction first here at the group level, scaling the left and right sides of the detector equally. Then, in the \texttt{OneOverF} step, we pass the initially generated bad pixel masks and time series estimate, and use the \texttt{scale-achromatic} method with a 40~pixel inner mask width. This method utilizes a median of all unmasked pixels per column to estimate the 1/f value \citep{radica2023_wasp96}. Following this, we run the \texttt{LinearityStep} to correct for group-to-group differences. Next, in Stage 2, we repeat background subtraction at the integration level. At the end of Stage 2, we perform a Principal Component Analysis to search for significant systematic artifacts such as thermal beating patterns or trace shifts, following the method of \cite{radica_super-solar_2026}. However, we do not identify any such components, potentially because the level of stellar variability is dominant. The only notable component identified several additional hot pixels, which we removed via the \texttt{PCAReconstructStep}. 

We extracted the time series stellar spectra via box extraction with a half-width of 40 pixels. In parallel, we extracted flux-calibrated stellar spectra using the \texttt{flux\_calibrate\_soss} step and the relevant JWST calibration files provided by \texttt{exoTEDRF}. We conservatively identified integrations 0--420 as being out-of-transit. There may be several post-transit integrations as well, but these are difficult to disentangle from the egress of V1298~Tau~b prior to light curve fitting.

\subsection{LCOGT Observations and Data Reduction} \label{subsec:obs_LCOdatareduction}

We also monitored V1298~Tau across multiple rotation periods with the Las Cumbres Observatory Global Telescope network (LCOGT) as part of programs LCO2025A-002 and LCO2025B-006. The objective of these observations was to provide priors on the global stellar heterogeneities. These observations used SDSS g$\prime$ and r$\prime$ filters, and were obtained from 2025 August 28--September 16. The data were automatically reduced by the LCOGT \texttt{BANZAI} pipeline \citep{BANZAI_phot_SPIE} and downloaded from the LCOGT Archive\footnote{\href{https://archive.lco.global/}{https://archive.lco.global/}}. \texttt{BANZAI} performs bad-pixel masking, bias subtraction, dark subtraction, and flat field correction.

We aligned the images using the \texttt{python} package \texttt{twirl} \citep{twirl_algo, twirl_implement}. \texttt{twirl} rapidly extracts the image sky coordinates based on Gaia-cataloged stars within the field of view, and computes the World Coordinate System (WCS) solution. We developed a custom script to perform this alignment on the several hundred LCOGT Sinistro images, which can be accessed on \texttt{PyPi}\footnote{\href{https://pypi.org/project/lco-aligner/}{https://pypi.org/project/lco-aligner/}}. 

While \texttt{BANZAI} does extract sources, we chose to perform our own aperture photometry using AstroImageJ \citep[AIJ,][]{AstroImageJ}, which performs differential photometry using multiple sources in the field of view. We used UCAC4 551-008569 and UCAC4 551-008564 as our comparison stars due to their proximity to V1298~Tau and similar brightness to each other. Due to the crowdedness of the field, we enforced a small aperture of 25\,pixels with inner/outer background annuli of 35/55\,pixels. We removed images for which the earlier \texttt{twirl} alignment failed, or AIJ was unable to compute the stellar centroids. This resulted in 3.4\% of the g$\prime$ and 2.2\% of the r$\prime$ images being removed from our analysis.

\begin{table*}[tb]
    \caption{Stellar properties derived from the 0.8--2.83\,$\mu$m stellar spectrum of V1298~Tau from two visits.}
    \centering
    \begin{tabular}{c|c|c|c|c|c|c|c|c}  
    N & $R_\star$ ($R_\odot$) & $T_{\mathrm{phot}}$ (K)& $T_{\mathrm{cool}}$ (K)& $f_{\mathrm{cool}}$ & $T_{\mathrm{hot}}$ (K)& $f_{\mathrm{hot}}$ & $A_V$ & $\Delta \log Z$ \\ \hline 
    \multicolumn{8}{c}{\textit{Visit 1: 2025 September 06}} \\ \hline 
    1 & 1.223 $\pm$ 0.001 & 5098 $\pm$ 4 & - & -&-&-& 0.074 (fixed) & 0 \\
    2 & 1.285 $\pm$ 0.005 & 5236 $\pm$ 15 & 3446 $\pm$ 36 & 0.22 $\pm$ 0.02 & - & - & 0.074 (fixed) & 71 \\
    3 & 1.282 $\pm$ 0.005 & 4891 $\pm$ 26 & 3403 $\pm$ 30 & 0.19 $\pm$ 0.02 & 5662 $\pm$ 54 & 0.32 $\pm$ 0.03 & 0.074 (fixed) & 128 \\ \hline 
    \multicolumn{8}{c}{\textit{Visit 2: 2025 September 10}} \\ \hline 
    1 &1.229 $\pm$ 0.001 & 5112 $\pm$ 3 & -&-&-&-&0.074 (fixed) & 0 \\
    2 & 1.279 $\pm$ 0.004 & 5224 $\pm$ 11 &3471 $\pm$ 38 & 0.18 $\pm$ 0.01 & - & - &  0.074 (fixed) & 82 \\
    3 & 1.276 $\pm$ 0.004 & 4866 $\pm$ 33 & 3405 $\pm$ 25 & 0.15 $\pm$ 0.01 & 5673 $\pm$ 41 & 0.33 $\pm$ 0.03 & 0.074 (fixed) & 118 \\ \hline
    \end{tabular}
    \label{tab:star_specproperties}
    \tablecomments{The results for Visit 1 are from the analysis of \cite{Murphy2026}. N refers to the number of components allowed in the fit. $\Delta \log Z$ refers to the difference in Bayesian evidence relative to the corresponding N=1 case, and higher $\Delta \log Z$ indicates a preference for that model. These fits each assume a fixed [Fe/H] = 0.1 \citep{suarez_mascareno_2022}, distance of 107.6\,pc \citep{gaiaDR3}, extinction $A_v$=0.074 \citep{david19_b}, and $\log\left( \mathrm{g/[cm/s^2]}\right)$ = 4.25 \citep{david19_b}.}
\end{table*}

\begin{figure}
    \centering
    \includegraphics[width=1\linewidth]{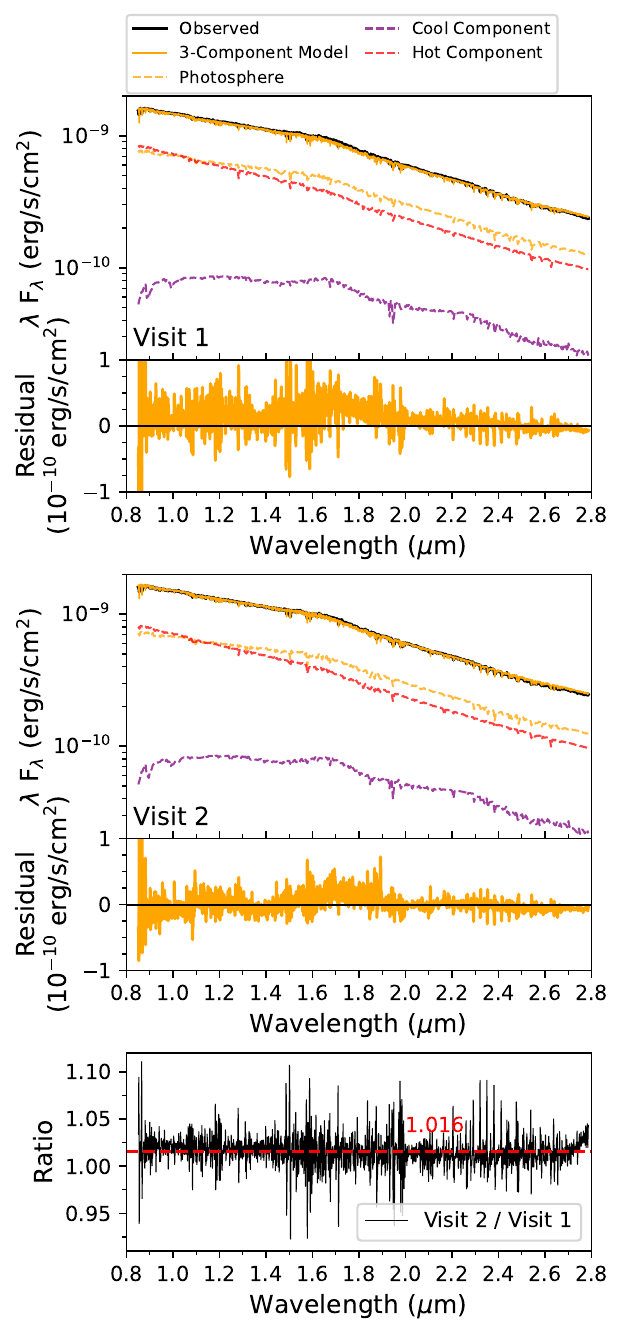}
    \caption{Median out-of-transit stellar spectra, best-fit 3-component models, and data--model residuals for Visit 1 (top) and Visit 2 (middle). The bottom panel shows the ratio of the observed spectra. V1298~Tau was $\sim$1.016$\times$ brighter during Visit 2. \href{https://github.com/kronos-jwst/KRONOS-II-V1298-Tau-Starspots/blob/main/Figure01.ipynb}{\githubicon}}
    \label{fig:stellarspectra}
\end{figure}

\section{Stellar Spectrum Analysis} \label{sec:stellarspectrum}

To understand the global heterogeneity properties of V1298~Tau, we first fit the flux-calibrated out-of-transit JWST stellar spectra. \cite{Murphy2026} previously presented the Visit 1 stellar spectrum observed on 2025 September 06, and best-fitting 1, 2, and 3-component models. We repeat these fits to the Visit 2 stellar spectrum. 

We construct the stellar spectrum as the median of out-of-transit integrations, and propagate the uncertainty as the standard error on the median. For Visit 1, following \cite{Murphy2026}, these are integrations 0--520 and 1800-2390. For Visit 2, the only definitively out-of-transit integrations in Visit 2 are pre-transit, and we conservatively use integrations 0--420.

To fit the observed spectrum, we draw model spectra from the NewEra grid \citep{NewEraGrid} using \texttt{speclib} \citep{Rackham2023, Rackham2024}. For consistency with \cite{Murphy2026}, we assume a fixed [Fe/H] = 0.1 \citep{suarez_mascareno_2022} and $\log\left( \mathrm{g/[cm/s^2]}\right)$ = 4.25 \citep{david19_b}, interpolated from the grid points at 4.0 and 4.5, but use the updated distance of 107.6\,pc \citep{gaiaDR3}. We use the nested sampling Monte Carlo algorithm \texttt{MLFriends} \citep{ultranest_algorithm1, ultranest_algorithm2} in \texttt{UltraNest} \citep{ultranest}, using 200 live points to balance efficiency with exploration of the parameter spaces. To account for surface heterogeneities, we test constructing the model spectrum using up to three successively added components: (i) the photosphere, (ii) a cool component (e.g., starspots), and (iii) a hot component (e.g., faculae or plages). We fit for the photosphere temperature $T_{\mathrm{phot}}$, the temperatures $T_i$ and covering fractions $f_i$ of the additional components, the stellar radius $R_\star$, and the interstellar extinction $A_V$. During the fit, the extinction value is converted to a wavelength-dependent attenuation factor based on \cite{CCM89}. We enforce Normal priors on the stellar radius of $R_\star$ = 1.278 $\pm$ 0.07\,$R_\odot$ and photosphere $T_{\mathrm{phot}}$ = 5050 $\pm$ 100\,K \citep{suarez_mascareno_2022}, and limit all temperature parameters via uniform priors to the range (3000\,K, 6000\,K) and all non-photosphere coverage fractions to (0, 0.5). We fix the interstellar extinction to $A_V$ = 0.074 based on previous interstellar dust map-derived estimates \citep{david19_b}.

We additionally include an uncertainty scaling parameter in the fits. Following the method of \cite{Rackham2024}, this scaler $f$ is flux-dependent so that, if $\sigma_{F_{\lambda}}$ is the unscaled flux uncertainty, the scaled uncertainty $\sigma^{\prime}_{F_{\lambda}}$ is
\begin{equation}
    \sigma^{\prime}_{F_{\lambda}} = \sqrt{ \left(\sigma_{F_{\lambda}}\right)^2 + \left( f~F_\lambda \right)^2}.
\end{equation}
To aid convergence, we parameterize this term as $\log_{10}\left(f\right)$ in the sampling and enforce a wide uniform prior range of [-4, -1]. This scaling is meant to account for potentially underestimated uncertainty in our stellar spectrum, particularly due to systematics like the native instrumental flux calibration.

\subsection{Stellar Spectrum Fit Results} \label{subsec:stellarspectrum_results}

The results from fitting the stellar spectra are listed in Table~\ref{tab:star_specproperties}. Figure~\ref{fig:stellarspectra} shows the observed spectra and best-fit models, including the individual components scaled by their covering fractions, for both Visit 1 and Visit 2. In both cases, the 3-component model (photosphere with hot and cool components) is strongly preferred over both the 1- and 2-component models by the Bayesian evidence. The stellar properties derived from Visit 1 and Visit 2 are consistent. Together, these results suggest that the number and distribution of faculae and/or plages on the visible disk of V1298~Tau remained very similar between visits, but the number of starspots decreased. 

Over the $\sim$4 days between visits, $f_{\mathrm{cool}}$ decreased in absolute value by 4\%, though the difference is only significant at the 1.8$\sigma$ level. This suggests there may have been fewer total starspots on the visible surface of V1298~Tau during Visit 2. The temperature of this starspot component $T_{\mathrm{cool}}$ is nearly identical between visits. The bottom panel of Figure~\ref{fig:stellarspectra} shows the ratio of the Visit 2 and Visit 1 spectra. The Visit 2 spectrum is brighter, and the ratio between them is fairly constant with wavelength with a median value of 1.016. All else held equal, this would be qualitatively consistent with Visit 2 having a lower spot coverage fraction. On the other hand, the properties of the hot component $T_{\mathrm{hot}}$ and $f_{\mathrm{hot}}$ are each consistent between visits within 0.24$\sigma$, suggesting nearly identical faculae/plage coverage.

\begin{figure*}[t]
    \centering
    \includegraphics[width=\linewidth]{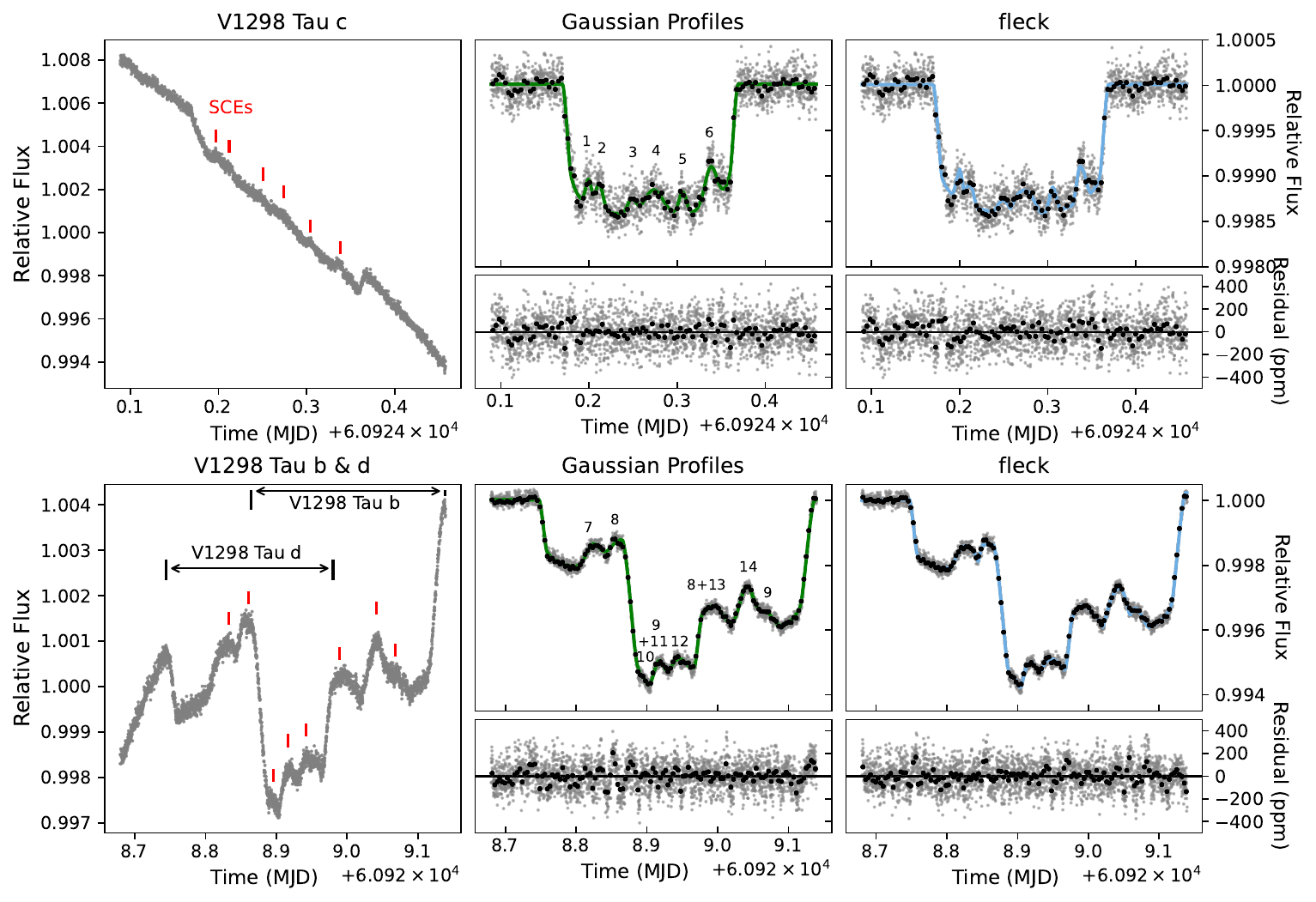}
    \caption{Broadband NIRISS/SOSS Order 1 ($\sim$0.8--2.83\,$\mu$m) light curves of V1298~Tau. The top row shows a transit of V1298~Tau~c during our first visit on 2025 September 06, and the bottom row shows a simultaneous transit of V1298~Tau~d and b during our second visit on 2025 September 10. The left-hand columns show the observed light curve, with starspot crossing events (SCEs) indicated in red. The middle and right-hand columns show the detrended light curves and best-fit models using two methods: modeling the SCEs as Gaussian Profiles, and modeling them using \texttt{fleck} \citep{fleck}. We identify crossings of at least 14 distinct starspots by these three planets, labeled in the middle columns. \href{https://github.com/kronos-jwst/KRONOS-II-V1298-Tau-Starspots/blob/main/Figure02.ipynb}{\githubicon}
    }
    \label{fig:bbfits}
\end{figure*}

\section{JWST Broadband Fits}\label{sec:broadband_fits}

\subsection{Light Curve Fit Setup}\label{subsec:obs_lightcurvefitting}

In this work, we perform a new fit to the Visit 2 transit light curve and adopt the fits from \cite{Murphy2026} for Visit 1. We apply the same light curve fitting methodology from \cite{Murphy2026} for Visit 2. We test different parameterizations for modeling the starspot crossing events (SCEs) during the two transits. Because we are focused on deriving and comparing the properties of the starspots, we restrict this analysis to two model types: 1) modeling the SCEs as Gaussian profiles, and 2) modeling the SCEs using the \texttt{fleck} starspot model \citep{fleck}. 

For both methods, we use a custom \texttt{python} routine to model each light curve as the product of a transit model, $T(\theta, t)$, out-of-transit baseline model, $S(\theta, t)$, and an SCE model, $F_\mathrm{SCE}(\theta, t)$. Each is a function of astrophysical and systematic parameters, $\theta$, and time, $t$. 

For the Gaussian Profiles method, we use the \texttt{batman} package \citep{batman} to generate the transit model. Then, we model the SCEs as:
\begin{equation} \label{eqn:gaussians}
    F_{{\mathrm{SCE}, 1}} (t) = 1 + \sum_{i} A_i \,\, \textrm{exp}\bigg({-\frac{\big(t - b_i\big)^2}{2 \sigma_i^2}}\bigg)\,. \label{eqn:GaussianFunc}
\end{equation}
Here, $A_i$ is the relative flux amplitude, $b_i$ is the mid-point time, and $\sigma_i$ is the width or duration for the $i$th SCE. 
To model the out-of-transit baseline of the Visit 2 light curve, we use an exponential form: 
\begin{equation}
    S(t) = a\, \textrm{exp}\left( \frac{-(t-t_\mathrm{med})}{\tau} \right) + c_\mathrm{offset}. \label{eqn:expbaseline}
\end{equation}
Here, $a$, $\tau$, and $c_\mathrm{offset}$ are each coefficients that are allowed to freely vary. $t_{\mathrm{med}}$ refers to the visit-median time. As shown later in Section~\ref{sec:lcophotometry}, the overall shape and sign of this baseline is consistent with the flux trend measured by our contemporaneous LCOGT/Sinistro photometry. This is also true for the Visit 1 baseline, which exhibits a downward curve well fit by a polynomial \citep{Murphy2026}.

For the second method, we use \texttt{fleck}'s built-in transit model based on the \texttt{batman} framework. The relative flux variations induced by SCEs are automatically included in this model, and it is assumed that each spot is circular. Each spot is prescribed a radius $r_{\mathrm{spot}}$ and position (latitude $lat$ and longitude $lon$) on the stellar surface, and all spots are enforced to have the same contrast $\alpha$. For consistency with \cite{Murphy2026}, we assume an edge-on stellar inclination $i_\star$ = 90$^\circ$ which is broadly consistent with previous estimates for V1298~Tau \citep[$i_\star$=85.1$^\circ$ $\pm$ 3.6$^\circ$;][]{johnson2022}.

We initially fit the broadband light curve to narrow the possible range of the spot and baseline parameters. In both cases, we freely fit for the time of conjunction of each planet ($t_{c, d}$ and $t_{c, b}$), planet-star radius ratios ($R_{p,d}/R_\star$ and $R_{p,b}/R_\star$), quadratic limb darkening coefficients $u_1$ and $u_2$, baseline model parameters $a$, $\tau$, and $c_{\mathrm{offset}}$, and a relative flux uncertainty scaler $\sigma_{scaler}$. For the fits using Gaussian profiles, we additionally fit for $A$, $b$, and $\sigma$ of each spot (Equation~\ref{eqn:gaussians}). For the fits using \texttt{fleck}, we fit for the shared contrast $\alpha$, and the $r_{\mathrm{spot}}$, $lat$, and $lon$ of each spot. Due to the complexity of this observation and the scarcity of contamination-free transit, we fix the orbital parameters (orbital period, semi-major axis, and inclination) of both d and b during the fit to values from \cite{david19_b}. 

We enforce loose, uniform priors on each parameter to keep them within physical bounds: we limit $R_p/R_\star$ to (0,1), the limb darkening coefficients to [0,1], and spot radii $r_{spot}$ to positive values. We performed these fits using Markov Chain Monte Carlo sampling with the \texttt{emcee} package \citep{emcee}, running each for a 10,000 step burn-in period followed by a 25,000 production run (25$\times$ the mean autocorrelation time). 

\subsection{Broadband Fit Results}

Figure~\ref{fig:bbfits} shows the observed and detrended light curves, and best-fit models for both the Gaussian and \texttt{fleck} methods for both visits. Table~\ref{tab:transittimes} lists the derived transit times for each planet. We achieved good fits to both light curves using both methods. 
For Visit 1, the reduced $\chi^2$ values without uncertainty scaling are 0.98 for the Gaussian method and 1.04 for \texttt{fleck}. The data--model residuals are Gaussian distributed with standard deviations of 138\, ppm for both methods, and pass the Anderson-Darling test for a normal distribution. 
For Visit 2, the reduced $\chi^2$ values without uncertainty scaling are 1.21 for the Gaussian method and 1.15 for \texttt{fleck}. The data--model residuals are again Gaussian distributed with standard deviations of 120\,ppm and 117\,ppm, respectively, and pass Anderson-Darling tests. 
We measure the broadband transit depths $\delta$ of each planet to be $\delta_c$ = 1298 $\pm$ 21\,ppm, $\delta_d$ = 1930 $\pm$ 2\,ppm, and $\delta_b$ = 3866 $\pm$ 2\,ppm from the Gaussian fits, and $\delta_c$ = 1268 $\pm$ 24\,ppm, $\delta_d$ = 1870 $\pm$ 13\,ppm, and $\delta_b$ = 3727$^{+9}_{-4}$\, ppm from the \texttt{fleck} fits. Assuming $R_\star$ = 1.276 $\pm$ 0.004\,$R_\odot$ (Table~\ref{tab:star_specproperties}) and the more conservative \texttt{fleck} results, these correspond to planetary radii of $R_{p,c}$ = 4.95 $\pm$ 0.05\,$R_\oplus$, $R_{p,d}$ = 6.02 $\pm$ 0.03\,$R_\oplus$, and $R_{p,b}$ = 8.50 $\pm$ 0.02\,$R_\oplus$. These are each consistent within 1.6$\sigma$ with the recent measurements of \cite{livingston2026_v1298tau}. 

\begin{table}[]
    \centering
    \caption{Best-fit times of conjunction for V1298~Tau~c, d, and b.}
    \begin{tabular}{c|c|c}
    Planet & Method & $t_c$ (BJD) \\ \hline 
    c & Gaussian & 2460924.7690 $\pm$ 0.0001 \\
    c & \texttt{fleck} & 2460924.7691 $\pm$ 0.0001 \\ \hline 
    d & Gaussian & 2460929.36229 $\pm$ 0.00009\\
    d & \texttt{fleck} & 2460929.36156 $\pm$ 0.00009 \\ \hline 
    b & Gaussian & 2460929.50032 $\pm$ 0.00008\\
    b & \texttt{fleck} & 2460929.50052 $\pm$ 0.00006 
    \end{tabular}
    \tablecomments{The times for V1298~Tau~c are copied from \cite{Murphy2026}.}
    \label{tab:transittimes}
\end{table}

\subsection{Broadband Starspot Properties} \label{sec:results_bbfits}

\begin{figure}
    \centering
    \includegraphics[width=\linewidth]{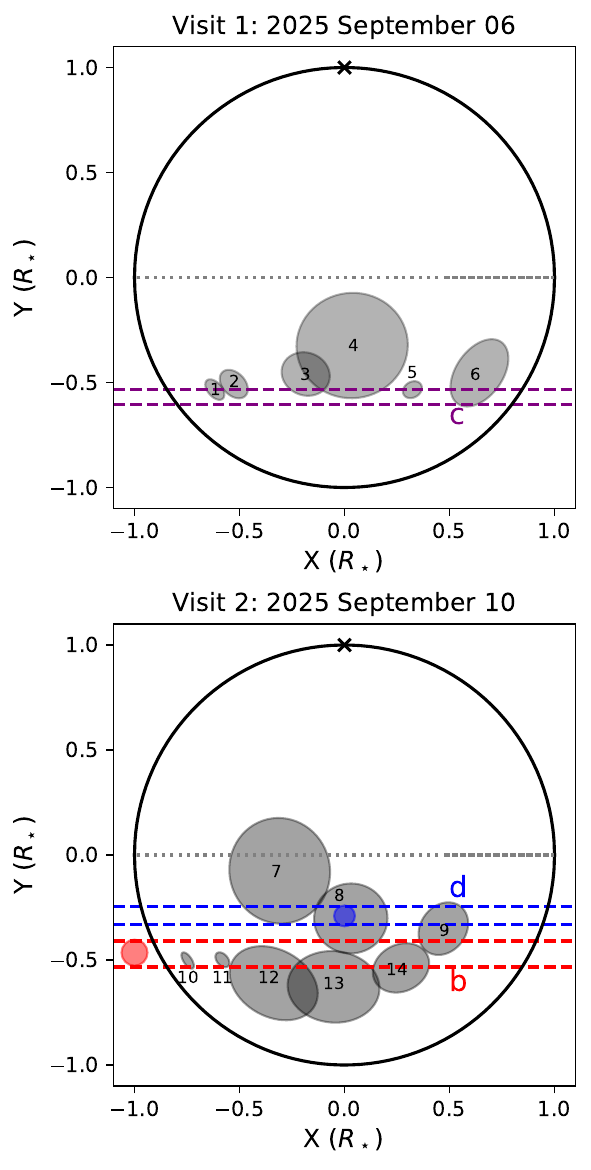}
    \caption{Occulted starspot distributions of V1298~Tau during the transit of planet c in Visit 1 (top) and transits of d and b in Visit 2 (bottom) derived from our broadband light curve fits with \texttt{fleck}, presented in Figure~\ref{fig:bbfits}. The purple, blue, and red dashed regions represent the transit chords of V1298~Tau~c, d, and b, respectively. All planets are assumed to transit from left to right, which is consistent with obliquity measurements for planets c and b \citep{feinstein2021_v1298tauGemini, johnson2022}. For Visit 2, we denote the relative transiting timing of planets d and b as colored circles. In both panels, the cross near the top represents the stellar pole assuming $i_\star=90^\circ$. Starspots are labeled by the adopted numbering scheme of this work. \href{https://github.com/kronos-jwst/KRONOS-II-V1298-Tau-Starspots/blob/main/Figure03.ipynb}{\githubicon}}
    \label{fig:fleckspotmap}
\end{figure}

Figure~\ref{fig:fleckspotmap} shows the best-fit occulted starspot map during Visits 1 and 2 from the broadband \texttt{fleck} fits. The starspots and transit chords are, by construction, always assumed to be in the lower hemisphere of the star as seen in Figure~\ref{fig:fleckspotmap}. At our achieved precision, we find Visit 1 is best-fit by six individual SCEs that are well-distributed along the transit chord of V1298~Tau~c \citep{Murphy2026}. Four days later, we identify at least eight additional starspots in Visit 2. Of these, one starspot solely intersected the transit chord of V1298~Tau~$d$, five solely intersected the chord of $b$, and two intersected both chords. We enumerate the spots across each visit as highlighted in Figure~\ref{fig:bbfits}. For reference, we also illustrate the approximate relative positions of V1298~Tau~d and b during Visit 2. 

Table~\ref{tab:spot_properties} lists the best-fit radii and positions of each starspot from our \texttt{fleck} fits. The radii of spots crossed during Visit 1 range from $R_\mathrm{spot, 1-6} = 0.04-0.27\,R_\star$ and from $R_\mathrm{spot,7-14} = 0.041-0.25\,R_\star$ for Visit 2. It is important to note that since \texttt{fleck} assumes all starspots are circular, whereas sunspots can often take more irregular shapes \citep[e.g.,][]{solanki2003_sunspots}, these radius values may be biased. If, for instance, an elongated cluster of smaller spots were present along the edge of the transit chord, our model could fit it using a single spot of large radius that is sufficiently offset from the chord so that the occulted area is the same. Therefore, our measured radii should be considered approximate with respect to the true sizes of these active regions. The relative radii of each planet may also have an impact on the inferred spot radii, as a larger planet occulting a given spot will create a larger apparent SCE.

A more grounded estimate of the physical starspot size can be derived from the FWHM crossing times of the Gaussian Profile fits, which we also list in Table~\ref{tab:spot_properties}. We measured FWHMs ranging from 9.6--29.4\,min, 19.7--48.4\,min, and 6.0--45.2\,min for the starspots occulted by c, d, and b, respectively. We convert the FWHM into approximate lengths $\ell$ of the occulted spot region based on the estimated orbital speed of each planet $v_p$. Based on the orbital parameters of \cite{livingston2026_v1298tau}, and assuming for simplicity that the velocity across the disk is always equal to the circular orbit tangential velocity, we estimate each planet's speed to be $v_{p,c}$=108.7\,km/s, $v_{p,d}$ = 94.8\,km/s, and $v_{p,b}$=75.9\,km/s. We multiply these speeds by the crossing time to get $\ell$. The results, in units of $R_\oplus$, are listed in Table~\ref{tab:spot_properties}. 

From this approximation, we find that the occulted regions span lengths of $\ell =$\,4.3--34.6\,$R_\oplus$. 58\% of the occulted regions are $> 20\,R_\oplus$. By comparison, the largest individual sunspots on the Sun are $\sim$9.5\,$R_\oplus$ \citep{solanki2003_sunspots}, though clusters of multiple closely-spaced sunspots can achieve significantly larger total lengths. However, it is likely these lengths for V1298~Tau are biased towards larger values by the relatively fast rotation period of V1298~Tau \citep[\prot=2.97\,day;][]{feinstein2022_v1298tauTESS} effectively smearing the length during the crossing event. This effect is most significant for starspots nearest 0$^\circ$ longitude where the projected arc length is greatest for fixed rotation, and with the longest crossing times. To correct for this, we estimate the apparent elongation $\Delta \ell$ due to the star rotating by an amount $\theta$ as:
\begin{equation}
    \Delta \ell = R_\star \left( \sin\left( lon + \theta\right) - \sin\left( lon \right) \right). \label{eqn:elongation}
\end{equation}
This $\theta$ can be determined as:
\begin{equation}
\theta = 360^\circ \times \left(\mathrm{FWHM} / P_{\mathrm{rot}}\right).    \label{eqn:theta}
\end{equation}
Assuming $R_\star$ = 1.276\,R$_\odot$ (Table~\ref{tab:star_specproperties}), we calculate $\Delta \ell$ for each starspot and subtract it from $\ell$ to derive a corrected length $\ell_{corrected}$. These values are also listed in Table~\ref{tab:spot_properties}, and correspond to reductions in $\ell$ by $\sim$1--10\,$R_\oplus$. With this correction, the maximum occulted region length is reduced to 24.7\,$R_\oplus$, which is still larger than seen in sunspots. Therefore, V1298~Tau either has significantly larger starspots than typically observed on the modern Sun, or we are observing the occultation of multiple closely-spaced spot clusters whose boundaries we cannot resolve. 

\begin{table*}[]
    \centering
    \caption{Occulted starspot geometrical properties from our broadband \texttt{fleck} and Gaussian Profile fits.}
    \begin{tabular}{c|c|c|c|c|c|c|c|c}
    Starspot & Occulter &  r$_{\mathrm{spot}}$ ($R_\star$) & $lat$ ($^\circ$) & $lon$ ($^\circ$) & $A$ (ppm) & FWHM (min) & $\ell$ ($R_\oplus$) & $\ell_{\mathrm{corrected}}$ ($R_\oplus$) \\ \hline 
    \multicolumn{8}{c}{\textit{Visit 1: 2025 September 06}} \\ \hline 
     1 & c & 0.05$^{+0.04}_{-0.02}$  & -32.7$^{+4.3}_{-4.0}$  & -47.2$^{+1.2}_{-1.0}$ & 282 $\pm$ 29 & 9.6$^{+1.5}_{-1.3}$ & 9.8 & 8.4 \\
     2 & c & 0.08$^{+0.04}_{-0.03}$ & -30.2$^{+3.6}_{-3.4}$ & -37.7$\pm$0.5 & 293 $\pm$ 28 & 12.0$^{+2.3}_{-1.8}$& 12.3 & 10.3\\
     3 & c &  0.13$^{+0.07}_{-0.06}$ & -25.8$^{+4.8}_{-4.3}$ & -11.9$^{+0.7}_{-0.6}$ & 166 $\pm$ 27 & 13.2$^{+3.5}_{-2.9}$& 13.5 & 10.9 \\
     4 & c & 0.27$^{+0.02}_{-0.03}$  & -18.8$^{+2.3}_{-3.1}$ & 2.2$\pm$0.4 & 271 $\pm$ 24 & 29.4$^{+2.9}_{-2.7}$& 30.0 & 24.0 \\
     5 & c & 0.04$^{+0.03}_{-0.01}$  & -32.6$^{+3.2}_{-2.7}$ & 22.5$\pm$0.4 & 237 $\pm$ 28 & 11.6$^{+2.3}_{-2.1}$& 11.9 & 9.7 \\
     6 & c & 0.09$^{+0.06}_{-0.02}$  & -30.9$^{+3.3}_{-3.0}$ & 46.1$\pm$0.4 & 434 $\pm$ 22 & 19.8 $\pm$ 1.8 & 20.2 & 17.4 \\ \hline 
     \multicolumn{8}{c}{\textit{Visit 2: 2025 September 10}} \\ \hline 
     7 & d & 0.25$\pm$0.01 & -4.3$\pm$0.6 & -18.0$\pm$0.1 & 687$\pm$17 & 27.3$\pm$0.7 & 24.4 & 19.1 \\
     8 & d & 0.175$^{+0.002}_{-0.001}$ & -17.6$\pm$0.1 & 1.8$\pm$0.1 & 877$\pm$15 & 48.4$\pm$1.3 & 34.6 & 24.7 \\
     8 & b & $\uparrow$                & $\uparrow$    & $\uparrow$  & --         & --           & N/A & N/A \\
     9 & d & 0.134$\pm$0.002 & -20.5$\pm$0.1 & 30.2$\pm$0.1 & 880$\pm$31 & 19.7$\pm$0.6 & N/A & N/A \\
     9 & b & $\uparrow$ & $\uparrow$ & $\uparrow$ & 601$\pm$21 & 28.6$\pm$1.7 & 25.5 & 20.5 \\ 
     10 & b & 0.045$^{+0.011}_{-0.003}$ & -30.2$^{+0.6}_{-1.2}$ & -59.7$\pm$0.4 & 105$\pm$32 & 6.0$^{+2.0}_{-12.1}$ & 4.3 & 3.7 \\
     11 & b & 0.041$^{+0.009}_{-0.002}$ & -29.9$^{+0.6}_{-1.1}$ & -42.1$\pm$0.3 & -- & -- & N/A & N/A \\
     12 & b & 0.223$\pm$0.007 & -37.6$\pm$0.4 & -25.2$\pm$0.1 & 1207$\pm$18 & 43.2$\pm$1.4 & 30.9 & 22.8 \\
     13 & b & 0.22$\pm$0.01 & -28.9$\pm$0.6 & -3.8$\pm$0.1 & 938$\pm$19 & 45.2$\pm$1.4 & N/A & N/A \\
     14 & b & 0.139$\pm$0.004 & -32.5$\pm$0.3 & -18.61$\pm$0.07 & 1490$\pm$19 & 29.8$\pm$0.8 & 21.3 & 15.5
    \end{tabular}
    \label{tab:spot_properties}
    \tablecomments{All values here are measured from the band-integrated NIRISS/SOSS light curves of their respective visits. Some parameters, such as the amplitudes $A$, have further wavelength-dependence. The $A$ and FWHM for spot 11 (b occultation) is not listed as it was simultaneous with spot 9 (d occultation), so we fit them as a single Gaussian profile and exclude them from the $\ell$ calculations. The same is true for spots 8 (b occultation) and 13 (b occultation). Latitudes are all assumed to be in the lower visible hemisphere ($<$0$^\circ$), so their values should only be interpreted with respect to the equator. Longitudes are relative to the visible disk at that particular visit, and are $\sim$196$^\circ$ offset between visits based on \prot=2.97\,day \citep{feinstein2022_v1298tauTESS}.}
\end{table*}

The starspot latitudes are poorly constrained by our fits, as there is symmetry about both the stellar equator and the transit chord. The latter is particularly true for Visit 1 with only one transit. Visit 2 has more constraining power in this respect, as a starspot crossed by one planet but not the other helps constrain its relative position. Since the orbits of each planet are not oblique, these relative latitudes should not be significantly biased by rotation. As shown in Figure~\ref{fig:fleckspotmap}, the chord of d is nearer to the stellar equator than the chord of b. Based on this, we are more confident in spots 7--9 being nearer to the stellar equator than spots 10--14. Considering the ambiguity in true $R_\mathrm{spot}$, we use the lower boundary of the transit chord of b as a limit to estimate the overall range in latitudes that these starspots reside in. Our results suggest that V1298~Tau has abundant starspots within $\pm$33$^\circ$ of its equator. This is analogous to the latitudes along which sunspots preferentially form and remain on the Sun \citep[][]{maunder1922, solanki2003_sunspots}. However, it is important to note that our results are biased to the extent of transit chords, as these observations cannot detect starspots at higher latitudes than the transit chords even if they are abundant. 

We can also estimate the presence of spots at latitudes outside the transit chords based on the best-fit radii. Neglecting overlap, we convert the best-fit \texttt{fleck} radii into areas and sum them together for each visit to determine the coverage fraction $f_{\mathrm{occulted}}$ of these occulted spots on the visible hemisphere of V1298~Tau. We find $f_{\mathrm{occulted}}$=0.06$\pm$0.01 for the Visit 1 spots and $f_{\mathrm{occulted}}$=0.117$\pm$0.004 for the Visit 2 spots. Compared to the $f_{\mathrm{cool}}$ values derived from the stellar spectrum (Table~\ref{tab:star_specproperties}), these occulted spots represent $\sim$27--31\% of all spots on the surface during Visit 1, and $\sim$64--80\% during Visit 2. 

\section{Starspot Contrast Spectra}\label{sec:contrast}

\subsection{Spectroscopic Light Curve Fit Setup}
For the spectroscopic light curve fits, we fix the time of conjunctions to the broadband-derived values and again fix the orbital parameters. We also fix the $b$ and $\sigma$ of each SCE in the Gaussian Profile fits, and $r_{spot}$, $lat$, and $lon$ of each SCE in the \texttt{fleck} fits. Then, we fit for the planet-star radius ratios, limb darkening coefficients, baseline parameters, uncertainty scaler, and either the SCE Gaussian amplitudes $A$ or the \texttt{fleck} contrast $\alpha$. We fit each spectroscopic light curve independently, running each fit for a 2,000 step burn-in period and a 2,500 step production run (30$\times$ the average autocorrelation time). In this work, we focus on the resulting chromatic starspot properties. The transmission spectra will be presented in \cite{Murphy2026}, Mukherjee et al. (in preparation), Petz et al. (in preparation), and Shivkumar et al. (in preparation).

\subsection{Light Curve Fit Results}
\begin{figure}
    \centering
    \includegraphics[width=\linewidth]{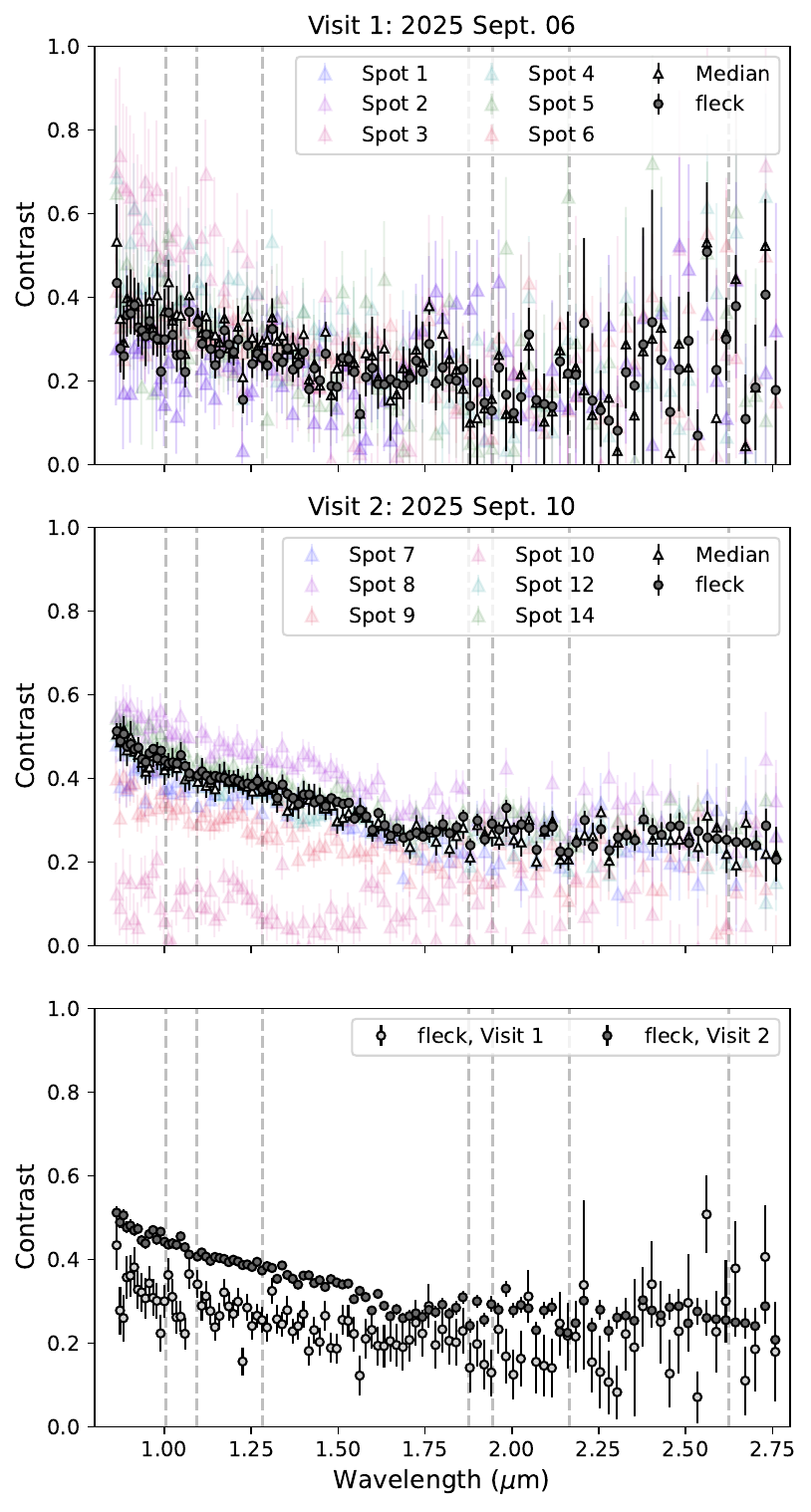}
    \caption{Observed starspot contrasts as a function of wavelength from the SCEs during Visit 1 (top) and Visit 2 (middle). Dark gray circles represent the \texttt{fleck} spectra, which assume all spots have the same contrast. Colored triangles represent the Gaussian-derived contrasts, and the gray triangles their median.  The bottom panel compares the \texttt{fleck} results for both visits. We indicate various hydrogen spectral lines as the vertical gray lines for reference. The contrast spectra from each visit show comparable structure: a slope at $\lambda \leq 1.7\mu$m and relatively flat at $\lambda > 1.7\mu$m. Comparing the \texttt{fleck}-derived spectra between Visits 1 and 2, we find that $\alpha_\textrm{visit 2} \sim \alpha_\textrm{visit 1} + 0.1$ across all wavelengths. The increased precision in measured contrast spectra in Visit 2 is due to the larger SCE amplitudes. \href{https://github.com/kronos-jwst/KRONOS-II-V1298-Tau-Starspots/blob/main/Figure04.ipynb}{\githubicon}
    }
    \label{fig:contrasts}
\end{figure}

By measuring the SCE properties as a function of wavelength, we derive starspot contrast spectra. Figure~\ref{fig:contrasts} shows our derived starspot contrast spectra at $R = \frac{\lambda}{\Delta \lambda} = 100$ from the Gaussian fits and \texttt{fleck}; we show both as the \texttt{fleck} fits assume the same contrast across all spots while the Gaussian fits are agnostic. The derivation of contrast spectra from the Gaussian fits follows the methods of \cite{murray2025_SCEeffects}, considering just the maximum overlap of the planet's shadow and the spot estimated from Figure~\ref{fig:fleckspotmap}. 
We exclude spot 11 as its crossing occurred simultaneously with spot 9 (by V1298~Tau~d), though spot 9 was later crossed again by V1298~Tau~d providing an isolated measurement. We similarly exclude spot 13 as it was crossed simultaneously with spot 8 by V1298~Tau~b. The contrast spectra for Visit 2 are $\sim 3-4\times$ more precise than Visit 1 due to the larger amplitudes of the SCEs in this observation. The overall shape of the contrast spectra are consistent between visits, particularly in the slope at $\lambda \lesssim 1.7\,\mu$m. Comparing the \texttt{fleck}-derived spectra across visits, we find that Visit 2 has a consistently larger contrast by $\Delta \alpha \sim0.1$ compared to Visit 1. 

\begin{figure}
    \centering
    \includegraphics[width=\linewidth]{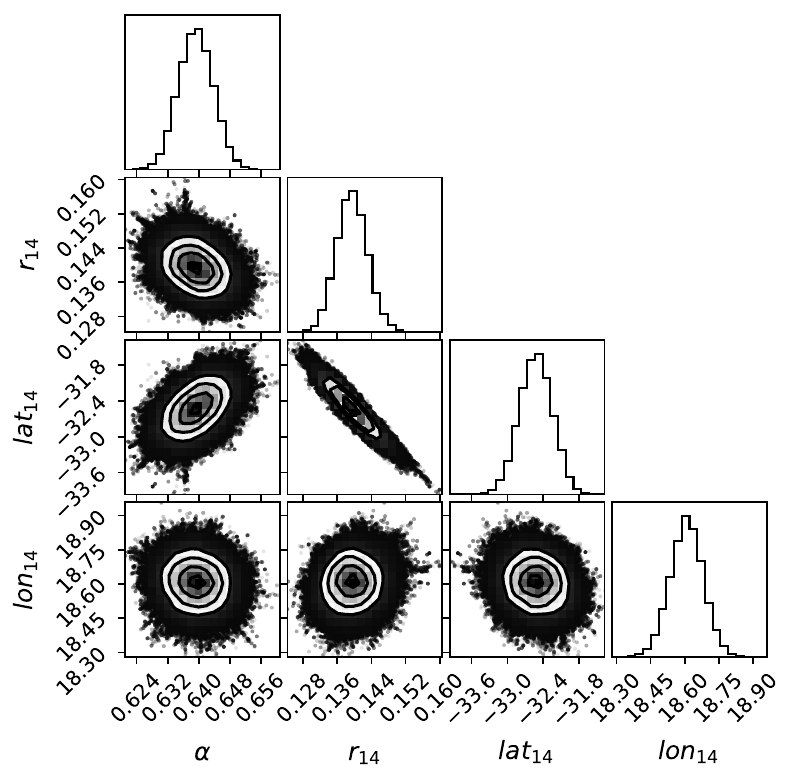}
    \caption{Corner plot for a subset of starspot parameters from our \texttt{fleck} fit to the Visit 2 broadband light curve. The radius and latitude of starspot 14 exhibit weak correlation with the starspot contrast $\alpha$. \href{https://github.com/kronos-jwst/KRONOS-II-V1298-Tau-Starspots/blob/main/Figure05.ipynb}{\githubicon}}
    \label{fig:spotcorrelations}
\end{figure}

Previous JWST observations of a persistent starspot on the star TOI-3884 measured visit-to-visit variations in absolute contrast of a similar magnitude \citep{murray2026}. These can be attributed to both astrophysical causes, such as differing contributions from unresolved umbra, penumbra, and faculae, or systematic causes, such as degeneracies with the spot positions and other light curve parameters \citep{murray2026}. We found that most of the starspot parameters display no strong correlation with the contrast, except for the radius and latitude of spot 14  which exhibit weak correlation (Figure~\ref{fig:spotcorrelations}). It is therefore possible that either cause, or combination of the two, drives the observed contrast offset. Regardless of the offsets, we are able to derive physically-informative properties of the photosphere of V1298~Tau.

We compare the median Gaussian-derived contrast spectra to the \texttt{fleck} spectra. For Visit 2, we exclude spot 10 from the median calculation due to its anomalously low and featureless contrast spectrum, likely due to its small size and timing near the ingress of V1298~Tau~b. While the overall shapes are consistent with the \texttt{fleck} spectra, the Gaussian-derived spectra are less precise by a factor of $\sim$2. This suggests that by enforcing all spots to have the same contrast, \texttt{fleck} is able to provide a more precise median starspot contrast. However, we do find variations between the individual Gaussian-derived starspot spectra. In Visit 1, we see variation in both the shape and absolute contrast, particularly at $\lambda\lesssim$1.7\,$\mu$m. Some of the spots are consistent with the median behavior, but some exhibit significantly stronger slopes (e.g. spots 3 and 5). This may be due to the smaller amplitudes of the SCEs, and thus lower signal-to-noise ratios, in the Visit 1 light curve. There is also considerable overlap between the profiles of spots 1 and 2, and spots 3 and 4 (Figure~\ref{fig:bbfits}) that may bias their individual results, as this overlap makes it more difficult to determine the precise width and amplitude of the individual profiles. On the other hand, in Visit 2, the individual contrast spectra exhibit the same shape as the median and vary only in absolute contrast. The only exception is spot 10. 

We also indicate the wavelengths of several prominent Hydrogen spectral lines in Figure~\ref{fig:contrasts} as the vertical gray lines. These include the Paschen series and Brackett $n$=6--9. There are no elevated contrasts near any of these wavelengths, making a flare-origin unlikely.

\subsection{Modeling Starspot Contrast Spectra} \label{subsec:contrast_modelsetup}

To model the observed starspot contrasts, we assume the underlying contrast spectra can be represented by combinations of stellar spectra of the appropriate temperature. We again draw stellar spectra from the NewEra grid assuming fixed $\log\left(g\right)$ = 4.25 and [Fe/H] = 0.1. First, following the method typically used in exoplanet analyses \citep[e.g.,][]{Fu2022, murray2026}, we test a 2-component (spot and photosphere) model with parameters $T_{\mathrm{phot}}$ and $T_{\mathrm{spot}}$, computing the model contrast spectrum $\alpha_\lambda$ at each step as:
\begin{equation}
    \alpha_\lambda = 1 - \frac{F_\mathrm{spot, \lambda}}{F_\mathrm{phot, \lambda}}. \label{eqn:contrast_2comp}
\end{equation}
In addition, motivated by the knowledge that sunspots cannot be described by a single temperature, we test a 3-component (photosphere, spot umbra, spot penumbra) model with parameters $T_{\mathrm{phot}}$, \tumb, \tpen, and an umbral area fraction $f_{\mathrm{umb}}$. In this case, the model contrast is given by:
\begin{equation}
    \alpha_\lambda = 1 - \frac{f_\mathrm{umb}  F_\mathrm{umb, \lambda} + \left(1 - f_\mathrm{umb} \right)F_\mathrm{pen, \lambda}}{F_\mathrm{phot, \lambda}}. \label{eqn:contrast_3comp}
\end{equation}


We fit each model using the MCMC method with \texttt{emcee}, running for 25000 steps after a 2500 step burn-in period ($>$85$\times$ the average autocorrelation time for the 3-component fits). We enforce a Normal prior on $T_\mathrm{phot}$ = 4866$\pm$33\,K, based on the Visit 2 out-of-transit stellar spectrum (Table~\ref{tab:star_specproperties}), and enforce that \tumb\,$<$\,\tpen\,$< T_{\mathrm{phot}}$, that all temperatures remain in a wide range (2600\,K, 6000\,K) of plausible values, and that $f_{\mathrm{umb}}$ remain in the range (0,1). We focus just on the \texttt{fleck} results due to their lower uncertainties, and the good agreement between \texttt{fleck} and the median Gaussian-derived results. We return to fitting the individual Gaussian results later.

\subsection{Contrast-Derived Starspot Properties} \label{subsec:contrast_modelresults}

\begin{figure*}
    \centering
    \includegraphics[width=\linewidth]{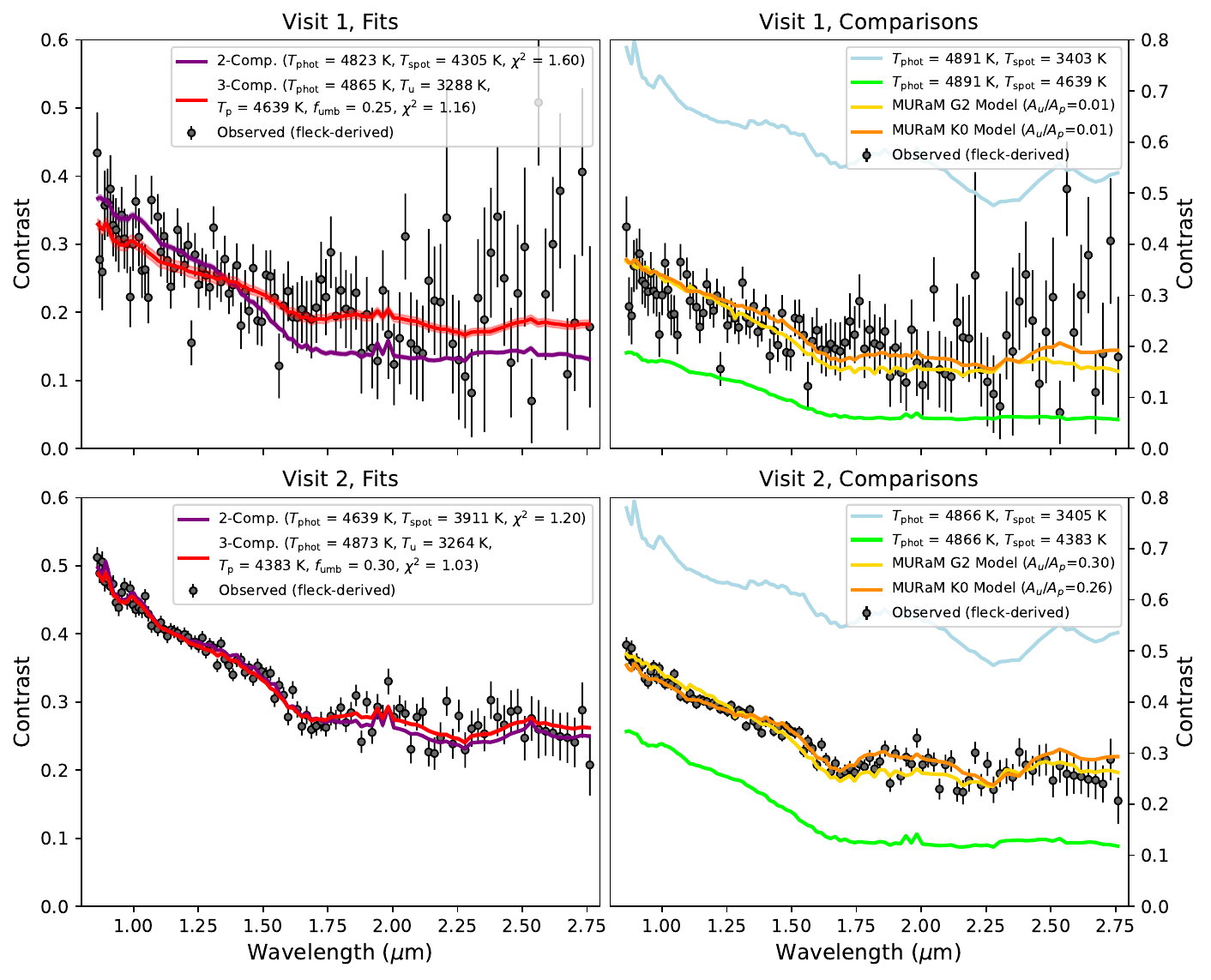}
    \caption{Multi-component fits (left) and model comparisons (right) to the \texttt{fleck}-derived starspot contrasts from Visit 1 (top) and 2 (bottom). The observed contrasts from both JWST visits are best-fit by 3-component models which include umbral and penumbral components that are 3264--3288\,K and 4383--4639\,K, respectively (red). The photospheric temperature is consistent within $\Delta T = 8$\,K between both visits. The 2-component fits (purple) poorly reproduce the slope at $\lambda < 1.7\mu$m for Visit 1 and the shape between 1.7--2.5\,$\mu$m for both visits, and have inconsistent temperatures. We additionally compare our observations to models based on the out-of-transit spectral fits (blue, green) and the 3D MURaM stellar models (orange/yellow). Our observations agree particularly well with the MURaM model for a K0 star \citep{Smitha2024_MURaMspotmodels}.  \href{https://github.com/kronos-jwst/KRONOS-II-V1298-Tau-Starspots/blob/main/Figure06.ipynb}{\githubicon}}
    \label{fig:contrast_fits}
\end{figure*}

We present our best-fit 2- and 3-component contrast models in Figure~\ref{fig:contrast_fits}. The best-fit temperatures for the 2-temperature component models are $T_{\mathrm{phot}}$ = 4823$\pm$40\,K and $T_{\mathrm{spot}}$ = 4305$^{+32}_{-36}$\,K for Visit 1, and $T_{\mathrm{phot}}$ = 4639$\pm$11\,K and $T_{\mathrm{spot}}$ = 3911$\pm$8\,K for Visit 2. These $T_{\mathrm{spot}}$ values are each $>$20$\sigma$ discrepant from the stellar spectrum-fit values (Table~\ref{tab:star_specproperties}), and are $>10\sigma$ discrepant with each other. In addition to this self-inconsistency, such variation in spot temperature is likely physically implausible. Moreover, the 8\,K uncertainty on Visit 2's spot temperature is implausibly small. The $T_{\mathrm{phot}}$ values are 4.4$\sigma$ discrepant between visits, which is also unphysical. The corresponding photosphere-spot temperature difference is 518$\pm$52\,K for Visit 1, and 728$\pm$13\,K for Visit 2. Particularly for Visit 1, the 2-component model overpredicts the slope in contrast at $\lambda\lesssim1.7\,\mu$m. For both visits, the 2-component model underpredicts the contrast between 1.7--2.5$\mu$m.

Additionally, we compare our observed contrast spectra to 2-component forward models generated based on the derived properties of our out-of-transit stellar spectrum fit. The blue forward models in the right-hand panel of Figure~\ref{fig:contrast_fits} are generated with the corresponding $T_{\mathrm{phot}}$ and $T_{\mathrm{spot}}$=$T_{\mathrm{cool}}$ derived from the out-of-transit stellar spectra. These models significantly over-predict the contrast. The discrepancy between the out-of-transit stellar and the contrast spectra fits is further explored in Section~\ref{subsec:discussion_SCEspecdiffs}.

\begin{table}[]
    \centering
    \caption{3-component starspot contrast fit results.}
    \begin{tabular}{c|c|c|c|c}
    Starspot & $T_{\mathrm{phot}}$ (K)  & \tumb\ (K) & \tpen\ (K) & $f_{\mathrm{umb}}$   \\ \hline 
    \multicolumn{5}{c}{\textit{Visit 1: 2025 September 06}} \\ \hline 
    All & 4865$\pm$34 & 3288$^{+355}_{-513}$ & 4639$^{+95}_{-80}$ & 0.25$^{+0.14}_{-0.07}$ \\ 
    1 & 4854$\pm$33 & 3040$^{+264}_{-266}$ & 4748$\pm$57 & 0.26$^{+0.05}_{-0.04}$ \\
    2 & 4857$\pm$33 & 3023$^{+269}_{-258}$ & 4718$\pm$66 & 0.33$^{+0.06}_{-0.05}$ \\
    3 & 4883$\pm$33 & 4122$^{+60}_{-81}$ & 4596$^{+101}_{-104}$ & 0.90$^{+0.07}_{-0.15}$ \\
    4 & 4872$^{+34}_{-32}$ & 3976$^{+124}_{-213}$ & 4585$^{+105}_{-102}$ & 0.71$^{0.18}_{-0.22}$ \\
    5 & 4870$\pm$32 & 3628$^{+323}_{-453}$ & 4600$^{+89}_{-86}$ & 0.36$^{+0.19}_{-0.14}$ \\ 
    6& 4863$^{+33}_{-32}$ & 3115$^{+310}_{-320}$ & 4666$^{+63}_{-59}$ & 0.30$^{+0.09}_{-0.05}$ \\ \hline 
    \multicolumn{5}{c}{\textit{Visit 2: 2025 September 10}} \\ \hline 
    All & 4873$\pm$33 & 3264$^{+205}_{-246}$ & 4383$^{+55}_{-46}$ & 0.30$^{+0.09}_{-0.06}$ \\ 
    7 & 4863$\pm$31&3171$^{+246}_{-236}$&4412$\pm$41&0.24$^{+0.07}_{-0.04}$ \\
    8 & 4873$\pm$32&3404$^{+159}_{-275}$& 4371$^{+53}_{-49}$ & 0.48$\pm$0.12 \\ 
    9 & 4843$\pm$30 & 3172$^{+197}_{-201}$&4472$\pm$35 & 0.18$\pm$0.04 \\
    10& 4798$^{+30}_{-31}$ & 3237$^{+230}_{-231}$ & 4662$\pm$30&0.03$\pm$0.02 \\
    12 & 4866$^{+32}_{-30}$ & 3344$^{+188}_{-231}$ & 4399$^{+41}_{-39}$ & 0.28$^{+0.08}_{-0.06}$ \\
    14 & 4880$\pm$29 & 3322$^{+153}_{-197}$ & 4366$^{+41}_{-39}$ & 0.34$^{+0.08}_{-0.06}$ \\
    \end{tabular}
    \label{tab:spot_contrastproperties}
    \tablecomments{\tumb\ refers to the spot umbra and \tpen\ refers to the spot penumbra. The ``all'' result refers to the fit of the \texttt{fleck}-derived contrasts as it enforces all starspots to have the same properties. The individual starspot values were derived from applying normal priors on $T_{\mathrm{phot}}$,  \tumb, and \tpen\ based on the ``all'' result for that visit.
    }
\end{table}

This discrepancy at nearly all wavelengths further motivates increasing the complexity of our fit to the sunspot-inspired 3-temperature component model, consisting of $T_{\mathrm{phot}}$, \tumb\, and \tpen\ components. The results of these fits are tabulated in Table~\ref{tab:spot_contrastproperties}. For Visit 1, we find $T_{\mathrm{phot}}$ = 4865$\pm$34\,K, \tumb\,=3288$^{+355}_{-513}$\,K, \tpen\,=4639$^{+95}_{-80}$\,K, and $f_{\mathrm{umb}}$= 0.25$^{+0.14}_{-0.07}$. For Visit 2, we find $T_{\mathrm{phot}}$ = 4873$\pm$33\,K, \tumb\,=3264$^{+205}_{-246}$\,K, \tpen\,=4383$^{+55}_{-46}$\,K, and $f_{\mathrm{umb}}$=0.30$^{+0.09}_{-0.06}$. These values correspond to temperature differences between the photosphere and spot umbra of $\Delta T_{\mathrm{umb}}$ = 1577$^{+514}_{-357}$\,K for Visit 1 and 1609$^{+248}_{-208}$\,K for Visit 2. The differences between the photosphere and spot penumbra are $\Delta T_{\mathrm{pen}}$ = 226$^{+87}_{-101}$\,K for Visit 1 and 490$^{+57}_{-64}$\,K for Visit 2. 

\begin{figure}
    \centering
    \includegraphics[width=\linewidth]{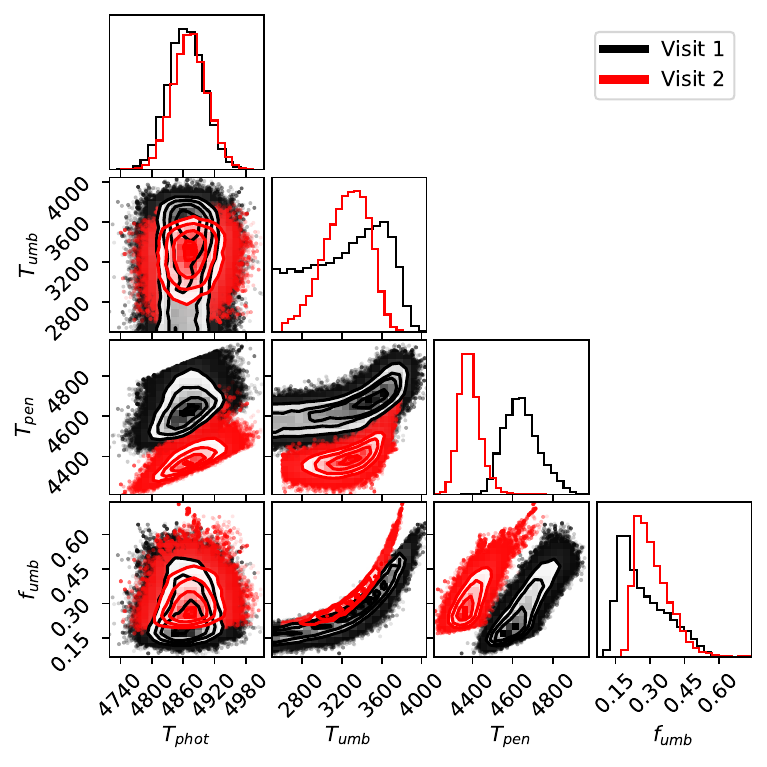}
    \caption{Posterior distributions from fitting the starspot contrasts measured from Visit 1 (black) and Visit 2 (red) with a 3-temperature component model. The spot components are correlated in both cases. For Visit 1, the lower limit of \tumb\ is not well constrained, though the posterior peaks strongly near that of Visit 2. With the better precision of Visit 2, each component is well constrained. \href{https://github.com/kronos-jwst/KRONOS-II-V1298-Tau-Starspots/blob/main/Figure07.ipynb}{\githubicon}}
    \label{fig:spottempposteriors}
\end{figure}

Figure~\ref{fig:spottempposteriors} shows the joint posteriors for these 3-component models. For Visit 1, the lower bound of \tumb\ is not well constrained, though the posterior peaks strongly near 3600\,K, higher than the median value of 3288\,K. This is likely due to the poorer precision of the contrasts measured during Visit 1. The other parameters are well constrained. On the other hand, for Visit 2, all parameters including \tumb\ are well constrained. In both cases, there is correlation between each of the spot properties. Despite this, particularly for Visit 2, we are able to constrain the umbral and penumbral properties from their combined contrast.

When using the 3-component model, we find $T_{\mathrm{phot}}$ to be in excellent agreement between visits and with the out-of-transit spectrum. The spot component temperatures (\tumb\, and \tpen) are up to $\Delta T$=256\,K colder for Visit 2; this difference is driven by the offset in contrast spectra between Visits 1 and 2. The \tumb\ posteriors are relatively consistent, considering the lower tail for Visit 1. The \tpen\ values differ by 2.6$\sigma$, while the umbra-to-penumbra ratios are in good agreement. Our derived \tumb\ values are similar to the $T_{\mathrm{cool}}$ component in the out-of-transit stellar spectrum fit, suggesting that this component may be dominated by starspot umbrae. Our 3-component models fit the data well at all wavelengths with a reduced $\chi^2$ value of 1.16 and 1.03 for Visits 1 and 2, respectively, with 102 degrees of freedom. By comparison, the 2-component model fits were relatively poorer with reduced $\chi^2$ values of 1.6 and 1.2 for Visits 1 and 2, respectively, with 104 degrees of freedom. Additionally, between 1.7--2.5$\mu$m, the median residual for the 2-component model is 4.4$\times$ and 3$\times$ higher than that of the 3-component model for Visits 1 and 2, respectively.

The 3-component contrast models are also statistically preferred over the 2-component models. Between the best-fit 3- and 2-component models, we find $\Delta|$BIC$|$ = 38.6 for Visit 1 and $\Delta|$BIC$|$ = 10.5 for Visit 2, both in favor of the 3-component model. These approximately correspond to Bayes factors of $\ln\left(B\right)$ = 19.3 and 5.2 \citep{bayarri2019}, meeting the typical threshold for ``very strong'' evidence \citep[e.g.,][]{kass_bayes_1995}.

For comparison, we generate additional 2-component forward models again assuming $T_{\mathrm{phot}}$ derived from the out-of-transit spectral fit but $T_{\mathrm{spot}}$ = \tpen\ derived from these new 3-component contrast fits. The resulting contrast spectra underpredict the observed contrast (Figure~\ref{fig:contrast_fits}, lime green). We therefore conclude that no 2-temperature component model can explain the observed starspot contrasts while attaining $T_{\mathrm{phot}}$ that is both consistent between visits and with the out-of-transit stellar spectrum observed within the same visit. The inability of the 2-component models to fully explain either data set, combined with the inconsistency between the best-fit parameters, suggests that our observed starspot contrasts are sensitive to the contributions of the spot umbrae and penumbrae. 

To verify the robustness of our result to the specific stellar model grid used, we repeated these fits using model spectra from the Phoenix grid \citep{Husser2013_phoenix}. The results are in agreement with the NewEra-based results. From the 2-component fits, we find $T_{\mathrm{phot}}$=4817$\pm$40\,K and $T_{\mathrm{spot}}$=4301$\pm$39\,K for Visit 1, and $T_{\mathrm{phot}}$=4661$\pm$12\,K and $T_{\mathrm{spot}}$=3937$\pm$9\,K for Visit 2. From the 3-component fits, we find $T_{\mathrm{phot}}$=4867$\pm$33\,K, \tumb\,=3419$^{+271}_{-450}$\,K, \tpen\,=4624$^{+101}_{-67}$\,K, and $f_\mathrm{umb}$=0.26$^{+0.13}_{-0.08}$ for Visit 1, and $T_{\mathrm{phot}}$=4881$\pm$32\,K, \tumb\,=3151$^{+218}_{-268}$\,K, \tpen\,=4367$^{+41}_{-39}$\,K, and $f_\mathrm{umb}$=0.26$^{+0.07}_{-0.04}$ for Visit 2. The individual values are generally consistent with the NewEra results, and these convey the same qualitative result. The 2-component temperatures are scattered, physically inconsistent, and do not well explain the observed contrasts. The umbral and penumbral components are needed.

Beyond our empirically-derived contrast fits, we compare our observations to the starspot models of \cite{Smitha2024_MURaMspotmodels}, generated using the 3D radiative magnetohydrodynamics code MURaM \citep{Vogler2005_muram}. The available MURaM grid contains precomputed photosphere, umbral, and penumbral spectra for G2V and K0V stars.\footnote{The MURaM models are available at the Max Planck Digital Library \citep[doi=10.17617/3.HS2EE6;][]{MURaM_onlinespectrarepo}.} We converted these into contrast spectra following Equation~\ref{eqn:contrast_3comp} along a grid of $f_\mathrm{umb}$ values between 0.01 and 0.99 in steps of 0.01. For each spectral type, we calculate the $\chi^2$ value between each model and the observed spectrum for each visit to determine the values of $f_\mathrm{umb}$ that minimize the $\chi^2$. The results are shown as the yellow and orange models for the G2 and K0 spectra, respectively, in Figure~\ref{fig:contrast_fits}.

For the Visit 1 data, the best-fit $f_\mathrm{umb}$ = 0.01 for both K0 and G2 models. Both provide similar, adequate fits to the data with reduced $\chi^2$ = 1.33, 1.41 for the K0 and G2 models respectively. The derived $f_\mathrm{umb}$ is $3.4\sigma$ lower than the value derived from fitting the contrasts with the \texttt{NewEra} models ($f_\mathrm{umb} =0.25^{+0.14}_{-0.07}$). On the other hand, for Visit 2, the best-fit $f_\mathrm{umb}$ = 0.26, 0.30 for the K0 and G2 models, respectively; these values are consistent with the \texttt{NewEra}-based fit values within $\sim$0.7$\sigma$. With this $f_\mathrm{umb}$, the K0 model fits the observed contrasts from Visit 2 well with reduced $\chi^2$ = 1.4, and a median absolute residual of 0.65$\sigma$. The model begins to slightly underpredict the slope at $\lambda \lesssim 1.25\,\mu$m, and overpredicts the contrast at $\lambda \gtrsim 2.5\,\mu$m. By contrast, the G2 model overpredicts the slope $\lambda \lesssim 1.7\,\mu$m, and underpredicts the contrasts at longer wavelengths. The better fit of the K0 model is unsurprising given that, due to its young age, V1298~Tau is also spectral type K0--K1 \citep{david19_b, suarez_mascareno_2022} despite being Solar mass \citep[$M_\star$=1.095\,$M_\odot$;][]{feinstein2022_v1298tauTESS}. The origin of the discrepancy in $f_\mathrm{umb}$ for Visit 1 is unclear. However, the consistency in $f_\mathrm{umb}$ between these different modeling approaches for Visit 2 further raises our confidence that our observed contrasts are truly sensitive to the starspot umbral and penumbral regions.

We note a potential caveat that although our contrast spectra are best fit when including umbra and penumbral components, the underlying \texttt{fleck} light curve model assumes the spots are uniform (i.e., a single ``contrast'' component). This may potentially affect the contrast values that we determine from the SCEs, though it is unclear whether the crossings of umbral and penumbral regions could be differentiated in the light curve given current data precision. We encourage exploration of this problem in future work. 

\subsection{Testing the Robustness of the Derived Spot Properties}

As seen in Figure~\ref{fig:bbfits}, there is a limited amount of post-transit baseline in Visit 2. This may affect the measured transit depth of V1298 Tau b and the derived SCE amplitudes as the out-of-transit baseline-model parameters have limited data to fit to. To check for potential biases in our derived SCE amplitudes, we refit the Visit 1 observations assuming a limited post-transit baseline of the same duration as what is available in Visit 2. Specifically, we refit Visit 1 clipping all out-of-transit integrations except the 420 integrations immediately preceding the transit ingress of V1298~Tau~c. We repeat the same fitting technique and analysis as previously described. We find the derived spot contrast and latitudinal/longitudinal locations are $\sim1\sigma$ consistent with the results presented in Table~\ref{tab:spot_properties}. We note that the $R_p/R_\star$ increased by $\sim4\sigma$. We conclude that while the absolute transit depths of V1298~Tau~d and b may be biased by the limited post-transit baseline, the inferred properties of the starspots are robust to the conditions of the data.

We also checked how our priors affect the results of the 3-component fits. When increasing the standard deviation of our $T_{\mathrm{phot}}$ by 3$\times$ and rerunning the fit for Visit 2, we recover consistent values for each parameter: $T_{\mathrm{phot}}$=4926$^{+83}_{-92}$\,K, \tumb\,=3239$^{+208}_{-271}$\,K, \tpen\,=4435$^{+91}_{-104}$\,K, and $f_\mathrm{umb}$=0.29$^{+0.09}_{-0.06}$. We also performed a prior-predictive check by drawing 1000 random values for each parameter from the assumed prior distributions, computing the corresponding model contrast, and comparing these to our data. These drawn models filled the volume of physically-allowed contrasts between [0,1] at each wavelength, and only 0.6\% of these draws yielded a reduced $\chi^2 < 2$. Altogether, these tests show that our assumed priors do not drive our results.

We verified that we are not underestimating the uncertainties on the individual contrast measurements. We added a multiplicative uncertainty scaling term to the contrast fits which acts uniformly across wavelengths, then repeated each fit. We recover very consistent results from this test. The best-fit temperatures and area fractions are each within 0.4$\sigma$ with the results from the unscaled fits (Table~\ref{tab:spot_contrastproperties}), though the uncertainties on each value do increase by up to $\sim$20\%. The best-fit scalers are 1.27$\pm$0.09 and 1.07$^{+0.08}_{-0.07}$ for Visit 1's 2- and 3-component fits, respectively, and 1.12$^{+0.09}_{-0.08}$ and 1.0$\pm$0.07 for Visit 2, respectively.

Finally, we combined the baseline check with the uncertainty scaling test. When fitting the Visit 1 light curve assuming the truncated baseline of Visit 2, the best-fit broadband contrast changed by $0.02$, which was a 0.7$\sigma$ difference. We added this change as an error inflation term in quadrature to the individual Visit 2 contrast values, then repeated the 3-component fit including the uncertainty scaling term. The results were again each consistent and the best-fit scaling term was $0.60 \pm 0.04$, which effectively undid the error inflation.

\section{Global Starspot Analysis} \label{sec:lcophotometry}

\begin{figure}
    \centering
    \includegraphics[width=\linewidth]{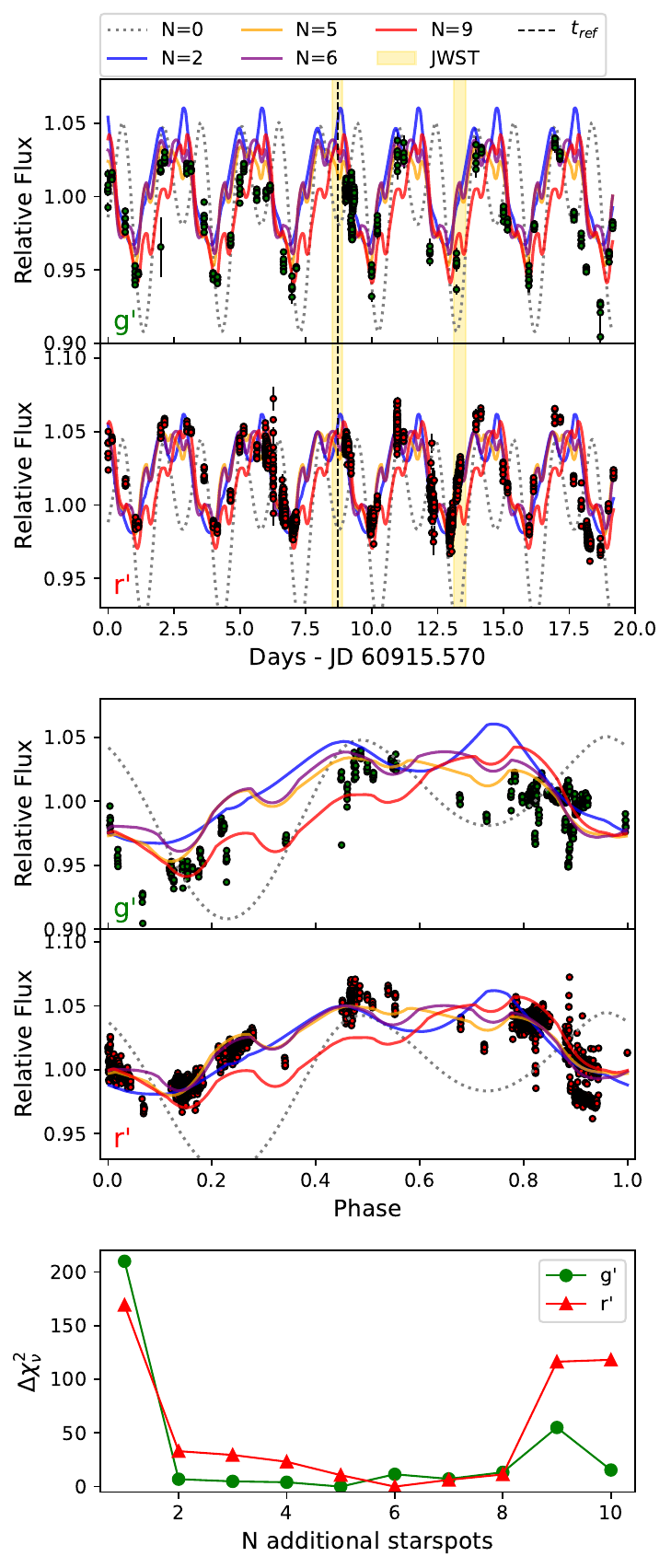}
    \caption{Rotational light curves of V1298~Tau in g$\prime$ and r$\prime$ filters from LCOGT/Sinistro between 2025 August 28--September 16 chronologically (top subfigure) and phase-folded (middle subfigure). The yellow regions indicate contemporaneous JWST observations. We fit these with \texttt{fleck}, fixing the occulted starspots ($N$=0) and adding $N$ additional spots, showing results for select $N$ showing well-fitting values and ends of each range. The bottom subfigure shows the relative reduced $\chi^2$ values as a function of $N$. The fits achieve minimum $\chi^2_\nu$ with $N$=5 for g$\prime$ and N=6 for r$\prime$. \href{https://github.com/kronos-jwst/KRONOS-II-V1298-Tau-Starspots/blob/main/Figure08.ipynb}{\githubicon}
    }
    \label{fig:lco_fits}
\end{figure}

While our transit observations can provide insights into the heterogeneity properties along the transit chord, we can contextualize these properties with a global view of V1298~Tau using the contemporaneous multi-band LCOGT/Sinistro photometry surrounding JWST Visits 1 and 2 \citep{brown2013}. Figure~\ref{fig:lco_fits} shows the g$\prime$ and r$\prime$ light curves observed with LCO. We highlight the times of Visits 1 and 2 with respect to each light curve. The light curves in both bands exhibit modulation at the expected \prot~=~2.97~day \citep{feinstein2022_v1298tauTESS}. 

Physically, this g$\prime$ and r$\prime$ modulation is driven by the rotation of starspots into and out of our line of sight \citep[e.g.,][]{Morris2020_spotsVSage}. Therefore, combined with our constraints on the occulted starspots from JWST, these data enable investigation into the approximate prevalence and distribution of additional unocculted starspots on all faces of V1298~Tau around the time of the JWST observations. 

To this end, we modeled the LCOGT time series using \texttt{fleck}. We set \prot = 2.97\,day \citep{feinstein2022_v1298tauTESS}. One key simplifying assumption is that the \texttt{fleck} framework does not model differential rotation. Previous empirical estimates for the equator--pole difference in rotation period range from $\sim$0.1\,day \citep{finociety_2023} to $\sim$0.71\,day \citep{maio2024}, based on which our assumed \prot\, would correspond to rotation in the mid-latitudes. As a fixed prior, we assume the occulted starspots from each JWST visit are distinct and exist simultaneously on the star throughout the entire span. The morphology of V1298~Tau's light curve has previously been seen to change on timescales $\sim4\times$~\prot \citep{david2019_v1298tau, david19_b}, which roughly corresponds to the duration of our LCOGT campaign. Thus, using the simplification that the spot distribution is not changing significantly is justified. We fix the Visit 1 starspots to their locations given in Table~\ref{tab:spot_properties}. We offset the longitudes of the Visit 2 starspots by 196$^\circ$ based on the difference between the times of conjunction for V1298~Tau~d and V1298~Tau~c relative to \prot, and fix all other starspot properties. We do not include additional fixed spots at similar longitudes as the JWST occulted spots, based on the difference between the total spot coverage fraction and occulted spot coverage fractions that we estimated in Section~\ref{sec:results_bbfits}. In addition to their unknown true number and distributions, they would have the same rotational signal as the occulted spots, and would therefore only slow down the modeling. 

\begin{figure}
    \centering
    \includegraphics[width=\linewidth]{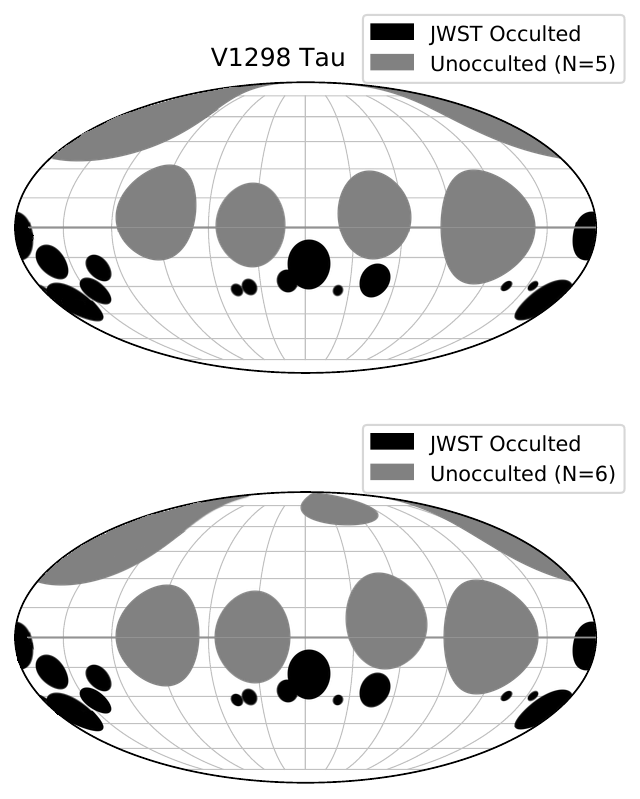}
    \caption{The starspot distribution on V1298~Tau based on our combined analysis of the JWST and LCOGT/Sinistro observations. Spots occulted during the JWST observations are indicated in black, and unocculted spots inferred based on the Sinistro data are in gray. We show the $N$=5 (top) and $N$=6 cases which yield the minimum reduced $\chi^2$ for the g$\prime$ and r$\prime$ bands, respectively. \href{https://github.com/kronos-jwst/KRONOS-II-V1298-Tau-Starspots/blob/main/Figure09.ipynb}{\githubicon}}
    \label{fig:lco_spotmaps}
\end{figure}

We fit for both the radii and positions of additional unocculted spots. New spots were added by drawing their initial radii, latitude, and longitude randomly from uniform distributions between (0 $R_\star$, 0.45$R_\star$), (1$^\circ$, 89$^\circ$), and (0$^\circ$, 359$^\circ$), respectively.  To avoid spatial overlap between the unocculted and occulted spots, we forced the unocculted spots to be in the opposite hemisphere (i.e., latitudes $>$ 0$^\circ$). This does not affect the shape of the resulting light curve since we assume $i_\star = 90^\circ$, so unresolved spots at equal latitudes of either hemisphere are identical. We run MCMC sampling using \texttt{emcee} for 10000 steps following a 10000 step burn-in period (10$\times$ the average autocorrelation time). Once the run for model $N$ has finished, we increase the number of spots to $N+1$ using the best-fit values for each of the $N$ spots that were in the previous model as the initial conditions for the $N+1$ spot model. We generated two \texttt{fleck} models with the same configuration but differing contrasts to represent the g$\prime$ and r$\prime$ light curves and fit the models simultaneously. We assumed contrasts fixed to the value predicted by the MURaM K0 model with its best-fit $f_\mathrm{umb}$ value (Figure~\ref{fig:contrast_fits}) at the corresponding wavelengths: $\alpha_g$ = 0.663 and $\alpha_r$ = 0.589. We evaluate how well the model fits the data by calculating the $\chi^2$.

The minimum $\chi^2$ is achieved for $N$=5 for the g$\prime$ band data and $N$=6 for the r$\prime$ data. The top subfigure of Figure~\ref{fig:lco_fits} shows the corresponding best-fit light curves in each bandpass, as well as for $N$=2 and $N$=9 for comparison. The middle subfigure shows these phase-folded according to \prot. The corresponding $\chi^2$ values are shown in the bottom subfigure. For reference, the $N$=0 case only includes the JWST occulted spots. The $N$=0 model provides a poor fit to the full time series, highlighting the need for additional unocculted spots. 

The corresponding spot maps for $N$=5 and 6 are shown in Figure~\ref{fig:lco_spotmaps}. The additional spots are each significantly larger than any of the occulted spots. It is likely these represent clusters of multiple spots or general active regions. For instance, several of the closely-spaced occulted spots could be combined into one of these active regions. 

The majority of the additional spots are preferentially located along the stellar equator, covering a comparable range of latitudes as the occulted starspots. They largely fill in the ``empty'' longitudes not probed by the transit observations. This suggests that, at this epoch, nearly all longitudes on the surface of V1298~Tau exhibited active regions, which were concentrated around the stellar equator. Furthermore, there are preferentially more additional starspots near the Visit 1 occulted spots, at the center of the map in Figure~\ref{fig:lco_spotmaps}, than near the Visit 2 spots. These may have been present on the visible disk during our observations. This distribution is qualitatively consistent with our finding that the occulted spots from Visit 1 can only account for a minority of the total coverage fraction, while those of Visit 2 were likely the majority. 

We note that these results assume no light curve contribution from faculae or other hot regions on the stellar surface. Our fits to the stellar spectrum suggest a significant coverage of these hot spots (Table~\ref{tab:star_specproperties}), but their size, shape, and distribution about the surface are completely unconstrained. To test how much these results may be biased by unaccounted-for hot spots, we generated additional light curve models including both hot and cool spots. For this test, we consider only the occulted starspots (the ``cool'' spots in this discussion) using the same 196$^\circ$ longitude offset as above, and use the cool-spot-only light curve as reference. First, we add two singular hot spots at the same offset longitudes as the cool spots, making each hot spot cover 30\% of the visible hemisphere as suggested by the spectral fits. Taking the difference between the combined hot+cool spot light curve and the cool-spot-only light curve, we find that these large, singular hot spots can modulate the light curve by up to $\sim$15\%, and thus have a similar rotational signature as the unocculted cool spots. This maximal case, however, is physically implausible. On the Sun, faculae are relatively small and numerous compared to individual sunspots. We find that when the hot spot coverage is divided into more numerous, physically smaller hot spots at random locations on the visible hemisphere--keeping the total coverage fraction the same--their net effect on the light curve decreases. For example, subdividing each large hot spot into three smaller hot spots decreases the light curve modulation to $\sim$3\%, and with ten hot spots the modulation is only $\sim$1\%. Residuals in our cool-spot-only fit may therefore be caused by additional unocculted hot spots.

\section{Discussion} \label{sec:discussion}
\subsection{Comparisons to the Sun} \label{subsec:discussion_sun}

V1298~Tau represents an analogue of the young Sun based on its mass \citep[$M_\star$=1.17\,M$_\odot$;][]{suarez_mascareno_2022}. Previous works have investigated the evolution of bulk spot properties, such as the total number of starspots, with age \citep[e.g.,][]{Morris2020_spotsVSage}, but comparatively little is known about the evolution of details like umbral and penumbral properties. Our observations enabled unique insight into the properties of spot umbrae and penumbrae on V1298~Tau, from which we can begin to gain insight into the evolution of starspot properties, and their substructures, on solar analogs as a function of age. 

Sunspot umbrae and penumbrae have effective temperatures spanning $T_{\mathrm{umb},\odot} = 3900-4800\,K$ and $T_{\mathrm{pen},\odot} = 5400-5500$\,K \citep{solanki2003_sunspots}. Assuming the quiet solar photosphere has $T_{\mathrm{phot},\odot}$ = 5772\,K \citep{mamajek2015iau}, each component therefore has a temperature deficit of $\Delta T_{\mathrm{umb},\odot}$ = 972--1872\,K and $\Delta T_{\mathrm{pen},\odot}$= 272--372\,K. By comparison, our fits find $\Delta$\,\tumb\, = 1577$^{+514}_{-357}$\,K, 1609$^{+248}_{-208}$\,K and $\Delta$\,\tpen\, = 226$^{+87}_{-101}$\,K, 490$^{+57}_{-64}$\,K for V1298~Tau from Visits 1 and 2, respectively. Our best-fit umbral temperature deficits are well within the range of those seen on the modern Sun. However, our penumbral temperature differences are slightly lower for Visit 1 and higher for Visit 2, compared to the modern Sun.

\begin{table}[]
    \centering
    \caption{Relative umbral and penumbral temperature measurements for three Sun-like stars at various ages.}
    \begin{tabular}{c|c|c|c}
    Star & Age (Gyr) & $\Delta T_{\mathrm{umb}}$ (K) & $\Delta T_{\mathrm{pen}}$ (K) \\ \hline 
    V1298~Tau$^{a}$ & 0.023$^{b}$ & 1577$^{+514}_{-357}$ & 226$^{+87}_{-101}$ \\
                    && 1609$^{+248}_{-208}$ & 490$^{+57}_{-64}$\\
    EK Draconis$^{c}$&0.050$^{d}$ & 990         & 180 \\
    Sun$^{e}$ & 4.6 & 972--1872 & 272--372 
    \end{tabular}
    \tablecomments{$\Delta T_i$ refers to the temperature relative to the quiet photosphere.}
    \tablerefs{a: this work, b: \cite{david19_b}, c: \cite{Jarvinen2018}, d: \cite{waite2017}, e: \cite{solanki2003_sunspots}.}
    \label{tab:allstarsdT}
\end{table}

\begin{figure}
    \centering
    \includegraphics[width=1\linewidth]{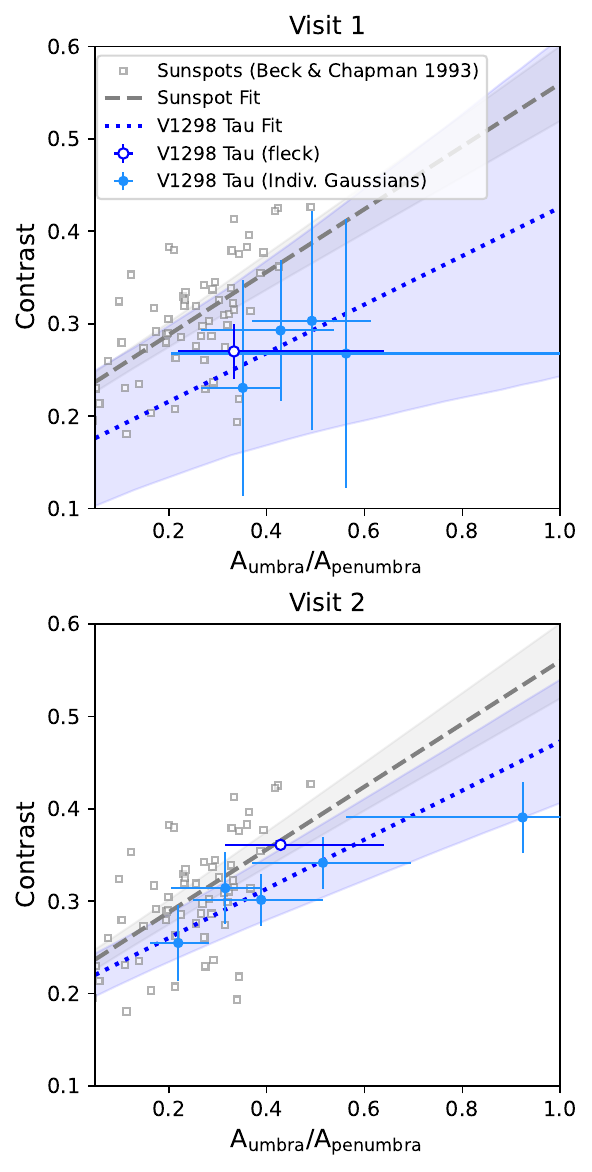}
    \caption{Contrast versus the ratio of umbral to penumbral area for starspots on V1298~Tau from each JWST Visit. Points with black borders are derived from our \texttt{fleck} fits, which model all spots together, and gray points are each starspot fit individually. The observations from both visits exhibit a linear trend, and the Visit 2 observations are very consistent with sunspot data \citep{Beck1993}. \href{https://github.com/kronos-jwst/KRONOS-II-V1298-Tau-Starspots/blob/main/Figure10.ipynb}{\githubicon}}
    \label{fig:contrast_vs_area}
\end{figure}

Empirical relations have been derived between sunspot contrasts $\alpha_\odot$ and the relative areas of umbrae and penumbrae. In particular, based on resolved observations of sunspots during Solar Cycle 22, \cite{Beck1993} demonstrated that sunspots generally follow the trend:
\begin{equation}
    \alpha_{\mathrm{\odot}} = (0.34\pm0.06) \frac{A_u}{A_p}_{\odot} + (0.22 \pm 0.02). \label{eqn:BC1993}
\end{equation}
Here, $A_u/A_p$ represents the ratio of umbral to penumbral areas for individual spots. This trend is shown in Figure~\ref{fig:contrast_vs_area} (gray dashed), along with the measurements from which it was fit (gray boxes), collected from Table~1 of \cite{Beck1993}. 

To investigate whether the same trend holds for V1298~Tau, we first modeled the individual Gaussian-derived contrast spectra to estimate the $f_\mathrm{umb}$ per individual spot, following the same 3-component fit method described in Section~\ref{subsec:contrast_modelsetup}. To circumvent degeneracy between \tumb\, and $f_\mathrm{umb}$, we apply normal priors on all temperature components based on the \texttt{fleck} results. This is equivalent to encouraging the individual temperatures to be consistent with their average values across all spots. The results are provided in Table~\ref{tab:spot_contrastproperties}. Then, we converted $f_\mathrm{umb}$ into $A_u/A_p$ as
\begin{equation}
    \frac{A_u}{A_p} = \frac{f_\mathrm{umb}}{1 - f_\mathrm{umb}}.
\end{equation}
These values are plotted against the median spot contrast in Figure~\ref{fig:contrast_vs_area}.

For two of the spots from Visit 1 (Spots 3 and 4), we find $f_\mathrm{umb} > 0.5$ which translates to $A_u/A_p > 1$. In other words, the occulted portion of these spots appear to be predominantly umbrae. This could potentially suggest these spots are asymmetrically shaped (e.g., the penumbra is not axisymmetric around the umbra) and/or the umbral region is more centered along the transit chord, or that these are clusters of umbral pores. Irregular sunspot shapes are commonly found on the Sun. It is also worth noting that the corresponding individual fits to these spots' contrasts returned relatively high $T_\mathrm{umb}$ compared to the others of that visit (Table~\ref{tab:spot_contrastproperties}). In any case, they are likely not representative of a full and ``typical'' starspot, so we exclude them from this comparison. One spot from Visit 2 (Spot 8) also has a high $f_u = 0.48 \pm 0.12$, consistent with equal portions of occulted umbra and penumbra. We plot this spot in Figure~\ref{fig:contrast_vs_area}, but similarly exclude it from the comparison.

For both visits, we see a similar linear correlation between the starspot contrast and $A_u/A_p$ to that of the sun. For Visit 1, we can fit the data with a line of:
\begin{equation}
    \alpha_1 = \left(0.26 \pm 0.25\right) \frac{A_u}{A_p} + \left(0.16 \pm 0.11\right). \label{eqn:alphaA_v1}
\end{equation}

For Visit 2, we can fit the data with a line of:

\begin{equation}
    \alpha_2 = \left(0.27 \pm 0.09\right) \frac{A_u}{A_p} + \left( 0.21 \pm 0.03\right). \label{eqn:alphaA_v2}
\end{equation}
In each fit, we define $\alpha_i$ as the median contrast for spots of the $i^{th}$ visit.

The best-fit slopes and intercepts for each visit are similar in magnitude. The Visit 1 slope is consistent with zero at 1$\sigma$, likely due to the higher individual uncertainties and overall scatter. The individual median values are generally of lower contrast and higher $A_u/A_p$ than corresponding sunspot values. On the other hand, both the slope and intercept for Visit 2 are consistent within 1$\sigma$ with the sunspot trend of \cite{Beck1993}. Three of the individual points lie well within the scatter of sunspot measurements.

We also performed a 2-dimensional Kolmogorov-Smirnov test using \texttt{ndtest} \citep{peacock1983_fitting, fasano1987_KStest, ndtest} to determine whether our measurements for V1298~Tau can be described as being from the same distribution as the sunspot measurements. We assume a typical p-value threshold of 0.05. For Visit 1, we find $p$=0.001, which rejects the null hypothesis that the Visit 1 values are from the sunspot distribution. However, for Visit 2, we find a relatively weak $p$=0.45, suggesting the Visit 2 values are indeed from the sunspot distribution.

Another similar quantity that has been studied for sunspots is the ratio of the total area of the sunspot to the umbral area, denoted as $r_A$. This is the inverse of our $f_\mathrm{umb}$. Measurements vary between $r_{A,\odot}=[4.0,5.9]$ \citep[e.g.,][]{osherovich1983, tandberg1956, gokhale1972}, owing to varied observational techniques over the last century. It has also been claimed that $r_{A,\odot}$ depends on both the size of the sunspot and where the observation falls within the solar activity cycle. Particularly, that larger sunspots have smaller $r_{A,\odot}$ \citep{steinegger1990, brandt1990, Beck1993}, and that $r_{A,\odot}$ is larger near solar minimum and smaller near solar maximum \citep{tandberg1956, Jensen1956, ringnes1964}. We cannot compare our measurements for V1298~Tau to its activity cycle, based on the limited baseline of our data and because its activity cycle has not been empirically constrained. For Visit 1 spots, we calculate a mean $r_{A,1} = 3.2 \pm 0.2$. The minimum and maximum values are $r_{A,1} = [2.8, 3.8]$. For Visit 2, we calculate a mean of $r_{A,2} = 4.1 \pm 0.5$, and minimum and maximum values of $r_{A,2} = [2.9, 5.6]$. While some of our measured $r_A$ are consistent with solar values, we see no significant correlations between $r_A$ and the spot radius (from Table~\ref{tab:spot_properties}), though this may be expected given our small sample size and uncertainty in the true spot radius.

Several observational studies of sunspots have also suggested a relation between the size of an umbral region and its intensity relative to the quiet photosphere. Larger umbrae are found to have lower relative intensity, and thus a lower temperature \citep{kopp1992, solanki1992, martinez1993, sobotka1993, collados1994, ruedi1995, solanki2003_sunspots}. On V1298~Tau, we found that several of the starspots occulted during Visit 2 appear to be larger than those during Visit 1, and our \texttt{fleck} fits also suggest the Visit 2 umbrae are cooler by $\sim$200\,K. This would be consistent with the behavior seen in sunspots. However, we do not see any such significant correlations between size and \tumb\ from the individual starspot fits. The lack of relationship between $r_A$, \tumb, and spot radius can be attributed to degeneracies in extrapolating the true spot radius from a single starspot crossing event. 

Altogether, our results suggest that the thermal and geometric properties of starspots present on the $\sim$23\,Myr solar analogue V1298~Tau are similar to those of sunspots on the modern Sun. This means that while the prevalence of starspots may be substantially higher for young stars, their physical properties may be very similar across the star's lifetime. Additional high-precision measurements of starspot contrasts on V1298~Tau and other young stars will be necessary to refine this comparison, and reveal the extent of variation in these properties on young stars. In particular, additional observations to determine V1298~Tau's activity cycle and how this cycle affects its starspot properties would provide an invaluable comparison to the Sun.

\subsection{Comparisons to EK Draconis}

Stepping towards older ages, we also compare our measured properties to those of EK Draconis (EK Dra), a similarly young ($\sim$50\,Myr) solar analogue \citep{waite2017}. EK Dra and V1298~Tau are estimated to have comparable magnetic field strengths on the order of 100s~G \citep{finociety_2023, Yamashita2025}. Using the Doppler imaging technique \citep[see e.g.,][]{strassmeier2009_starspots} with high-resolution spectroscopy from PEPSI on the Large Binocular Telescope, \cite{Jarvinen2018} inferred the properties of EK Dra's spot umbrae and penumbrae. 

\cite{Jarvinen2018} spectroscopically resolved four starspots or starspot groups on EK Dra, two of which exhibited morphology consistent with umbral and penumbral regions. For the most isolated and reliable of these spots, they reported a $\Delta$\,\tumb\,=\,990\,K and $\Delta$\,\tpen\,=\,180\,K, relative to $T_{\mathrm{phot}}$=5750\,K. These translate to \tumb\,=\,4760~K and \tpen\,=\,5570~K. This spot covered $\sim$5\% of the visible stellar hemisphere, comparable to the larger spots that we see on V1298~Tau. The measured temperature deficits are each slightly smaller than we observe for V1298~Tau. Table~\ref{tab:allstarsdT} compares the relative spot temperatures for these stars and the Sun.
These differences could be attributed to different measurements or intrinsically different spot properties observed during each observation. Additional observations may be used to understand the statistical significance of these differences. 

\subsection{Interpreting Differences between SCE and Spectral Fitting Properties} \label{subsec:discussion_SCEspecdiffs}

Starspot contrast spectra and the out-of-transit stellar spectrum offer independent measures of the properties of the stellar photosphere and heterogeneities. It is therefore important to understand any differences between the results obtained from each method. 

Each of our fits to the out-of-transit stellar spectra suggest they are best fit with a 3-component model: a cool component (i.e., starspots, $T_{\mathrm{cool}} \approx 3400$\,K), a hot component, and a photospheric component. $T_{\mathrm{cool}}$ is consistent with the umbrae temperature derived from our contrast spectra. This consistency implies that this cool spectral component is dominated by umbrae. However, as $T_{\mathrm{cool}}$ was not an average of \tumb\ and \tpen, it raises questions about whether the warmer penumbrae do not contribute significantly to the stellar spectrum, despite making up $\sim$70\% of the starspot by area (Table~\ref{tab:spot_contrastproperties}), or if we are simply missing their contribution with our fitting method.

To explore this discrepancy further, we tested fitting a 4-component model to the stellar spectra ($T_{\mathrm{phot}}, T_{\mathrm{hot}}$, \tumb, and \tpen). We enforced \tumb~$<$\,\tpen\,$<T_{\mathrm{phot}}<T_{\mathrm{hot}}$ and applied the same normal prior on $T_{\mathrm{phot}}$ as in the 1--3-component fits. We applied priors on \tumb\ and \tpen\ based on the contrast fits. This model was unable to constrain the four components. The temperature of the penumbral component was prior-dominated and its covering fraction, along with that of the hot component, both increased to upwards of 40--50\%. As a result, the coverage fraction of the photosphere became effectively zero, as it was mimicked by the increased prevalence of penumbrae and hot components.

\begin{figure*}
    \centering
    \includegraphics[width=\linewidth]{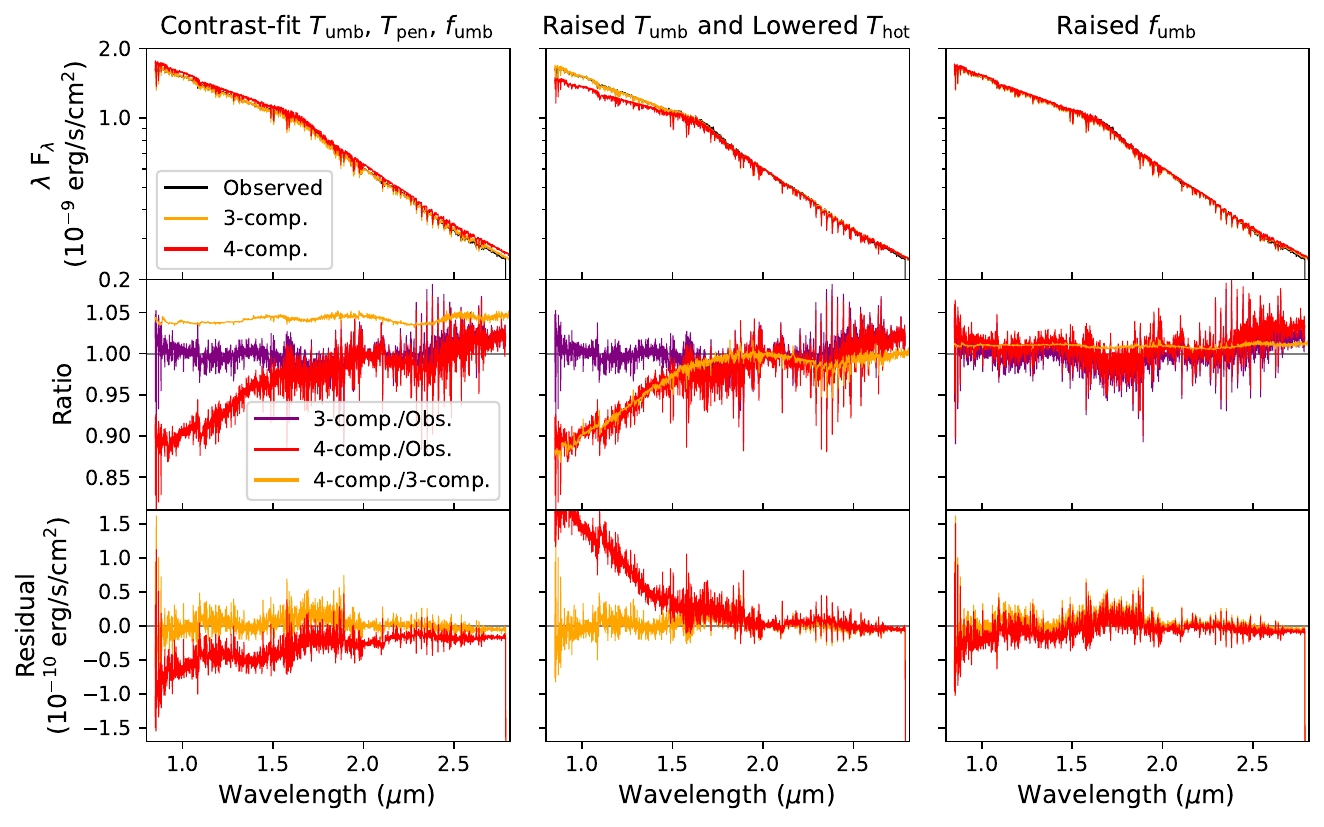}
    \caption{
    Comparing 4-component stellar forward model spectra, including umbral and penumbral contributions, to the observed Visit 2 spectrum and our best-fit 3-component model. The top rows show the spectra, the middle row shows the ratio between each, and the bottom row shows the residual for each model. Left column: We replace $T_{\mathrm{cool}}$ in the 3-component model with the umbra \tumb\ and add a penumbral \tpen\ component. This yields a brighter spectrum with a sloped residual at short wavelengths. Middle: We lower $T_{\mathrm{hot}}$ and raise \tumb. This reduces the brightness ratio, but a sloped residual remains. Right: We raise $f_\mathrm{umb}$ for the model in the left-hand column. This brings the 4-component model into agreement with the observations, but requires a likely implausibly high $f_\mathrm{umb} \approx 75\%$. \href{https://github.com/kronos-jwst/KRONOS-II-V1298-Tau-Starspots/blob/main/Figure11.ipynb}{\githubicon}}
    \label{fig:4compstellarmodel}
\end{figure*}

To speculate on this issue, we generate a synthetic 4-component stellar spectrum based on the contrast fit results. We conduct this exercise for Visit 2 only. Assuming the same extinction, we set $R_\star$ = 1.276\,$R_\odot$, $T_{\mathrm{phot}}$ = 4866\,K, $T_{\mathrm{hot}}$ = 5673\,K, and $f_{\mathrm{hot}}$ = 0.33 based on the spectrum fits (Table~\ref{tab:star_specproperties}), and set \tumb\,= 3264\,K and \tpen\,= 4383\,K (Table~\ref{tab:spot_contrastproperties}). We  assume the starspot coverage fraction is $f_{\mathrm{spot}} = f_{\mathrm{cool}}$ = 0.15, and partition this between the umbral and penumbral components based on $f_\mathrm{umb}$ = 0.30. The resulting model is shown in the left column of Figure~\ref{fig:4compstellarmodel}. This 4-component spectrum is uniformly $\sim$4\% brighter at all wavelengths than observed, and yields a data--model residual larger than the best-fit model. 

Part of this discrepancy may be due to degeneracies between \tumb\ and $T_\mathrm{hot}$. Our best-fit $T_\mathrm{hot}$ is 807\,K hotter than the photosphere, which is significantly larger than the $\sim$100--300\,K excess typical for solar faculae \citep[e.g.,][]{wang1998_faculae, sutterlin199, solovev2019_faculae}. If we reduce $T_\mathrm{hot}$ by 600\,K to be more consistent with solar faculae, in addition to raising \tumb\ by $\sim$200\,K to equal the best-fit $T_{cool}$, then the resulting 4-component model becomes near-identical to the 3-component model at wavelengths longer than $\lambda \gtrsim1.5\,\mu$m. However, the spectral slope becomes inconsistent at shorter wavelengths (middle column, Figure~\ref{fig:4compstellarmodel}).  

An alternate solution could be that the $f_\mathrm{umb}$ derived from the contrast fits do not represent the average $f_\mathrm{umb}$ present across the entire stellar disk, as opposed to just along the transit chords. This could be the case if the planets are crossing non-axisymmetric starspots, which are common on the Sun \citep{solanki2003_sunspots}. If we leave \tumb, \tpen, and $T_{hot}$ at their initial values, we find that raising $f_\mathrm{umb}$ to $\sim 0.75$ yields a 4-component model that is consistent with the 3-component best-fit one (right column, Figure~\ref{fig:4compstellarmodel}). The resulting spectrum is uniformly $\sim$1\% brighter than observed, but this can be reconciled by reducing $R_\star$ by 2--3$\sigma$. However, if $f_\mathrm{umb}$ were this high in reality, it seems unlikely that the large majority of SCEs would involve crossing regions with a 2--3$\times$ lower umbral area fraction. Furthermore, such a high ratio of umbra to penumbra is rarely seen for sunspots once they are fully formed. 

On the Sun, umbrae form before penumbrae, typically as numerous smaller umbra-like ``pores'' that coalesce and grow \citep{solanki2003_sunspots}. The penumbra does not begin to form until a critical size of $\sim$3500\,km or $\sim$0.5\,$R_{\oplus}$ is reached \citep{Bray1964}. Therefore, this explanation may be plausible if there is an overabundance of umbral pores on the surface of V1298~Tau that have not coalesced into full starspots. On the Sun, over 2$\times$ more total pores have been counted than fully formed sunspots between 2010 and 2022 \citep{tlatov2023_sunspotporelifetimes}.
Crossings of such small regions would be difficult to detect in our light curves. Additional starspot crossing observations of the V1298~Tau planets may help us better understand the true $f_\mathrm{umb}$ on the photosphere. Our work demonstrates that both techniques can yield complementary results, but should be considered carefully in the context of one another.

\subsection{Starspots as the Source of Measured Radius Discrepancies}

The V1298~Tau planetary system, particularly V1298~Tau bcd, has been observed several times over the last decade with a variety of instruments \citep[e.g.,][]{livingston2026_v1298tau}. One surprising result of these observations is that the radii of V1298~Tau bcd measured by TESS between 2021 September--November are systematically smaller than those measured by K2 between 2015 February--April \citep{david2019_v1298tau, feinstein2022_v1298tauTESS}. One suggested explanation for this discrepancy is stellar heterogeneity, either in the form of unocculted starspots inducing the transit light source effect (TLSE) or occulted starspots biasing the light curve fits \citep{feinstein2022_v1298tauTESS}. Given the prevalence of SCEs presented here, either explanation could be likely. 

\begin{figure}
    \centering
    \includegraphics[width=\linewidth]{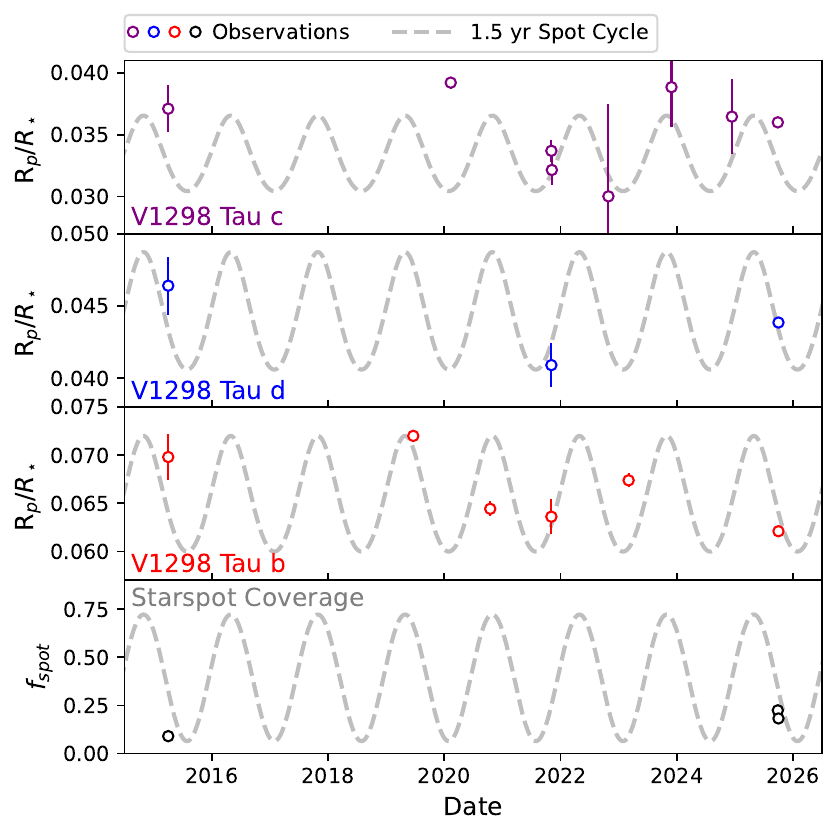}
    \caption{Variations in the observed planet-star radius ratio $R_p/R_\star$ for V1298~Tau~c, d, and b between 2015--2025. These variations can generally be explained by variability in the starspot coverage fraction $f_{spot}$ over a hypothesized 1.5\,yr stellar activity cycle, which induces changes in $R_p/R_\star$ via the transit light source effect. \href{https://github.com/kronos-jwst/KRONOS-II-V1298-Tau-Starspots/blob/main/Figure12.ipynb}{\githubicon}}
    \label{fig:rprsvariations}
\end{figure}

Figure~\ref{fig:rprsvariations} shows archival $R_p/R_\star$ measurements for V1298~Tau cdb collected from \cite{david2019_v1298tau, johnson2022, feinstein2021_v1298tauGemini, suarez_mascareno_2022, feinstein2022_v1298tauTESS, barat2024_bHST, barat2024_v1298taubc, barat2025_v1298taub, livingston2026_v1298tau, Murphy2026}, and this work. Also shown are the spot coverage fractions from Table~\ref{tab:star_specproperties}, and from \cite{Morris2020_spotsVSage} based on the 2015 Kepler observations. Each planet exhibits statistically significant radius variation over the $\sim$10\,yr baseline. Though these data are a conglomerate of measurements within different bandpasses, these variations exceed what would be expected from wavelength dependent opacity alone.

We construct a forward model to demonstrate that the system-wide radius variability can be explained by time variability in the starspot coverage fraction. We follow the transit light source effect formulation of \cite{rackham2018_tlseM}, and consider only contamination from starspots. For any planet, the observed planet-star radius ratio $R_{p, \, \mathrm{obs}}/R_\star$ is modified from its true value $R_{p, 0}/R_\star$ by a multiplicative contamination factor:

\begin{equation}
    \frac{R_{p,obs}}{R_\star} \left(t, \lambda \right) = \sqrt{f_\mathrm{contam}\left(t, \lambda\right)}~ \frac{R_{p,0}}{R_\star} \left(\lambda\right).  \label{eqn:rprs_over_t}
\end{equation}

This contamination is driven by starspots present outside of the transit chord:

\begin{equation}
    f_\mathrm{contam} \left(t, \lambda\right) = \frac{1}{1 - f_\mathrm{spot} \left(t\right) \left(1 - \frac{F_\mathrm{spot}\left(\lambda\right)}{F_\mathrm{phot} \left(\lambda\right)}\right)}. \label{eqn:fcontam}
\end{equation}

This formalism assumes that the stellar surface within the transit chord is entirely described by $F_{\mathrm{phot}}$. Based on the observed JWST SCEs, in our case this formalism should be thought of in terms of a simple difference in the number of starspots within and without the transit chord. 

Motivated by the near-sinusoidal behavior of the number of sunspots over the course of the solar activity cycle \citep{solanki2003_sunspots}, we assume that $f_\mathrm{spot}$ varies smoothly and sinusoidally in time:

\begin{equation}
    f_{spot} \left(t\right) = f_\mathrm{spot,0} + A \sin\left(\frac{2 \pi}{P_\mathrm{activity}} \left(t - t_0\right)\right). \label{eqn:fspot_of_t}
\end{equation}
$A$ is a dimensionless amplitude, $f_\mathrm{spot, 0}$ is an offset term, $P_\mathrm{activity}$ is the period of the underlying stellar activity cycle, and $t_0$ is effectively a phase offset. For simplicity in this initial exploration, we neglect wavelength dependence of $R_{p,i}/R_\star$ and apply a single value of $F_\mathrm{spot}/F_\mathrm{phot}$ across all measurements. We calculate the latter using the MURaM K0 spot and photosphere spectra described in Section~\ref{subsec:contrast_modelresults} for an intermediate TESS-like bandpass between 0.6--1.05\,$\mu$m. 

The magnetic activity cycle of V1298~Tau is not currently constrained. Based on current empirical relations between $P_\mathrm{activity}$ and $P_\mathrm{rot}$, we estimate $P_\mathrm{activity}\sim$1.5\,yr \citep{olah2016}. Similarly short activity cycles have been inferred for other young, Sun-like, fast rotating stars \citep{metcalfe2010_iHorcycle, sanzforcada2013_iHorcycle, sanzforcada2019_iHorcycle, singh2024_ABDor, ayres2025_xraycycles}. We emphasize that this cycle length is an informed first guess, and we defer a more detailed investigation of the cycle length to future work. Nevertheless, it is likely that the archival measurements span at least one full cycle length. Using Eqn.~\ref{eqn:rprs_over_t}, we calculate the corresponding radius variation assuming $P_\mathrm{activity}$ = 1.5\,yr, $A$=0.33, $f_\mathrm{spot,0}$=0.39, $t_0$=0.8\,yr, and $R_{p,0}/R_\star$ = 0.03, 0.04, and 0.06 for planets c, d, and b, respectively. These values were tuned manually to provide an adequate fit, while having the $R_{p,0}/R_\star$ remain physically plausible. The resulting models for each $R_{p, obs}/R_\star$ as well as $f_\mathrm{spot} \left(t\right)$ are shown in Figure~\ref{fig:rprsvariations}. 

Despite our simplifying assumptions,  $P_\mathrm{activity}=1.5$\,yr can well explain the observed radius variability for V1298~Tau~bcd as well as the observed starspot covering fractions in 2015 and 2025. This model predicts $f_\mathrm{spot}$ to reach a large maximum value near 75\% which has not been observed for V1298~Tau, but is commonly seen on other young stars \citep[e.g.,][]{stauffer2003_spottyyoungKdwarfs, GullySantiago2017_youngspottystar}. Additional high-resolution observations of V1298~Tau within the expected peaks would be helpful to determine the true extent of $f_\mathrm{spot}$, and confirm, or refute, this 1.5\,yr cycle. The outstanding residuals in Figure~\ref{fig:rprsvariations} are most likely due to wavelength dependent transit depths. For instance, the high outlier for V1298~Tau~c in $\sim$2020 was measured within the narrow optical \ion{Ca}{2} lines that trace the upper planetary atmosphere, which may also respond more extremely to stellar activity \citep{allan2026}. We reserve a more rigorous modeling effort, which completely treats wavelength dependencies and the difference between the transit chord and outlying photosphere, to future work. 

\subsection{Elongation of Starspot Sizes due to Stellar Rotation} \label{subsec:elongation}

In Section~\ref{sec:results_bbfits}, we estimated the physical length of each occulted starspot region on V1298~Tau based on the duration of each SCE. We also argued how this length is biased by stellar rotation during the SCE, which we corrected for using Eqn.~\ref{eqn:elongation}. Here, we explore this problem more generally. For a given star, this elongation of the apparent spot length is strongest for larger spots, where the crossing times are naturally larger, and for spots nearest 0$^\circ$ longitude, where the displacement is most tangential to the observer line of sight. More generally, this elongation will also depend on the stellar rotation rate and radius. SCEs have been observed for a variety of stars ranging from relatively fast rotators (e.g., this work) to slow rotators \citep[e.g., WASP-107, $P_\mathrm{rot}$=17\,day; ][]{mocnik2017_wasp107, dai2017, sing2024_wasp107, Murphy2025_wasp107b}, as well as relatively small radii \citep[e.g., TOI-3884, $R_\star$=0.3\,$R_\odot$;][]{libbyroberts2023, murray2026} to large radii \citep[e.g., HIP 67522, $R_\star$=1.4\,$R_\odot$;][]{thao2024}. These motivate a more general description of this apparent elongation effect.  

We select representative stellar parameters: radii of $R_\star$=0.2, 0.5, 1, and 1.3\,$R_\sun$, as well as rotation periods of $P_\mathrm{rot}$=1, 3, 5, 10, and 20\,days. For each of these stars, we evaluate Eqn.~\ref{eqn:elongation} for a grid of starspot crossing times between 5 and 50 minutes, which are representative of the range of observed values. In reality, this crossing time will depend on the exact size of the spot and the orbital velocity of the planet. We perform this calculation for two initial spot positions: $lon = 0^\circ$ (the maximal effect) and $lon=45^\circ$ (halfway towards the limb). 

\begin{figure}
    \centering
    \includegraphics[width=\linewidth]{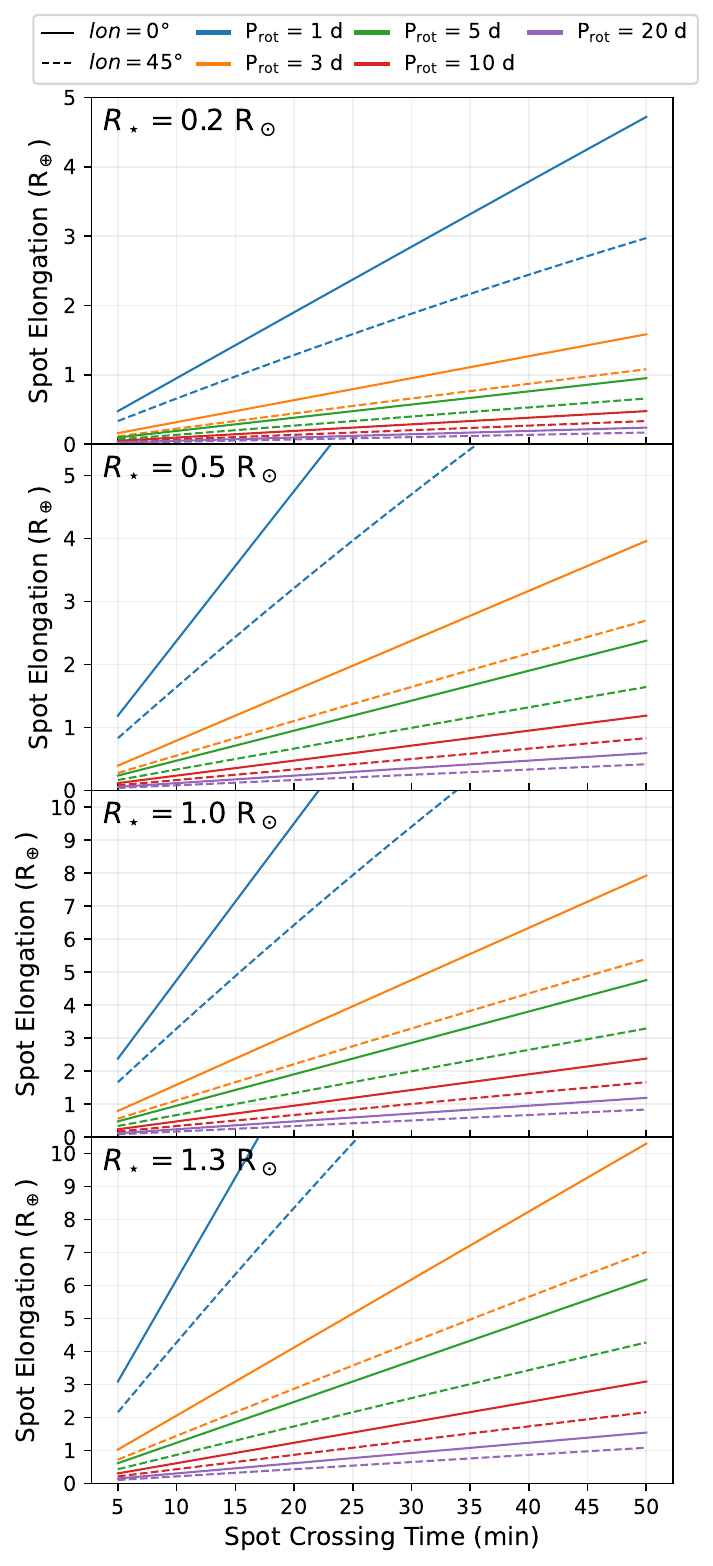}
    \caption{Elongation of the apparent size of occulted starspots, along the direction of the planet's motion, due to stellar rotation during the spot crossing event. For several representative stellar radii, we calculate the apparent elongation based on Eqn.~\ref{eqn:elongation} as a function of the crossing time for various values of \prot (colors), and for spots initially at $0^\circ$ longitude (solid lines) and $45^\circ$ longitude (dashed lines). This effect is maximized for spots nearest $0^\circ$ longitude on fast rotating, large stars. \href{https://github.com/kronos-jwst/KRONOS-II-V1298-Tau-Starspots/blob/main/Figure13.ipynb}{\githubicon}}
    \label{fig:elonggrids}
\end{figure}

Figure~\ref{fig:elonggrids} shows the resulting apparent elongation values ($\Delta \ell$) in units of Earth radii. For a given stellar radius, this effect is a strong function of \prot\ and the initial spot longitude. For example, for a 20\,minute spot crossing on a 1\,$R_\odot$ star, the elongation for a spot at $lon = 0^\circ$ may reach up to 10\,$R_\oplus$ for \prot=1\,day but is only 1\,$R_\oplus$ for \prot=10\,day. For a spot at $lon=45^\circ$, the effect is lessened by $\sim$20\%. For a given crossing time and rotation period, the effect also depends directly on $R_\star$, though not as strongly as the other parameters. For the same 20\,minute crossing with \prot=3\,day and $lon = 0^\circ$, the elongation is $\sim$4\,$R_\oplus$ for a 1.3\,$R_\odot$ star but only $\sim$0.5\,$R_\oplus$ for the 0.2\,$R_\odot$ star. Therefore, this elongation effect is most important to consider for fast rotating, large stars similar to V1298~Tau or HIP 67522.

\section{Summary} \label{sec:summary}

We present one transit observation each of V1298~Tau c, d, and b using JWST NIRISS/SOSS, across two visits on 2025 September 06 and 2025 September 10. We leverage these data to measure the properties of starspots on V1298~Tau, a young ($\sim$20--30\,Myr) solar analog. We identified at least 14 individual SCEs across the two visits. We derived and modeled the 0.8--2.8\,$\mu$m starspot contrast spectra. We also derived and modeled the disk-integrated stellar spectrum from each visit. A summary of our key findings is as follows:
\begin{enumerate}
    \item The stellar spectrum of V1298~Tau was uniformly $\sim$1.6\% brighter during our second visit. Both visits' spectra are best fit by 3-component models (photosphere with hotter and cooler components) with consistent temperatures. For Visit 2, we find $T_\mathrm{phot}$ = 4866 $\pm$33\,K, $T_\mathrm{cool}$ = 3405 $\pm$ 25\,K, and $T_\mathrm{hot}$ = 5673 $\pm$ 41\,K. The hot component (e.g., faculae) makes up $\sim$30\% of the disk, whereas the cool component (e.g., spots) decreases from $\sim$19\% to $\sim$15\% between our two visits.

    \item The observed starspot contrast spectra from each visit are consistent in shape: a slope at $\lambda < 1.7\mu$m and relatively flat at $1.7 < \lambda < 2.75\mu$m. While the shape is consistent, the spots in Visit 2 have a higher contrast than those in Visit 1. This difference can be explained by the spots during Visit 2 being cooler, but may also be due to the signal-to-noise ratio as the SCEs in Visit 2 have significantly larger amplitudes.
    
    
    \item The starspot contrast spectra for both visits are better fit by a 3-component (photosphere, spot umbra, spot penumbra) model than a 2-component (photosphere and single temperature spot) model. From these fits, we find \tumb\,$=3200-3400$\,K, and \tpen\,$=4400-4600$\,K, and an umbral area fraction of 25--30\%. Our best constraints on these components come from the transits of V1298~Tau~d and b during Visit 2.

    \item Our observed contrast spectra for Visit 2 are consistent with model starspot spectra for K0 stars from the MURaM stellar model \citep{Smitha2024_MURaMspotmodels} when assuming a similar (26\%) umbral fraction as our best-fit value. 

    \item The umbral temperatures we derive from the contrast spectra are similar to the temperature of the cool component of the disk-integrated out-of-transit spectrum. Therefore, this spectral component may be dominated by umbral flux. However, we cannot fully explain the apparent lack of a penumbral component. 

    \item We find a strong linear correlation between the contrast and umbral-to-penumbral area ratio of individual starspots, particularly for Visit 2. For Visit 2, this trend is nearly identical to the corresponding relation derived for sunspots \citep{Beck1993}.

    \item We combine the JWST transit observations with multi-band photometric monitoring from LCOGT to provide insights into the global properties of V1298~Tau. These fits assume only spots with no facular contribution. We find that the presence of the spots along the transit chords plus 5--6 additional large unocculted starspot regions can explain the observed photometric variability. The distribution of these unocculted spot regions is also qualitatively consistent with the contribution of the occulted starspots to the total coverage fraction inferred from the disk-integrated stellar spectrum. Outstanding residuals in the fit are likely caused by the unocculted hot spots that made up $\sim$30\% of the surface, but are not included in our light curve model. This highlights the synergy between photometric monitoring and measuring precise spot properties from SCEs.
\end{enumerate}

Altogether, our results suggest that the thermal and geometrical properties of starspots on V1298~Tau are similar to those of the Sun. Therefore, although young Sun-like stars have significantly higher prevalence of starspots, the underlying physical mechanisms driving them may not change between their pre-main-sequence and main-sequence phases. 
This work also demonstrates the unique ability of JWST exoplanet observations to constrain starspot properties in detail through occultations.

For stars other than the Sun, the properties and diversity of starspot substructures (i.e., umbrae and penumbrae) are largely unknown due to the difficulty in separately resolving them. \cite{Zaleski2025} recently proposed exoplanet transits as a method for detecting and studying this substructure. Our results demonstrate that spectroscopy of occulted starspots during exoplanet transits with JWST can be used to infer the properties of umbral and penumbral regions by measuring and modeling their combined thermal contrast. Future work is needed to determine to which other systems this may be applied.

Though starspot crossings have been reported during numerous other JWST transit observations, spanning a range of host star types \citep[e.g.,][]{Fu2022, sing2024_wasp107, LibbyRoberts2025, FournierTondreau2025, Roy2025, triantafillides2026}, these data have either not been tested for or not been sensitive to starspot substructure. Additional work is necessary to determine the limits of this technique, and for what other systems this can be applied. Recent theoretical work has identified such a need for empirical constraints on umbral and penumbral properties across different stellar types to benchmark stellar models \citep{Smitha2024_MURaMspotmodels}. In turn, these constraints would improve our ability to disentangle bias from stellar activity in exoplanetary transmission spectra.

\begin{acknowledgments}

The authors would like to thank the co-Is who contributed to the KRONOS proposal: Lili Alderson, Jonathan Brande, Ian Crossfield, N\'estor Espinoza, Kevin France, Peter Gao, Giannina Guzman Caloca, Garrett Levine, Jonathan Lunine, Andrew Mann, Sagnick Mukherjee, James Owen, Keighley Rockcliffe, Leslie Rogers, Sara Seager, Alexander Shapiro, Pa Chia Thao, and Shreyas Vissapragada. We also thank our anonymous reviewer, whose insightful comments and suggestions helped improve our analysis.

This work is based on observations made with the NASA/ESA/CSA JWST. The data were obtained from the Mikulski Archive for Space Telescopes at the Space Telescope Science Institute, which is operated by the Association of Universities for Research in Astronomy, Inc., under NASA contract NAS 5-03127 for JWST. These observations are associated with the program JWST GO 5959. Support for program JWST GO 5959 was provided by NASA through a grant from the Space Telescope Science Institute. The JWST data are available at \dataset[doi:10.17909/vjpc-v080]{\doi{10.17909/vjpc-v080}} and \dataset[doi:10.17909/e600-p678]{\doi{10.17909/e600-p678}}. A Zenodo repository will also be created upon final acceptance.

This work makes use of observations from the Sinistro cameras on the Las Cumbres Observatory global telescope network \citep{brown2013} from Proposal LCO2025B-006. LCOGT observations were taken at SAAO, CTIO, Teide Observatory, Siding Spring Observatory, \& McDonald Observatory. This paper is based on observations made with the MuSCAT3 instrument, developed by Astrobiology Center and under financial supports by JSPS KAKENHI (JP18H05439) and JST PRESTO (JPMJPR1775), at Faulkes Telescope North on Maui, HI, operated by the Las Cumbres Observatory.
\end{acknowledgments}

\begin{contribution}

M.M.M. led the data analysis and wrote this article. A.D.F. led the original proposal for these data, and contributed significantly to their interpretation as well as writing this article. M.E.S. led the reduction of the LCOGT observations, and contributed to writing this article. B.V.R. aided in the fitting and interpretation of the data. L.W. led the original proposal for these data, and contributed to their interpretation. D.Z.S., G.M.D., and E-M.A. contributed to the preparation of the manuscript. J.H.L. contributed new measurements of the planetary radii. All other authors contributed to the overall interpretation of the results.
\end{contribution}

\facilities{JWST (NIRISS), LCOGT (Sinistro/MuSCAT)}

\software{AstroImageJ \citep{AstroImageJ}, 
\texttt{BANZAI} \citep{banzai_zenodo}, 
\texttt{batman} \citep{batman},
\texttt{emcee} \citep{emcee},
\texttt{ExoTEDRF} \citep{exotedrf},
\texttt{fleck} \citep{fleck},
\texttt{SciPy} \citep{scipy},
\texttt{speclib} \citep{Rackham2023},
\texttt{twirl} \citep{twirl_implement},
\texttt{ndtest} \citep{ndtest},
\texttt{UltraNest} \citep{ultranest}
}

\bibliography{refs}{}

@ARTICLE{AstroImageJ,
       author = {{Collins}, Karen A. and {Kielkopf}, John F. and {Stassun}, Keivan G. and {Hessman}, Frederic V.},
        title = "{AstroImageJ: Image Processing and Photometric Extraction for Ultra-precise Astronomical Light Curves}",
      journal = {\aj},
         year = 2017,
        month = feb,
       volume = {153},
       number = {2},
          eid = {77},
        pages = {77},
          doi = {10.3847/1538-3881/153/2/77},
archivePrefix = {arXiv},
       eprint = {1701.04817},
 primaryClass = {astro-ph.IM},
       adsurl = {https://ui.adsabs.harvard.edu/abs/2017AJ....153...77C}
}

@ARTICLE{twirl_algo,
       author = {{Lang}, Dustin and {Hogg}, David W. and {Mierle}, Keir and {Blanton}, Michael and {Roweis}, Sam},
        title = "{Astrometry.net: Blind Astrometric Calibration of Arbitrary Astronomical Images}",
      journal = {\aj},
         year = 2010,
        month = may,
       volume = {139},
       number = {5},
        pages = {1782-1800},
          doi = {10.1088/0004-6256/139/5/1782},
archivePrefix = {arXiv},
       eprint = {0910.2233},
 primaryClass = {astro-ph.IM},
       adsurl = {https://ui.adsabs.harvard.edu/abs/2010AJ....139.1782L}
}

@ARTICLE{twirl_implement,
       author = {{Garcia}, Lionel J. and {Timmermans}, Mathilde and {Pozuelos}, Francisco J. and {Ducrot}, Elsa and {Gillon}, Micha{\"e}l and {Delrez}, Laetitia and {Wells}, Robert D. and {Jehin}, Emmanu{\"e}l},
        title = "{PROSE: a PYTHON framework for modular astronomical images processing}",
      journal = {\mnras},
         year = 2022,
        month = feb,
       volume = {509},
       number = {4},
        pages = {4817-4828},
          doi = {10.1093/mnras/stab3113},
archivePrefix = {arXiv},
       eprint = {2111.02814},
 primaryClass = {astro-ph.IM},
       adsurl = {https://ui.adsabs.harvard.edu/abs/2022MNRAS.509.4817G}
}

@INPROCEEDINGS{BANZAI_phot_SPIE,
       author = {{McCully}, Curtis and {Volgenau}, Nikolaus H. and {Harbeck}, Daniel-Rolf and {Lister}, Tim A. and {Saunders}, Eric S. and {Turner}, Monica L. and {Siiverd}, Robert J. and {Bowman}, Mark},
        title = "{Real-time processing of the imaging data from the network of Las Cumbres Observatory Telescopes using BANZAI}",
    booktitle = {Software and Cyberinfrastructure for Astronomy V},
         year = 2018,
       editor = {{Guzman}, Juan C. and {Ibsen}, Jorge},
       series = {Society of Photo-Optical Instrumentation Engineers (SPIE) Conference Series},
       volume = {10707},
        month = jul,
          eid = {107070K},
        pages = {107070K},
          doi = {10.1117/12.2314340},
archivePrefix = {arXiv},
       eprint = {1811.04163},
 primaryClass = {astro-ph.IM},
       adsurl = {https://ui.adsabs.harvard.edu/abs/2018SPIE10707E..0KM}
}

@misc{mamajek2015iau,
      title={IAU 2015 Resolution B3 on Recommended Nominal Conversion Constants for Selected Solar and Planetary Properties}, 
      author={E. E. Mamajek and A. Prsa and G. Torres and P. Harmanec and M. Asplund and P. D. Bennett and N. Capitaine and J. Christensen-Dalsgaard and E. Depagne and W. M. Folkner and M. Haberreiter and S. Hekker and J. L. Hilton and V. Kostov and D. W. Kurtz and J. Laskar and B. D. Mason and E. F. Milone and M. M. Montgomery and M. T. Richards and J. Schou and S. G. Stewart},
      year={2015},
      eprint={1510.07674},
      archivePrefix={arXiv},
      primaryClass={astro-ph.SR},
      url={https://arxiv.org/abs/1510.07674}, 
}

@ARTICLE{solanki1992,
       author = {{Solanki}, S.~K. and {Rueedi}, I. and {Livingston}, W.},
        title = "{Infrared lines as probes of solar magnetic features. V - The magnetic structure of a simple sunspot and its canopy}",
      journal = {\aap},
         year = 1992,
        month = sep,
       volume = {263},
       number = {1-2},
        pages = {339-350},
       adsurl = {https://ui.adsabs.harvard.edu/abs/1992A&A...263..339S}
}

@ARTICLE{kopp1992,
       author = {{Kopp}, Greg and {Rabin}, Douglas},
        title = "{A Relation Between Magnetic Field Strength and Temperature in Sunspots}",
      journal = {\solphys},
         year = 1992,
        month = oct,
       volume = {141},
       number = {2},
        pages = {253-265},
          doi = {10.1007/BF00155178},
       adsurl = {https://ui.adsabs.harvard.edu/abs/1992SoPh..141..253K}
}

@ARTICLE{martinez1993,
       author = {{Martinez Pillet}, V. and {Vazquez}, M.},
        title = "{The continuum intensity-magnetic field relation in sunspot umbrae}",
      journal = {\aap},
         year = 1993,
        month = mar,
       volume = {270},
       number = {1-2},
        pages = {494-508},
       adsurl = {https://ui.adsabs.harvard.edu/abs/1993A&A...270..494M}
}

@ARTICLE{sobotka1993,
       author = {{Sobotka}, Michal and {Bonet}, Jose A. and {Vazquez}, Manuel},
        title = "{A High-Resolution Study of Inhomogeneities in Sunspot Umbrae}",
      journal = {\apj},
         year = 1993,
        month = oct,
       volume = {415},
        pages = {832},
          doi = {10.1086/173205},
       adsurl = {https://ui.adsabs.harvard.edu/abs/1993ApJ...415..832S}
}

@ARTICLE{collados1994,
       author = {{Collados}, M. and {Martinez Pillet}, V. and {Ruiz Cobo}, B. and {del Toro Iniesta}, J.~C. and {Vazquez}, M.},
        title = "{Observed differences between large and small sunspots.}",
      journal = {\aap},
         year = 1994,
        month = nov,
       volume = {291},
        pages = {622-634},
       adsurl = {https://ui.adsabs.harvard.edu/abs/1994A&A...291..622C}
}

@ARTICLE{ruedi1995,
       author = {{Rueedi}, I. and {Solanki}, S.~K. and {Livingston}, W.},
        title = "{Infrared lines as probes of solar magnetic features. XI. Structure of a sunspot umbra with a light bridge.}",
      journal = {\aap},
         year = 1995,
        month = oct,
       volume = {302},
        pages = {543},
       adsurl = {https://ui.adsabs.harvard.edu/abs/1995A&A...302..543R}
}

@ARTICLE{tandberg1956,
       author = {{Tandberg-Hanssen}, Einar},
        title = "{A Study of the Penumbra-Umbra Ratio of Sunspot Pairs}",
      journal = {Astrophysica Norvegica},
         year = 1956,
        month = jan,
       volume = {5},
        pages = {207},
       adsurl = {https://ui.adsabs.harvard.edu/abs/1956ApNr....5..207T}
}

@ARTICLE{gokhale1972,
       author = {{Gokhale}, M.~H. and {Zwaan}, C.},
        title = "{The Structure of Sunspots. I: Observational Constraints: Current Sheet Models}",
      journal = {\solphys},
         year = 1972,
        month = sep,
       volume = {26},
       number = {1},
        pages = {52-75},
          doi = {10.1007/BF00155105},
       adsurl = {https://ui.adsabs.harvard.edu/abs/1972SoPh...26...52G}
}

@ARTICLE{ilin21,
       author = {{Ilin}, Ekaterina and {Schmidt}, Sarah J. and {Poppenh{\"a}ger}, Katja and {Davenport}, James R.~A. and {Kristiansen}, Martti H. and {Omohundro}, Mark},
        title = "{Flares in open clusters with K2. II. Pleiades, Hyades, Praesepe, Ruprecht 147, and M 67}",
      journal = {\aap},
         year = 2021,
        month = jan,
       volume = {645},
          eid = {A42},
        pages = {A42},
          doi = {10.1051/0004-6361/202039198},
archivePrefix = {arXiv},
       eprint = {2010.05576},
 primaryClass = {astro-ph.SR},
       adsurl = {https://ui.adsabs.harvard.edu/abs/2021A&A...645A..42I}
}

@ARTICLE{ilin19,
       author = {{Ilin}, Ekaterina and {Schmidt}, Sarah J. and {Davenport}, James R.~A. and {Strassmeier}, Klaus G.},
        title = "{Flares in open clusters with K2 . I. M 45 (Pleiades), M 44 (Praesepe), and M 67}",
      journal = {\aap},
         year = 2019,
        month = feb,
       volume = {622},
          eid = {A133},
        pages = {A133},
          doi = {10.1051/0004-6361/201834400},
archivePrefix = {arXiv},
       eprint = {1812.06725},
 primaryClass = {astro-ph.SR},
       adsurl = {https://ui.adsabs.harvard.edu/abs/2019A&A...622A.133I}
}

@ARTICLE{feinstein24,
       author = {{Feinstein}, Adina D. and {Seligman}, Darryl Z. and {France}, Kevin and {Gagn{\'e}}, Jonathan and {Kowalski}, Adam},
        title = "{Evolution of Flare Activity in GKM Stars Younger Than 300 Myr over Five Years of TESS Observations}",
      journal = {\aj},
         year = 2024,
        month = aug,
       volume = {168},
       number = {2},
          eid = {60},
        pages = {60},
          doi = {10.3847/1538-3881/ad4edf},
archivePrefix = {arXiv},
       eprint = {2405.00850},
 primaryClass = {astro-ph.SR},
       adsurl = {https://ui.adsabs.harvard.edu/abs/2024AJ....168...60F}
}

@ARTICLE{ealy24,
       author = {{Ealy}, Jordan N. and {Schlieder}, Joshua E. and {Komacek}, Thaddeus D. and {Gilbert}, Emily A.},
        title = "{Flaring Activity for Low-mass Stars in the {\ensuremath{\beta}} Pictoris Moving Group}",
      journal = {\aj},
         year = 2024,
        month = oct,
       volume = {168},
       number = {4},
          eid = {173},
        pages = {173},
          doi = {10.3847/1538-3881/ad6b7e},
archivePrefix = {arXiv},
       eprint = {2408.04624},
 primaryClass = {astro-ph.SR},
       adsurl = {https://ui.adsabs.harvard.edu/abs/2024AJ....168..173E}
}

@ARTICLE{mamonova25,
       author = {{Mamonova}, E. and {Shan}, Y. and {Kowalski}, A.~F. and {Wedemeyer}, S. and {Werner}, S.~C.},
        title = "{Flare frequency in M dwarfs belonging to young moving groups}",
      journal = {\aap},
         year = 2025,
        month = aug,
       volume = {700},
          eid = {A53},
        pages = {A53},
          doi = {10.1051/0004-6361/202554614},
archivePrefix = {arXiv},
       eprint = {2506.04465},
 primaryClass = {astro-ph.SR},
       adsurl = {https://ui.adsabs.harvard.edu/abs/2025A&A...700A..53M}
}

@ARTICLE{tran26,
       author = {{Tran}, Andrew and {Song}, Inseok},
        title = "{Stellar Flare Study of nearby Young Moving Group Members with TESS Data}",
      journal = {\apj},
         year = 2026,
        month = mar,
       volume = {999},
       number = {2},
          eid = {252},
        pages = {252},
          doi = {10.3847/1538-4357/ae44f2},
archivePrefix = {arXiv},
       eprint = {2602.20402},
 primaryClass = {astro-ph.SR},
       adsurl = {https://ui.adsabs.harvard.edu/abs/2026ApJ...999..252T}
}

@ARTICLE{boyle26,
       author = {{Boyle}, Andrew W. and {Bouma}, Luke G. and {Mann}, Andrew W.},
        title = "{The TESS All-Sky Rotation Survey: Periods for 1,046,317 Stars Within 500 pc}",
      journal = {arXiv e-prints},
         year = 2026,
        month = mar,
          eid = {arXiv:2603.05586},
        pages = {arXiv:2603.05586},
          doi = {10.48550/arXiv.2603.05586},
archivePrefix = {arXiv},
       eprint = {2603.05586},
 primaryClass = {astro-ph.SR},
       adsurl = {https://ui.adsabs.harvard.edu/abs/2026arXiv260305586B}
}

@ARTICLE{waalkes24,
       author = {{Waalkes}, William C. and {Berta-Thompson}, Zachory K. and {Newton}, Elisabeth R. and {Mann}, Andrew W. and {Gao}, Peter and {Wakeford}, Hannah R. and {Alderson}, Lili and {Plavchan}, Peter},
        title = "{Quantifying the Transit Light Source Effect: Measurements of Spot Temperature and Coverage on the Photosphere of AU Microscopii with High-resolution Spectroscopy and Multicolor Photometry}",
      journal = {\apj},
         year = 2024,
        month = feb,
       volume = {962},
       number = {1},
          eid = {97},
        pages = {97},
          doi = {10.3847/1538-4357/ad0bed},
archivePrefix = {arXiv},
       eprint = {2310.17043},
 primaryClass = {astro-ph.EP},
       adsurl = {https://ui.adsabs.harvard.edu/abs/2024ApJ...962...97W}
}

@ARTICLE{biagini24,
       author = {{Biagini}, Alfredo and {Petralia}, Antonino and {Di Maio}, Claudia and {Betti}, Lorenzo and {Pace}, Emanuele and {Micela}, Giuseppina},
        title = "{Spot modelling through multi-band photometry: Analysis of V1298 Tau}",
      journal = {\aap},
         year = 2024,
        month = oct,
       volume = {690},
          eid = {A386},
        pages = {A386},
          doi = {10.1051/0004-6361/202450036},
archivePrefix = {arXiv},
       eprint = {2409.11034},
 primaryClass = {astro-ph.SR},
       adsurl = {https://ui.adsabs.harvard.edu/abs/2024A&A...690A.386B}
}

@ARTICLE{mori24,
       author = {{Mori}, Mayuko and {Ikuta}, Kai and {Fukui}, Akihiko and {Narita}, Norio and {de Leon}, Jerome P. and {Livingston}, John H. and {Ikoma}, Masahiro and {Kawai}, Yugo and {Kawauchi}, Kiyoe and {Murgas}, Felipe and {Palle}, Enric and {Parviainen}, Hannu and {Fern{\'a}ndez Rodr{\'\i}guez}, Gareb and {Terada}, Yuka and {Watanabe}, Noriharu and {Tamura}, Motohide},
        title = "{Characterization of starspots on a young M-dwarf K2-25: multiband observations of stellar photometric variability and planetary transits}",
      journal = {\mnras},
         year = 2024,
        month = may,
       volume = {530},
       number = {1},
        pages = {167-189},
          doi = {10.1093/mnras/stae841},
archivePrefix = {arXiv},
       eprint = {2403.13946},
 primaryClass = {astro-ph.EP},
       adsurl = {https://ui.adsabs.harvard.edu/abs/2024MNRAS.530..167M}
}

@ARTICLE{brown2013,
       author = {{Brown}, T.~M. and {Baliber}, N. and {Bianco}, F.~B. and {Bowman}, M. and {Burleson}, B. and {Conway}, P. and {Crellin}, M. and {Depagne}, {\'E}. and {De Vera}, J. and {Dilday}, B. and {Dragomir}, D. and {Dubberley}, M. and {Eastman}, J.~D. and {Elphick}, M. and {Falarski}, M. and {Foale}, S. and {Ford}, M. and {Fulton}, B.~J. and {Garza}, J. and {Gomez}, E.~L. and {Graham}, M. and {Greene}, R. and {Haldeman}, B. and {Hawkins}, E. and {Haworth}, B. and {Haynes}, R. and {Hidas}, M. and {Hjelstrom}, A.~E. and {Howell}, D.~A. and {Hygelund}, J. and {Lister}, T.~A. and {Lobdill}, R. and {Martinez}, J. and {Mullins}, D.~S. and {Norbury}, M. and {Parrent}, J. and {Paulson}, R. and {Petry}, D.~L. and {Pickles}, A. and {Posner}, V. and {Rosing}, W.~E. and {Ross}, R. and {Sand}, D.~J. and {Saunders}, E.~S. and {Shobbrook}, J. and {Shporer}, A. and {Street}, R.~A. and {Thomas}, D. and {Tsapras}, Y. and {Tufts}, J.~R. and {Valenti}, S. and {Vander Horst}, K. and {Walker}, Z. and {White}, G. and {Willis}, M.},
        title = "{Las Cumbres Observatory Global Telescope Network}",
      journal = {\pasp},
         year = 2013,
        month = sep,
       volume = {125},
       number = {931},
        pages = {1031},
          doi = {10.1086/673168},
archivePrefix = {arXiv},
       eprint = {1305.2437},
 primaryClass = {astro-ph.IM},
       adsurl = {https://ui.adsabs.harvard.edu/abs/2013PASP..125.1031B}
}

@ARTICLE{Sagynbayeva26,
       author = {{Sagynbayeva}, Sabina and {Farr}, Will M.},
        title = "{Polka-dotted Stars. II. Starspots and Obliquities of Kepler-17 and Kepler-63}",
      journal = {\apj},
         year = 2026,
        month = feb,
       volume = {997},
       number = {2},
          eid = {297},
        pages = {297},
          doi = {10.3847/1538-4357/ae2bdb},
archivePrefix = {arXiv},
       eprint = {2510.07130},
 primaryClass = {astro-ph.EP},
       adsurl = {https://ui.adsabs.harvard.edu/abs/2026ApJ...997..297S}
}

@ARTICLE{Sagynbayeva25,
       author = {{Sagynbayeva}, Sabina and {Farr}, Will M. and {Morris}, Brett M. and {Luger}, Rodrigo},
        title = "{Polka-dotted Stars: A Hierarchical Model for Mapping Stellar Surfaces Using Occultation Light Curves and the Case of TOI-3884}",
      journal = {\apj},
         year = 2025,
        month = sep,
       volume = {990},
       number = {1},
          eid = {32},
        pages = {32},
          doi = {10.3847/1538-4357/adf6be},
archivePrefix = {arXiv},
       eprint = {2504.21852},
 primaryClass = {astro-ph.EP},
       adsurl = {https://ui.adsabs.harvard.edu/abs/2025ApJ...990...32S}
}

@ARTICLE{osherovich1983,
       author = {{Osherovich}, V.~A. and {Lawrence}, J.~K.},
        title = "{Elaboration of the New Magnetohydrostatic Sunspot Theory - Double Return Flux Model}",
      journal = {\solphys},
         year = 1983,
        month = oct,
       volume = {88},
       number = {1-2},
        pages = {117-135},
          doi = {10.1007/BF00196182},
       adsurl = {https://ui.adsabs.harvard.edu/abs/1983SoPh...88..117O}
}

@ARTICLE{steinegger1990,
       author = {{Steinegger}, M. and {Brandt}, P.~N. and {Pap}, J. and {Schmidt}, W.},
        title = "{Sunspot Photometry and the Total Solar Irradiance Deficit Measured in 1980 BY ACRIM}",
      journal = {\apss},
         year = 1990,
        month = aug,
       volume = {170},
       number = {1-2},
        pages = {127-133},
          doi = {10.1007/BF00652658},
       adsurl = {https://ui.adsabs.harvard.edu/abs/1990Ap&SS.170..127S}
}

@ARTICLE{brandt1990,
       author = {{Brandt}, P.~N. and {Schmidt}, W. and {Steinegger}, M.},
        title = "{On the Umbra / Penumbra Area Ratio of Sunspots}",
      journal = {\solphys},
         year = 1990,
        month = sep,
       volume = {129},
       number = {1},
        pages = {191-194},
          doi = {10.1007/BF00154373},
       adsurl = {https://ui.adsabs.harvard.edu/abs/1990SoPh..129..191B}
}

@ARTICLE{ringnes1964,
       author = {{Ringnes}, Truls S.},
        title = "{Secular variations of long-lived sunspots}",
      journal = {Astrophysica Norvegica},
         year = 1964,
        month = jan,
       volume = {8},
        pages = {303},
       adsurl = {https://ui.adsabs.harvard.edu/abs/1964ApNr....8..303R}
}

@ARTICLE{wang1998_faculae,
       author = {{Wang}, Haimin and {Spirock}, Thomas and {Goode}, Philip R. and {Lee}, Chikyin and {Zirin}, Harold and {Kosonocky}, Walter},
        title = "{Contrast of Faculae at 1.6 Microns}",
      journal = {\apj},
         year = 1998,
        month = mar,
       volume = {495},
       number = {2},
        pages = {957-964},
          doi = {10.1086/305311},
       adsurl = {https://ui.adsabs.harvard.edu/abs/1998ApJ...495..957W}
}

@ARTICLE{solovev2019_faculae,
       author = {{Solov'ev}, A.~A. and {Kirichek}, E.~A.},
        title = "{Structure of solar faculae}",
      journal = {\mnras},
         year = 2019,
        month = feb,
       volume = {482},
       number = {4},
        pages = {5290-5301},
          doi = {10.1093/mnras/sty3050},
       adsurl = {https://ui.adsabs.harvard.edu/abs/2019MNRAS.482.5290S}
}

@ARTICLE{sutterlin199,
       author = {{S{\"u}tterlin}, P. and {Wiehr}, E. and {Stellmacher}, G.},
        title = "{Continuum photometry of solar white-light faculae}",
      journal = {\solphys},
         year = 1999,
        month = oct,
       volume = {189},
       number = {1},
        pages = {57-68},
          doi = {10.1023/A:1005243302461},
       adsurl = {https://ui.adsabs.harvard.edu/abs/1999SoPh..189...57S}
}

@BOOK{bray1964,
       author = {{Bray}, R.~J. and {Loughhead}, R.~E.},
        title = "{Sunspots}",
         year = 1964,
       adsurl = {https://ui.adsabs.harvard.edu/abs/1964suns.book.....B}
}

@article{maio2024,
	title = {The {GAPS} programme at {TNG} - {LII}. {Spot} modelling of {V1298} {Tau} using the {SpotCCF} tool},
	volume = {683},
	copyright = {© The Authors 2024},
	issn = {0004-6361, 1432-0746},
	url = {https://www.aanda.org/articles/aa/abs/2024/03/aa48223-23/aa48223-23.html},
	doi = {10.1051/0004-6361/202348223},
	language = {en},
	urldate = {2026-05-01},
	journal = {Astronomy \& Astrophysics},
	publisher = {EDP Sciences},
	author = {Maio, C. Di and Petralia, A. and Micela, G. and Lanza, A. F. and Rainer, M. and Malavolta, L. and Benatti, S. and Affer, L. and Maldonado, J. and Colombo, S. and Damasso, M. and Maggio, A. and Biazzo, K. and Bignamini, A. and Borsa, F. and Boschin, W. and Cabona, L. and Cecconi, M. and Claudi, R. and Covino, E. and Fabrizio, L. Di and Gratton, R. and Lorenzi, V. and Mancini, L. and Messina, S. and Molinari, E. and Molinaro, M. and Nardiello, D. and Poretti, E. and Sozzetti, A.},
	month = mar,
	year = {2024},
	pages = {A239},
}

@ARTICLE{wittmann1987,
       author = {{Wittmann}, A.~D. and {Xu}, Z.~T.},
        title = "{A catalogue of sunspot observations from 165 BC to AD 1684}",
      journal = {\aaps},
         year = 1987,
        month = jul,
       volume = {70},
       number = {1},
        pages = {83-94},
       adsurl = {https://ui.adsabs.harvard.edu/abs/1987A&AS...70...83W}
}

@ARTICLE{yau1988,
       author = {{Yau}, K.~K.~C. and {Stephenson}, F.~R.},
        title = "{A revised catalogue of Far Eastern observations of sunspots (165 BC to AD 1918)}",
      journal = {\qjras},
         year = 1988,
        month = jun,
       volume = {29},
        pages = {175-197},
       adsurl = {https://ui.adsabs.harvard.edu/abs/1988QJRAS..29..175Y}
}

@ARTICLE{eddy1989,
       author = {{Eddy}, John A. and {Stephenson}, F. Richard and {Yau}, Kevin K.~C.},
        title = "{On pre-telescopic sunspot records}",
      journal = {\qjras},
         year = 1989,
        month = mar,
       volume = {30},
        pages = {65-73},
       adsurl = {https://ui.adsabs.harvard.edu/abs/1989QJRAS..30...65E}
}

@ARTICLE{dai2017,
       author = {{Dai}, Fei and {Winn}, Joshua N.},
        title = "{The Oblique Orbit of WASP-107b from K2 Photometry}",
      journal = {\aj},
         year = 2017,
        month = may,
       volume = {153},
       number = {5},
          eid = {205},
        pages = {205},
          doi = {10.3847/1538-3881/aa65d1},
archivePrefix = {arXiv},
       eprint = {1702.04734},
 primaryClass = {astro-ph.EP},
       adsurl = {https://ui.adsabs.harvard.edu/abs/2017AJ....153..205D}
}

@ARTICLE{libbyroberts2023,
       author = {{Libby-Roberts}, Jessica E. and {Schutte}, Maria and {Hebb}, Leslie and {Kanodia}, Shubham and {Ca{\~n}as}, Caleb I. and {Stef{\'a}nsson}, Gu{\dh}mundur and {Lin}, Andrea S.~J. and {Mahadevan}, Suvrath and {Parts}, Winter and {Powers}, Luke and {Wisniewski}, John and {Bender}, Chad F. and {Cochran}, William D. and {Diddams}, Scott A. and {Everett}, Mark E. and {Gupta}, Arvind F. and {Halverson}, Samuel and {Kobulnicky}, Henry A. and {Kowalski}, Adam F. and {Larsen}, Alexander and {Monson}, Andrew and {Ninan}, Joe P. and {Parker}, Brock A. and {Ramsey}, Lawrence W. and {Robertson}, Paul and {Schwab}, Christian and {Swaby}, Tera N. and {Terrien}, Ryan C.},
        title = "{An In-depth Look at TOI-3884b: A Super-Neptune Transiting an M4Dwarf with Persistent Starspot Crossings}",
      journal = {\aj},
         year = 2023,
        month = jun,
       volume = {165},
       number = {6},
          eid = {249},
        pages = {249},
          doi = {10.3847/1538-3881/accc2f},
       adsurl = {https://ui.adsabs.harvard.edu/abs/2023AJ....165..249L}
}

@ARTICLE{charbonneau2000_hd209458,
       author = {{Charbonneau}, David and {Brown}, Timothy M. and {Latham}, David W. and {Mayor}, Michel},
        title = "{Detection of Planetary Transits Across a Sun-like Star}",
      journal = {\apjl},
         year = 2000,
        month = jan,
       volume = {529},
       number = {1},
        pages = {L45-L48},
          doi = {10.1086/312457},
archivePrefix = {arXiv},
       eprint = {astro-ph/9911436},
 primaryClass = {astro-ph},
       adsurl = {https://ui.adsabs.harvard.edu/abs/2000ApJ...529L..45C}
}

@article{tlatov2023_sunspotporelifetimes,
	title = {Lifetime of {Sunspots} and {Pores}},
	volume = {298},
	issn = {1573-093X},
	url = {https://doi.org/10.1007/s11207-023-02186-7},
	doi = {10.1007/s11207-023-02186-7},
	language = {en},
	number = {7},
	urldate = {2026-06-02},
	journal = {Solar Physics},
	author = {Tlatov, Andrey G.},
	month = jul,
	year = {2023},
	pages = {93},
}

@article{Husser2013_phoenix,
   title={A new extensive library of PHOENIX stellar atmospheres and synthetic spectra},
   volume={553},
   ISSN={1432-0746},
   url={http://dx.doi.org/10.1051/0004-6361/201219058},
   DOI={10.1051/0004-6361/201219058},
   journal={Astronomy \& Astrophysics},
   publisher={EDP Sciences},
   author={Husser, T.-O. and Wende-von Berg, S. and Dreizler, S. and Homeier, D. and Reiners, A. and Barman, T. and Hauschildt, P. H.},
   year={2013},
   month=Apr, pages={A6} }

@ARTICLE{scipy,
  author  = {Virtanen, Pauli and Gommers, Ralf and Oliphant, Travis E. and
            Haberland, Matt and Reddy, Tyler and Cournapeau, David and
            Burovski, Evgeni and Peterson, Pearu and Weckesser, Warren and
            Bright, Jonathan and {van der Walt}, St{\'e}fan J. and
            Brett, Matthew and Wilson, Joshua and Millman, K. Jarrod and
            Mayorov, Nikolay and Nelson, Andrew R. J. and Jones, Eric and
            Kern, Robert and Larson, Eric and Carey, C J and
            Polat, {\.I}lhan and Feng, Yu and Moore, Eric W. and
            {VanderPlas}, Jake and Laxalde, Denis and Perktold, Josef and
            Cimrman, Robert and Henriksen, Ian and Quintero, E. A. and
            Harris, Charles R. and Archibald, Anne M. and
            Ribeiro, Ant{\^o}nio H. and Pedregosa, Fabian and
            {van Mulbregt}, Paul and {SciPy 1.0 Contributors}},
  title   = {{{SciPy} 1.0: Fundamental Algorithms for Scientific
            Computing in Python}},
  journal = {Nature Methods},
  year    = {2020},
  volume  = {17},
  pages   = {261--272},
  adsurl  = {https://rdcu.be/b08Wh},
  doi     = {10.1038/s41592-019-0686-2},
}

@ARTICLE{roettenbacher2026,
       author = {{Roettenbacher}, Rachael M. and {Monnier}, John D. and {Korhonen}, Heidi and {Henry}, Gregory W. and {Kotnik}, Cliff and {Pepper}, Joshua and {Seli}, B{\'a}lint and {Vida}, Kriszti{\'a}n and {B{\'o}di}, Attila and {Cseh}, Borb{\'a}la and et al.},
        title = "{Interferometric Images of the Starspot Evolution of {\ensuremath{\zeta}} Andromedae}",
      journal = {\apj},
         year = 2026,
        month = may,
       volume = {1002},
       number = {2},
          eid = {139},
        pages = {139},
          doi = {10.3847/1538-4357/ae5a38},
archivePrefix = {arXiv},
       eprint = {2603.18144},
 primaryClass = {astro-ph.SR},
       adsurl = {https://ui.adsabs.harvard.edu/abs/2026ApJ..1002..139R}
}

@ARTICLE{silva2003,
       author = {{Silva}, Adriana V.~R.},
        title = "{Method for Spot Detection on Solar-like Stars}",
      journal = {\apjl},
         year = 2003,
        month = mar,
       volume = {585},
       number = {2},
        pages = {L147-L150},
          doi = {10.1086/374324},
       adsurl = {https://ui.adsabs.harvard.edu/abs/2003ApJ...585L.147S}
}

@ARTICLE{rabus2009,
       author = {{Rabus}, M. and {Alonso}, R. and {Belmonte}, J.~A. and {Deeg}, H.~J. and {Gilliland}, R.~L. and {Almenara}, J.~M. and {Brown}, T.~M. and {Charbonneau}, D. and {Mandushev}, G.},
        title = "{A cool starspot or a second transiting planet in the TrES-1 system?}",
      journal = {\aap},
         year = 2009,
        month = jan,
       volume = {494},
       number = {1},
        pages = {391-397},
          doi = {10.1051/0004-6361:200811110},
archivePrefix = {arXiv},
       eprint = {0812.1799},
 primaryClass = {astro-ph},
       adsurl = {https://ui.adsabs.harvard.edu/abs/2009A&A...494..391R}
}

@ARTICLE{silva2010,
       author = {{Silva-Valio}, A. and {Lanza}, A.~F. and {Alonso}, R. and {Barge}, P.},
        title = "{Properties of starspots on CoRoT-2}",
      journal = {\aap},
         year = 2010,
        month = feb,
       volume = {510},
          eid = {A25},
        pages = {A25},
          doi = {10.1051/0004-6361/200911904},
archivePrefix = {arXiv},
       eprint = {0909.4055},
 primaryClass = {astro-ph.SR},
       adsurl = {https://ui.adsabs.harvard.edu/abs/2010A&A...510A..25S}
}

@ARTICLE{tregloan2013,
       author = {{Tregloan-Reed}, Jeremy and {Southworth}, John and {Tappert}, C.},
        title = "{Transits and starspots in the WASP-19 planetary system}",
      journal = {\mnras},
         year = 2013,
        month = feb,
       volume = {428},
       number = {4},
        pages = {3671-3679},
          doi = {10.1093/mnras/sts306},
archivePrefix = {arXiv},
       eprint = {1211.0864},
 primaryClass = {astro-ph.EP},
       adsurl = {https://ui.adsabs.harvard.edu/abs/2013MNRAS.428.3671T}
}

@ARTICLE{sanchis2013,
       author = {{Sanchis-Ojeda}, Roberto and {Winn}, Joshua N. and {Marcy}, Geoffrey W. and {Howard}, Andrew W. and {Isaacson}, Howard and {Johnson}, John Asher and {Torres}, Guillermo and {Albrecht}, Simon and {Campante}, Tiago L. and {Chaplin}, William J. and et al.},
        title = "{Kepler-63b: A Giant Planet in a Polar Orbit around a Young Sun-like Star}",
      journal = {\apj},
         year = 2013,
        month = sep,
       volume = {775},
       number = {1},
          eid = {54},
        pages = {54},
          doi = {10.1088/0004-637X/775/1/54},
archivePrefix = {arXiv},
       eprint = {1307.8128},
 primaryClass = {astro-ph.EP},
       adsurl = {https://ui.adsabs.harvard.edu/abs/2013ApJ...775...54S}
}

@ARTICLE{mohler2013,
       author = {{Mohler-Fischer}, M. and {Mancini}, L. and {Hartman}, J.~D. and {Bakos}, G. {\'A}. and {Penev}, K. and {Bayliss}, D. and {Jord{\'a}n}, A. and {Csubry}, Z. and {Zhou}, G. and {Rabus}, M. and et al.},
        title = "{HATS-2b: A transiting extrasolar planet orbiting a K-type star showing starspot activity}",
      journal = {\aap},
         year = 2013,
        month = oct,
       volume = {558},
          eid = {A55},
        pages = {A55},
          doi = {10.1051/0004-6361/201321663},
archivePrefix = {arXiv},
       eprint = {1304.2140},
 primaryClass = {astro-ph.EP},
       adsurl = {https://ui.adsabs.harvard.edu/abs/2013A&A...558A..55M}
}

@ARTICLE{mocnik2016,
       author = {{Mo{\v{c}}nik}, T. and {Clark}, B.~J.~M. and {Anderson}, D.~R. and {Hellier}, C. and {Brown}, D.~J.~A.},
        title = "{Starspots on WASP-85}",
      journal = {\aj},
         year = 2016,
        month = jun,
       volume = {151},
       number = {6},
          eid = {150},
        pages = {150},
          doi = {10.3847/0004-6256/151/6/150},
archivePrefix = {arXiv},
       eprint = {1508.07281},
 primaryClass = {astro-ph.EP},
       adsurl = {https://ui.adsabs.harvard.edu/abs/2016AJ....151..150M}
}

@ARTICLE{mocnik2017_wasp107,
       author = {{Mo{\v{c}}nik}, T. and {Hellier}, C. and {Anderson}, D.~R. and {Clark}, B.~J.~M. and {Southworth}, J.},
        title = "{Starspots on WASP-107 and pulsations of WASP-118}",
      journal = {\mnras},
         year = 2017,
        month = aug,
       volume = {469},
       number = {2},
        pages = {1622-1629},
          doi = {10.1093/mnras/stx972},
archivePrefix = {arXiv},
       eprint = {1702.05078},
 primaryClass = {astro-ph.EP},
       adsurl = {https://ui.adsabs.harvard.edu/abs/2017MNRAS.469.1622M}
}

@ARTICLE{mocnik2017_qatar2,
       author = {{Mo{\v{c}}nik}, T. and {Southworth}, J. and {Hellier}, C.},
        title = "{Recurring sets of recurring starspot occultations on exoplanet host Qatar-2}",
      journal = {\mnras},
         year = 2017,
        month = oct,
       volume = {471},
       number = {1},
        pages = {394-403},
          doi = {10.1093/mnras/stx1557},
archivePrefix = {arXiv},
       eprint = {1608.07524},
 primaryClass = {astro-ph.EP},
       adsurl = {https://ui.adsabs.harvard.edu/abs/2017MNRAS.471..394M}
}

@ARTICLE{murray2026,
       author = {{Murray}, C.~A. and {Garcia}, L. and {Rackham}, B.~V. and {Berta-Thompson}, Z. and {Feinstein}, A.~D. and {Mercier}, S.~J. and {Charnay}, B. and {Hebb}, L. and {Libby-Roberts}, J.~E. and {Rotman}, Y. and {Stephens}, A. and {Timmermans}, M. and {Welbanks}, L. and {Barkaoui}, K. and {Canas}, Caleb I. and {Delamer}, M. and {Ducrot}, E. and {Kanodia}, S. and {Mahadevan}, S. and {Ninan}, J.~P. and {de Wit}, J.},
        title = "{A Panchromatic JWST Spectrum of a Giant Starspot on the Fully Convective M-dwarf TOI-3884}",
      journal = {arXiv e-prints},
         year = 2026,
        month = mar,
          eid = {arXiv:2603.15414},
        pages = {arXiv:2603.15414},
          doi = {10.48550/arXiv.2603.15414},
archivePrefix = {arXiv},
       eprint = {2603.15414},
 primaryClass = {astro-ph.EP},
       adsurl = {https://ui.adsabs.harvard.edu/abs/2026arXiv260315414M}
}

@ARTICLE{kochukov2020,
       author = {{Kochukhov}, O. and {Hackman}, T. and {Lehtinen}, J.~J. and {Wehrhahn}, A.},
        title = "{Hidden magnetic fields of young suns}",
      journal = {\aap},
         year = 2020,
        month = mar,
       volume = {635},
          eid = {A142},
        pages = {A142},
          doi = {10.1051/0004-6361/201937185},
archivePrefix = {arXiv},
       eprint = {2002.10469},
 primaryClass = {astro-ph.SR},
       adsurl = {https://ui.adsabs.harvard.edu/abs/2020A&A...635A.142K}
}

@ARTICLE{folsom2016,
       author = {{Folsom}, C.~P. and {Petit}, P. and {Bouvier}, J. and {L{\`e}bre}, A. and {Amard}, L. and {Palacios}, A. and {Morin}, J. and {Donati}, J.-F. and {Jeffers}, S.~V. and {Marsden}, S.~C. and {Vidotto}, A.~A.},
        title = "{The evolution of surface magnetic fields in young solar-type stars - I. The first 250 Myr}",
      journal = {\mnras},
         year = 2016,
        month = mar,
       volume = {457},
       number = {1},
        pages = {580-607},
          doi = {10.1093/mnras/stv2924},
archivePrefix = {arXiv},
       eprint = {1601.00684},
 primaryClass = {astro-ph.SR},
       adsurl = {https://ui.adsabs.harvard.edu/abs/2016MNRAS.457..580F}
}

@ARTICLE{hackman2016,
       author = {{Hackman}, T. and {Lehtinen}, J. and {Ros{\'e}n}, L. and {Kochukhov}, O. and {K{\"a}pyl{\"a}}, M.~J.},
        title = "{Zeeman-Doppler imaging of active young solar-type stars}",
      journal = {\aap},
         year = 2016,
        month = mar,
       volume = {587},
          eid = {A28},
        pages = {A28},
          doi = {10.1051/0004-6361/201527320},
archivePrefix = {arXiv},
       eprint = {1509.02285},
 primaryClass = {astro-ph.SR},
       adsurl = {https://ui.adsabs.harvard.edu/abs/2016A&A...587A..28H}
}

@ARTICLE{bouvier1993,
       author = {{Bouvier}, J. and {Cabrit}, S. and {Fernandez}, M. and {Martin}, E.~L. and {Matthews}, J.~M.},
        title = "{COYOTES I : the photometric variability and rotational evolution of T Tauri stars.}",
      journal = {\aap},
         year = 1993,
        month = may,
       volume = {272},
        pages = {176-206},
       adsurl = {https://ui.adsabs.harvard.edu/abs/1993A&A...272..176B}
}

@misc{triantafillides2026,
      title={The Identification of CS2 and Evidence for Carbon-Sulfur Chemical Coupling in a Warm Giant Exoplanet Atmosphere}, 
      author={Anastasia Triantafillides and Thomas G. Beatty and Matthew C. Nixon and Taylor J. Bell and Everett Schlawin and Luis Welbanks and Thomas P. Greene and Melinda Soares-Furtado and Jonathan J. Fortney Michael R. Line and Nishil Mehta and Sagnick Mukherjee and Matthew M. Murphy and Kazumasa Ohno and Vivien Parmentier and Yoav Rotman and Lindsey S. Wiser},
      year={2026},
      eprint={2604.13168},
      archivePrefix={arXiv},
      primaryClass={astro-ph.EP},
      url={https://arxiv.org/abs/2604.13168}, 
}

@ARTICLE{olah2016,
       author = {{Ol{\'a}h}, K. and {K{\H{o}}v{\'a}ri}, Zs. and {Petrovay}, K. and {Soon}, W. and {Baliunas}, S. and {Koll{\'a}th}, Z. and {Vida}, K.},
        title = "{Magnetic cycles at different ages of stars}",
      journal = {\aap},
         year = 2016,
        month = jun,
       volume = {590},
          eid = {A133},
        pages = {A133},
          doi = {10.1051/0004-6361/201628479},
archivePrefix = {arXiv},
       eprint = {1604.06701},
 primaryClass = {astro-ph.SR},
       adsurl = {https://ui.adsabs.harvard.edu/abs/2016A&A...590A.133O}
}

@ARTICLE{waite2017,
       author = {{Waite}, I.~A. and {Marsden}, S.~C. and {Carter}, B.~D. and {Petit}, P. and {Jeffers}, S.~V. and {Morin}, J. and {Vidotto}, A.~A. and {Donati}, J.-F. and {BCool Collaboration}},
        title = "{Magnetic fields on young, moderately rotating Sun-like stars - II. EK Draconis (HD 129333)}",
      journal = {\mnras},
         year = 2017,
        month = feb,
       volume = {465},
       number = {2},
        pages = {2076-2091},
          doi = {10.1093/mnras/stw2731},
archivePrefix = {arXiv},
       eprint = {1611.07751},
 primaryClass = {astro-ph.SR},
       adsurl = {https://ui.adsabs.harvard.edu/abs/2017MNRAS.465.2076W}
}

@software{banzai_zenodo,
  author       = {Curtis McCully and
                  Monica Turner and
                  N Volgenau and
                  Daniel Harbeck and
                  Stefano Valenti and
                  Austin Riba and
                  Etienne Bachelet and
                  Ira W. Snyder and
                  Brodie Kurczynski and
                  Martin Norbury and
                  Rachel Street},
  title        = {LCOGT/banzai: Initial Release},
  year         = 2018,
  publisher    = {Zenodo},
  doi          = {10.5281/zenodo.1257559},
  url          = {https://doi.org/10.5281/zenodo.1257559},
}

@software{ndtest,
  author = {{Li}, Z.},
  title = {ndtest},
  url = {https://github.com/syrte/ndtest},
  year = 2019
}

@ARTICLE{luhman2023,
       author = {{Luhman}, K.~L.},
        title = "{A Census of the Taurus Star-forming Region and Neighboring Associations with Gaia}",
      journal = {\aj},
         year = 2023,
        month = feb,
       volume = {165},
       number = {2},
          eid = {37},
        pages = {37},
          doi = {10.3847/1538-3881/ac9da3},
archivePrefix = {arXiv},
       eprint = {2211.09785},
 primaryClass = {astro-ph.GA},
       adsurl = {https://ui.adsabs.harvard.edu/abs/2023AJ....165...37L}
}

@ARTICLE{luhman2024,
       author = {{Luhman}, K.~L.},
        title = "{A Census of the {\ensuremath{\beta}} Pic Moving Group and Other Nearby Associations with Gaia}",
      journal = {\aj},
         year = 2024,
        month = oct,
       volume = {168},
       number = {4},
          eid = {159},
        pages = {159},
          doi = {10.3847/1538-3881/ad697d},
archivePrefix = {arXiv},
       eprint = {2409.06092},
 primaryClass = {astro-ph.GA},
       adsurl = {https://ui.adsabs.harvard.edu/abs/2024AJ....168..159L}
}

@article{kass_bayes_1995,
	title = {Bayes {Factors}},
	volume = {90},
	url = {https://doi.org/10.1080/01621459.1995.10476572},
	doi = {10.1080/01621459.1995.10476572},
	number = {430},
	journal = {Journal of the American Statistical Association},
	publisher = {Taylor \& Francis},
	author = {Kass, Robert E. and Raftery, Adrian E.},
	year = {1995},
	pages = {773--795},
}

@article{bayarri2019,
	title = {Prior-based {Bayesian} information criterion},
	volume = {3},
	url = {https://doi.org/10.1080/24754269.2019.1582126},
	doi = {10.1080/24754269.2019.1582126},
	number = {1},
	journal = {Statistical Theory and Related Fields},
	publisher = {Taylor \& Francis},
	author = {Bayarri, M. J. and Berger, James O. and Jang, Woncheol and Ray, Surajit and Pericchi, Luis R. and Visser, Ingmar},
	year = {2019},
	pages = {2--13},
}

@ARTICLE{gaiaDR3,
       author = {{Gaia Collaboration} and {Vallenari}, A. and {Brown}, A.~G.~A. and {Prusti}, T. and {de Bruijne}, J.~H.~J. and {Arenou}, F. and {Babusiaux}, C. and {Biermann}, M. and {Creevey}, O.~L. and {Ducourant}, C. and {Evans}, D.~W. and {Eyer}, L. and {Guerra}, R. and {Hutton}, A. and {Jordi}, C. and {Klioner}, S.~A. and {Lammers}, U.~L. and {Lindegren}, L. and {Luri}, X. and {Mignard}, F. and {Panem}, C. and {Pourbaix}, D. and {Randich}, S. and {Sartoretti}, P. and {Soubiran}, C. and {Tanga}, P. and {Walton}, N.~A. and {Bailer-Jones}, C.~A.~L. and {Bastian}, U. and {Drimmel}, R. and {Jansen}, F. and {Katz}, D. and {Lattanzi}, M.~G. and {van Leeuwen}, F. and {Bakker}, J. and {Cacciari}, C. and {Casta{\~n}eda}, J. and {De Angeli}, F. and {Fabricius}, C. and {Fouesneau}, M. and {Fr{\'e}mat}, Y. and {Galluccio}, L. and {Guerrier}, A. and {Heiter}, U. and {Masana}, E. and {Messineo}, R. and {Mowlavi}, N. and {Nicolas}, C. and {Nienartowicz}, K. and {Pailler}, F. and {Panuzzo}, P. and {Riclet}, F. and {Roux}, W. and {Seabroke}, G.~M. and {Sordo}, R. and {Th{\'e}venin}, F. and {Gracia-Abril}, G. and {Portell}, J. and {Teyssier}, D. and {Altmann}, M. and {Andrae}, R. and {Audard}, M. and {Bellas-Velidis}, I. and {Benson}, K. and {Berthier}, J. and {Blomme}, R. and {Burgess}, P.~W. and {Busonero}, D. and {Busso}, G. and {C{\'a}novas}, H. and {Carry}, B. and {Cellino}, A. and {Cheek}, N. and {Clementini}, G. and {Damerdji}, Y. and {Davidson}, M. and {de Teodoro}, P. and {Nu{\~n}ez Campos}, M. and {Delchambre}, L. and {Dell'Oro}, A. and {Esquej}, P. and {Fern{\'a}ndez-Hern{\'a}ndez}, J. and {Fraile}, E. and {Garabato}, D. and {Garc{\'\i}a-Lario}, P. and {Gosset}, E. and {Haigron}, R. and {Halbwachs}, J.-L. and {Hambly}, N.~C. and {Harrison}, D.~L. and {Hern{\'a}ndez}, J. and {Hestroffer}, D. and {Hodgkin}, S.~T. and {Holl}, B. and {Jan{\ss}en}, K. and {Jevardat de Fombelle}, G. and {Jordan}, S. and {Krone-Martins}, A. and {Lanzafame}, A.~C. and {L{\"o}ffler}, W. and {Marchal}, O. and {Marrese}, P.~M. and {Moitinho}, A. and {Muinonen}, K. and {Osborne}, P. and {Pancino}, E. and {Pauwels}, T. and {Recio-Blanco}, A. and {Reyl{\'e}}, C. and {Riello}, M. and {Rimoldini}, L. and {Roegiers}, T. and {Rybizki}, J. and {Sarro}, L.~M. and {Siopis}, C. and {Smith}, M. and {Sozzetti}, A. and {Utrilla}, E. and {van Leeuwen}, M. and {Abbas}, U. and {{\'A}brah{\'a}m}, P. and {Abreu Aramburu}, A. and {Aerts}, C. and {Aguado}, J.~J. and {Ajaj}, M. and {Aldea-Montero}, F. and {Altavilla}, G. and {{\'A}lvarez}, M.~A. and {Alves}, J. and {Anders}, F. and {Anderson}, R.~I. and {Anglada Varela}, E. and {Antoja}, T. and {Baines}, D. and {Baker}, S.~G. and {Balaguer-N{\'u}{\~n}ez}, L. and {Balbinot}, E. and {Balog}, Z. and {Barache}, C. and {Barbato}, D. and {Barros}, M. and {Barstow}, M.~A. and {Bartolom{\'e}}, S. and {Bassilana}, J.-L. and {Bauchet}, N. and {Becciani}, U. and {Bellazzini}, M. and {Berihuete}, A. and {Bernet}, M. and {Bertone}, S. and {Bianchi}, L. and {Binnenfeld}, A. and {Blanco-Cuaresma}, S. and {Blazere}, A. and {Boch}, T. and {Bombrun}, A. and {Bossini}, D. and {Bouquillon}, S. and {Bragaglia}, A. and {Bramante}, L. and {Breedt}, E. and {Bressan}, A. and {Brouillet}, N. and {Brugaletta}, E. and {Bucciarelli}, B. and {Burlacu}, A. and {Butkevich}, A.~G. and {Buzzi}, R. and {Caffau}, E. and {Cancelliere}, R. and {Cantat-Gaudin}, T. and {Carballo}, R. and {Carlucci}, T. and {Carnerero}, M.~I. and {Carrasco}, J.~M. and {Casamiquela}, L. and {Castellani}, M. and {Castro-Ginard}, A. and {Chaoul}, L. and {Charlot}, P. and {Chemin}, L. and {Chiaramida}, V. and {Chiavassa}, A. and {Chornay}, N. and {Comoretto}, G. and {Contursi}, G. and {Cooper}, W.~J. and {Cornez}, T. and {Cowell}, S. and {Crifo}, F. and {Cropper}, M. and {Crosta}, M. and {Crowley}, C. and {Dafonte}, C. and {Dapergolas}, A. and {David}, M. and {David}, P. and {de Laverny}, P. and {De Luise}, F. and {De March}, R.},
        title = "{Gaia Data Release 3. Summary of the content and survey properties}",
      journal = {\aap},
         year = 2023,
        month = jun,
       volume = {674},
          eid = {A1},
        pages = {A1},
          doi = {10.1051/0004-6361/202243940},
archivePrefix = {arXiv},
       eprint = {2208.00211},
 primaryClass = {astro-ph.GA},
       adsurl = {https://ui.adsabs.harvard.edu/abs/2023A&A...674A...1G}
}

@ARTICLE{peacock1983_fitting,
       author = {{Peacock}, J.~A.},
        title = "{Two-dimensional goodness-of-fit testing in astronomy.}",
      journal = {\mnras},
         year = 1983,
        month = feb,
       volume = {202},
        pages = {615-627},
          doi = {10.1093/mnras/202.3.615},
       adsurl = {https://ui.adsabs.harvard.edu/abs/1983MNRAS.202..615P}
}

@ARTICLE{fasano1987_KStest,
       author = {{Fasano}, G. and {Franceschini}, A.},
        title = "{A multidimensional version of the Kolmogorov-Smirnov test}",
      journal = {\mnras},
         year = 1987,
        month = mar,
       volume = {225},
        pages = {155-170},
          doi = {10.1093/mnras/225.1.155},
       adsurl = {https://ui.adsabs.harvard.edu/abs/1987MNRAS.225..155F}
}

@ARTICLE{ayres2025_xraycycles,
       author = {{Ayres}, Thomas},
        title = "{Landscape of Coronal X-Ray Variability and Cycles}",
      journal = {\aj},
         year = 2025,
        month = may,
       volume = {169},
       number = {5},
          eid = {281},
        pages = {281},
          doi = {10.3847/1538-3881/adc570},
       adsurl = {https://ui.adsabs.harvard.edu/abs/2025AJ....169..281A}
}

@ARTICLE{sanzforcada2013_iHorcycle,
       author = {{Sanz-Forcada}, J. and {Stelzer}, B. and {Metcalfe}, T.~S.},
        title = "{{\ensuremath{\i}}Horologi, the first coronal activity cycle in a young solar-like star}",
      journal = {\aap},
         year = 2013,
        month = may,
       volume = {553},
          eid = {L6},
        pages = {L6},
          doi = {10.1051/0004-6361/201321388},
archivePrefix = {arXiv},
       eprint = {1305.1132},
 primaryClass = {astro-ph.SR},
       adsurl = {https://ui.adsabs.harvard.edu/abs/2013A&A...553L...6S}
}

@ARTICLE{sanzforcada2019_iHorcycle,
       author = {{Sanz-Forcada}, J. and {Stelzer}, B. and {Coffaro}, M. and {Raetz}, S. and {Alvarado-G{\'o}mez}, J.~D.},
        title = "{Multi-wavelength variability of the young solar analog {\ensuremath{\i}} Horologii. X-ray cycle, star spots, flares, and UV emission}",
      journal = {\aap},
         year = 2019,
        month = nov,
       volume = {631},
          eid = {A45},
        pages = {A45},
          doi = {10.1051/0004-6361/201935703},
archivePrefix = {arXiv},
       eprint = {1909.01320},
 primaryClass = {astro-ph.SR},
       adsurl = {https://ui.adsabs.harvard.edu/abs/2019A&A...631A..45S}
}

@ARTICLE{metcalfe2010_iHorcycle,
       author = {{Metcalfe}, T.~S. and {Basu}, S. and {Henry}, T.~J. and {Soderblom}, D.~R. and {Judge}, P.~G. and {Kn{\"o}lker}, M. and {Mathur}, S. and {Rempel}, M.},
        title = "{Discovery of a 1.6 Year Magnetic Activity Cycle in the Exoplanet Host Star {\ensuremath{\i}} Horologii}",
      journal = {\apjl},
         year = 2010,
        month = nov,
       volume = {723},
       number = {2},
        pages = {L213-L217},
          doi = {10.1088/2041-8205/723/2/L213},
archivePrefix = {arXiv},
       eprint = {1009.5399},
 primaryClass = {astro-ph.SR},
       adsurl = {https://ui.adsabs.harvard.edu/abs/2010ApJ...723L.213M}
}

@ARTICLE{singh2024_ABDor,
       author = {{Singh}, Gurpreet and {Pandey}, J.~C.},
        title = "{AB Dor: Coronal Imaging and Activity Cycles}",
      journal = {\apj},
         year = 2024,
        month = may,
       volume = {966},
       number = {1},
          eid = {86},
        pages = {86},
          doi = {10.3847/1538-4357/ad2f2e},
archivePrefix = {arXiv},
       eprint = {2402.19083},
 primaryClass = {astro-ph.SR},
       adsurl = {https://ui.adsabs.harvard.edu/abs/2024ApJ...966...86S}
}

@ARTICLE{allan2026,
       author = {{Allan}, Andrew P. and {Vidotto}, Aline A. and {Sanz-Forcada}, Jorge and {Villarreal D'Angelo}, Carolina},
        title = "{The effects of stellar activity cycles on planetary atmospheric escape and the He I 1083 nm transit signature}",
      journal = {\mnras},
         year = 2026,
        month = jan,
       volume = {545},
       number = {2},
          eid = {staf1855},
        pages = {staf1855},
          doi = {10.1093/mnras/staf1855},
archivePrefix = {arXiv},
       eprint = {2510.23282},
 primaryClass = {astro-ph.EP},
       adsurl = {https://ui.adsabs.harvard.edu/abs/2026MNRAS.545f1855A}
}

@misc{Murphy2026,
      title={KRONOS I: The $1{-}2.8\mu$m JWST Transmission Spectrum of the 23 Myr V1298 Tau c}, 
      author={Matthew M. Murphy and Matthew C. Nixon and Adina D. Feinstein and Luis Welbanks and Girish M. Duvvuri and Saugata Barat and Benjamin V. Rackham and Darryl Z. Seligman and Michael Radica and Ian J. M. Crossfield and Kevin France and John H. Livingston and Jonathan Lunine and Rafael Luque and Catriona Murray and Sagnick Mukherjee and Biruk Nardos and Sydney Petz and Hinna Shivkumar},
      year={2026},
      eprint={2606.03740},
      archivePrefix={arXiv},
      primaryClass={astro-ph.EP},
      url={https://arxiv.org/abs/2606.03740}, 
}

@ARTICLE{solanki2003_sunspots,
       author = {{Solanki}, Sami K.},
        title = "{Sunspots: An overview}",
      journal = {\aapr},
         year = 2003,
        month = jan,
       volume = {11},
       number = {2-3},
        pages = {153-286},
          doi = {10.1007/s00159-003-0018-4},
       adsurl = {https://ui.adsabs.harvard.edu/abs/2003A&ARv..11..153S}
}

@ARTICLE{strassmeier2009_starspots,
       author = {{Strassmeier}, Klaus G.},
        title = "{Starspots}",
      journal = {\aapr},
         year = 2009,
        month = sep,
       volume = {17},
       number = {3},
        pages = {251-308},
          doi = {10.1007/s00159-009-0020-6},
       adsurl = {https://ui.adsabs.harvard.edu/abs/2009A&ARv..17..251S}
}

@ARTICLE{radica_super-solar_2026,
       author = {{Radica}, Michael and {Taylor}, Jake and {Rotman}, Yoav and {Blecic}, Jasmina and {Welbanks}, Luis and {Ahrer}, Eva-Maria and {Christie}, Duncan and {Coulombe}, Louis-Philippe and {Lowry}, Gillis and {Murphy}, Matthew M. and {Feinstein}, Adina D. and {Lafreni{\`e}re}, David and {MacDonald}, Ryan J. and {Mayne}, Nathan J. and {Tsai}, Shang-Min and {Zamyatina}, Maria},
        title = "{Supersolar Metallicity and Tentative Evidence for Photochemistry on WASP-96 b from JWST and Ground-based VLT Transmission Spectroscopy}",
      journal = {\aj},
         year = 2026,
        month = may,
       volume = {171},
       number = {5},
          eid = {314},
        pages = {314},
          doi = {10.3847/1538-3881/ae5b9f},
archivePrefix = {arXiv},
       eprint = {2604.05049},
 primaryClass = {astro-ph.EP},
       adsurl = {https://ui.adsabs.harvard.edu/abs/2026AJ....171..314R}
}

@article{Jensen1956, title={Variations in the relative size of penumbra and umbra of sunspots in the Years 1878-1954}, volume={19}, ISSN={0365-0499}, note={ADS Bibcode: 1956AnAp...19..165J}, journal={Annales d’Astrophysique}, author={Jensen, E. and Nordø, J. and Ringnes, T. S.}, year={1956}, month=jan, pages={165} }

@article{Beck1993, title={A study of the contrast of sunspots from photometric images}, volume={146}, ISSN={1573-093X}, DOI={10.1007/BF00662169}, abstractNote={The thermal contrast α, and the umbra-penumbraAu/Ap, were calculated for 63 sunspots of various sizes and morphologies. Contrary to the assumptions of the PSI model, α andAu/Ap were found to be quite variable. The values of α ranged from 0.1807 to 0.4266;Au/Ap ranged from 0.0089 to 0.4899. The values of α andAu/Ap correlated well (r = 0.6018;p<0.005) and the regression for α andAu/Ap was obtained: α = (0.220 ± 0.016) + (0.340 ± O.06)Au/Ap. The values of α andAu/Ap were then compared with complexity ratings, magnetic field strength, time, and μ. The quantities α andAu/Ap were found to be independent of the complexity, magnetic field strength, and time factors. The correlation between α andAu/Ap lead to the proposed division of α into an umbral thermal contrast αu, and a penumbral thermal contrast αp. These values were calculated from the photometric data: αu = 0.57 ± 0.01 and αp = 0.26 ± 0.006.}, number={1}, journal={Solar Physics}, author={Beck, John G. and Chapman, Gary A.}, year={1993}, month=july, pages={49–60}, language={en} }

@ARTICLE{finociety_2023,
       author = {{Finociety}, B. and {Donati}, J.-F. and {Cristofari}, P.~I. and {Moutou}, C. and {Cadieux}, C. and {Cook}, N.~J. and {Artigau}, E. and {Baruteau}, C. and {Debras}, F. and {Fouqu{\'e}}, P. and {Bouvier}, J. and {Alencar}, S.~H.~P. and {Delfosse}, X. and {Grankin}, K. and {Carmona}, A. and {Petit}, P. and {K{\'o}sp{\'a}l}, {\'A}. and {The Sls/Spice Consortium}},
        title = "{Monitoring the young planet host V1298 Tau with SPIRou: planetary system and evolving large-scale magnetic field}",
      journal = {\mnras},
         year = 2023,
        month = dec,
       volume = {526},
       number = {3},
        pages = {4627-4672},
          doi = {10.1093/mnras/stad3012},
archivePrefix = {arXiv},
       eprint = {2310.02613},
 primaryClass = {astro-ph.SR},
       adsurl = {https://ui.adsabs.harvard.edu/abs/2023MNRAS.526.4627F}
}

@ARTICLE{suarez_mascareno_2022,
       author = {{Su{\'a}rez Mascare{\~n}o}, A. and {Damasso}, M. and {Lodieu}, N. and {Sozzetti}, A. and {B{\'e}jar}, V.~J.~S. and {Benatti}, S. and {Zapatero Osorio}, M.~R. and {Micela}, G. and {Rebolo}, R. and {Desidera}, S. and {Murgas}, F. and {Claudi}, R. and {Gonz{\'a}lez Hern{\'a}ndez}, J.~I. and {Malavolta}, L. and {del Burgo}, C. and {D'Orazi}, V. and {Amado}, P.~J. and {Locci}, D. and {Tabernero}, H.~M. and {Marzari}, F. and {Aguado}, D.~S. and {Turrini}, D. and {Cardona Guill{\'e}n}, C. and {Toledo-Padr{\'o}n}, B. and {Maggio}, A. and {Aceituno}, J. and {Bauer}, F.~F. and {Caballero}, J.~A. and {Chinchilla}, P. and {Esparza-Borges}, E. and {Gonz{\'a}lez-{\'A}lvarez}, E. and {Granzer}, T. and {Luque}, R. and {Mart{\'\i}n}, E.~L. and {Nowak}, G. and {Oshagh}, M. and {Pall{\'e}}, E. and {Parviainen}, H. and {Quirrenbach}, A. and {Reiners}, A. and {Ribas}, I. and {Strassmeier}, K.~G. and {Weber}, M. and {Mallonn}, M.},
        title = "{Rapid contraction of giant planets orbiting the 20-million-year-old star V1298 Tau}",
      journal = {Nature Astronomy},
         year = 2021,
        month = dec,
       volume = {6},
        pages = {232-240},
          doi = {10.1038/s41550-021-01533-7},
archivePrefix = {arXiv},
       eprint = {2111.09193},
 primaryClass = {astro-ph.EP},
       adsurl = {https://ui.adsabs.harvard.edu/abs/2022NatAs...6..232S}
}

@ARTICLE{johnson2022,
       author = {{Johnson}, Marshall C. and {David}, Trevor J. and {Petigura}, Erik A. and {Isaacson}, Howard T. and {Van Zandt}, Judah and {Ilyin}, Ilya and {Strassmeier}, Klaus and {Mallonn}, Matthias and {Zhou}, George and {Mann}, Andrew W. and {Livingston}, John H. and {Luger}, Rodrigo and {Dai}, Fei and {Weiss}, Lauren M. and {Mo{\v{c}}nik}, Teo and {Giacalone}, Steven and {Hill}, Michelle L. and {Rice}, Malena and {Blunt}, Sarah and {Rubenzahl}, Ryan and {Dalba}, Paul A. and {Esquerdo}, Gilbert A. and {Berlind}, Perry and {Calkins}, Michael L. and {Foreman-Mackey}, Daniel},
        title = "{An Aligned Orbit for the Young Planet V1298 Tau b}",
      journal = {\aj},
         year = 2022,
        month = jun,
       volume = {163},
       number = {6},
          eid = {247},
        pages = {247},
          doi = {10.3847/1538-3881/ac6271},
archivePrefix = {arXiv},
       eprint = {2110.10707},
 primaryClass = {astro-ph.EP},
       adsurl = {https://ui.adsabs.harvard.edu/abs/2022AJ....163..247J}
}

@ARTICLE{luger21,
       author = {{Luger}, Rodrigo and {Foreman-Mackey}, Daniel and {Hedges}, Christina and {Hogg}, David W.},
        title = "{Mapping Stellar Surfaces. I. Degeneracies in the Rotational Light-curve Problem}",
      journal = {\aj},
         year = 2021,
        month = sep,
       volume = {162},
       number = {3},
          eid = {123},
        pages = {123},
          doi = {10.3847/1538-3881/abfdb8},
archivePrefix = {arXiv},
       eprint = {2102.00007},
 primaryClass = {astro-ph.SR},
       adsurl = {https://ui.adsabs.harvard.edu/abs/2021AJ....162..123L}
}

@ARTICLE{david19_b,
       author = {{David}, Trevor J. and {Cody}, Ann Marie and {Hedges}, Christina L. and {Mamajek}, Eric E. and {Hillenbrand}, Lynne A. and {Ciardi}, David R. and {Beichman}, Charles A. and {Petigura}, Erik A. and {Fulton}, Benjamin J. and {Isaacson}, Howard T. and {Howard}, Andrew W. and {Gagn{\'e}}, Jonathan and {Saunders}, Nicholas K. and {Rebull}, Luisa M. and {Stauffer}, John R. and {Vasisht}, Gautam and {Hinkley}, Sasha},
        title = "{A Warm Jupiter-sized Planet Transiting the Pre-main-sequence Star V1298 Tau}",
      journal = {\aj},
         year = 2019,
        month = aug,
       volume = {158},
       number = {2},
          eid = {79},
        pages = {79},
          doi = {10.3847/1538-3881/ab290f},
archivePrefix = {arXiv},
       eprint = {1902.09670},
 primaryClass = {astro-ph.EP},
       adsurl = {https://ui.adsabs.harvard.edu/abs/2019AJ....158...79D}
}

@ARTICLE{niriss1_overview,
       author = {{Doyon}, Ren{\'e} and {Willott}, Chris J. and {Hutchings}, John B. and {Sivaramakrishnan}, Anand and {Albert}, Lo{\"\i}c and {Lafreni{\`e}re}, David and {Rowlands}, Neil and {Bego{\~n}a Vila}, M. and {Martel}, Andr{\'e} R. and {LaMassa}, Stephanie and {Aldridge}, David and {Artigau}, {\'E}tienne and {Cameron}, Peter and {Chayer}, Pierre and {Cook}, Neil J. and {Cooper}, Rachel A. and {Darveau-Bernier}, Antoine and {Dupuis}, Jean and {Earnshaw}, Colin and {Espinoza}, N{\'e}stor and {Filippazzo}, Joseph C. and {Fullerton}, Alexander W. and {Gaudreau}, Daniel and {Gawlik}, Roman and {Goudfrooij}, Paul and {Haley}, Craig and {Kammerer}, Jens and {Kendall}, David and {Lambros}, Scott D. and {Ignat}, Luminita Ilinca and {Maszkiewicz}, Michael and {McColgan}, Ashley and {Morishita}, Takahiro and {Ouellette}, Nathalie N.-Q. and {Pacifici}, Camilla and {Philippi}, Natasha and {Radica}, Michael and {Ravindranath}, Swara and {Rowe}, Jason and {Roy}, Arpita and {Roy}, Niladri and {Saad}, Karl and {Sohn}, Sangmo Tony and {Talens}, Geert Jan and {Touahri}, Driss and {Thatte}, Deepashri and {Taylor}, Joanna M. and {Vandal}, Thomas and {Volk}, Kevin and {Wander}, Michel and {Warner}, Gerald and {Zheng}, Sheng-Hai and {Zhou}, Julia and {Abraham}, Roberto and {Beaulieu}, Mathilde and {Benneke}, Bj{\"o}rn and {Ferrarese}, Laura and {Jayawardhana}, Ray and {Johnstone}, Doug and {Kaltenegger}, Lisa and {Meyer}, Michael R. and {Pipher}, Judy L. and {Rameau}, Julien and {Rieke}, Marcia and {Salhi}, Salma and {Sawicki}, Marcin},
        title = "{The Near Infrared Imager and Slitless Spectrograph for the James Webb Space Telescope. I. Instrument Overview and In-flight Performance}",
      journal = {\pasp},
         year = 2023,
        month = sep,
       volume = {135},
       number = {1051},
          eid = {098001},
        pages = {098001},
          doi = {10.1088/1538-3873/acd41b},
archivePrefix = {arXiv},
       eprint = {2306.03277},
 primaryClass = {astro-ph.IM},
       adsurl = {https://ui.adsabs.harvard.edu/abs/2023PASP..135i8001D}
}

@ARTICLE{david2019_v1298tau,
       author = {{David}, Trevor J. and {Petigura}, Erik A. and {Luger}, Rodrigo and {Foreman-Mackey}, Daniel and {Livingston}, John H. and {Mamajek}, Eric E. and {Hillenbrand}, Lynne A.},
        title = "{Four Newborn Planets Transiting the Young Solar Analog V1298 Tau}",
      journal = {\apjl},
         year = 2019,
        month = nov,
       volume = {885},
       number = {1},
          eid = {L12},
        pages = {L12},
          doi = {10.3847/2041-8213/ab4c99},
archivePrefix = {arXiv},
       eprint = {1910.04563},
 primaryClass = {astro-ph.EP},
       adsurl = {https://ui.adsabs.harvard.edu/abs/2019ApJ...885L..12D}
}

@ARTICLE{exotedrf,
       author = {{Radica}, Michael},
        title = "{exoTEDRF: An EXOplanet Transit and Eclipse Data Reduction Framework}",
      journal = {The Journal of Open Source Software},
         year = 2024,
        month = aug,
       volume = {9},
       number = {100},
          eid = {6898},
        pages = {6898},
          doi = {10.21105/joss.06898},
archivePrefix = {arXiv},
       eprint = {2407.17541},
 primaryClass = {astro-ph.IM},
       adsurl = {https://ui.adsabs.harvard.edu/abs/2024JOSS....9.6898R}
}

@article{radica2023_wasp96,
   title={Awesome SOSS: transmission spectroscopy of WASP-96b with NIRISS/SOSS},
   volume={524},
   ISSN={1365-2966},
   url={http://dx.doi.org/10.1093/mnras/stad1762},
   DOI={10.1093/mnras/stad1762},
   number={1},
   journal={Monthly Notices of the Royal Astronomical Society},
   publisher={Oxford University Press (OUP)},
   author={Radica, Michael and Welbanks, Luis and Espinoza, Néstor and Taylor, Jake and Coulombe, Louis-Philippe and Feinstein, Adina D and Goyal, Jayesh and Scarsdale, Nicholas and Albert, Loïc and Baghel, Priyanka and Bean, Jacob L and Blecic, Jasmina and Lafrenière, David and MacDonald, Ryan J and Zamyatina, Maria and Allart1, Romain and Artigau, Étienne and Batalha, Natasha E and Cook, Neil James and Cowan, Nicolas B and Dang, Lisa and Doyon, René and Fournier-Tondreau, Marylou and Johnstone, Doug and Line, Michael R and Moran, Sarah E and Mukherjee, Sagnick and Pelletier, Stefan and Roy, Pierre-Alexis and Talens, Geert Jan and Filippazzo, Joseph and Pontoppidan, Klaus and Volk, Kevin},
   year={2023},
   month=jun, pages={835–856} }

@ARTICLE{feinstein2022_v1298tauTESS,
       author = {{Feinstein}, Adina D. and {David}, Trevor J. and {Montet}, Benjamin T. and {Foreman-Mackey}, Daniel and {Livingston}, John H. and {Mann}, Andrew W.},
        title = "{V1298 Tau with TESS: Updated Ephemerides, Radii, and Period Constraints from a Second Transit of V1298 Tau e}",
      journal = {\apjl},
         year = 2022,
        month = jan,
       volume = {925},
       number = {1},
          eid = {L2},
        pages = {L2},
          doi = {10.3847/2041-8213/ac4745},
archivePrefix = {arXiv},
       eprint = {2111.08660},
 primaryClass = {astro-ph.EP},
       adsurl = {https://ui.adsabs.harvard.edu/abs/2022ApJ...925L...2F}
}

@ARTICLE{feinstein2021_v1298tauGemini,
       author = {{Feinstein}, Adina D. and {Montet}, Benjamin T. and {Johnson}, Marshall C. and {Bean}, Jacob L. and {David}, Trevor J. and {Gully-Santiago}, Michael A. and {Livingston}, John H. and {Luger}, Rodrigo},
        title = "{H-alpha and Ca II Infrared Triplet Variations During a Transit of the 23 Myr Planet V1298 Tau c}",
      journal = {\aj},
         year = 2021,
        month = nov,
       volume = {162},
       number = {5},
          eid = {213},
        pages = {213},
          doi = {10.3847/1538-3881/ac1f24},
archivePrefix = {arXiv},
       eprint = {2107.01213},
 primaryClass = {astro-ph.EP},
       adsurl = {https://ui.adsabs.harvard.edu/abs/2021AJ....162..213F}
}

@ARTICLE{barat2025_v1298taub,
       author = {{Barat}, Saugata and {D{\'e}sert}, Jean-Michel and {Mukherjee}, Sagnick and {Goyal}, Jayesh M. and {Xue}, Qiao and {Kawashima}, Yui and {Vazan}, Allona and {Misener}, William and {Schlichting}, Hilke E. and {Fortney}, Jonathan J. and {Bean}, Jacob L. and {Avarsekar}, Swaroop and {Henry}, Gregory W. and {Baeyens}, Robin and {Line}, Michael R. and {Livingston}, John H. and {David}, Trevor and {Petigura}, Erik A. and {Sikora}, James T. and {Shivkumar}, Hinna and {Feinstein}, Adina D. and {Oklop{\v{c}}i{\'c}}, Antonija},
        title = "{A Metal-poor Atmosphere with a Hot Interior for a Young Sub-Neptune Progenitor: JWST/NIRSpec Transmission Spectrum of V1298 Tau b}",
      journal = {\aj},
         year = 2025,
        month = sep,
       volume = {170},
       number = {3},
          eid = {165},
        pages = {165},
          doi = {10.3847/1538-3881/adec89},
archivePrefix = {arXiv},
       eprint = {2507.08837},
 primaryClass = {astro-ph.EP},
       adsurl = {https://ui.adsabs.harvard.edu/abs/2025AJ....170..165B}
}

@ARTICLE{fleck,
       author = {{Morris}, Brett},
        title = "{fleck: Fast approximate light curves for starspot rotational modulation}",
      journal = {The Journal of Open Source Software},
         year = 2020,
        month = mar,
       volume = {5},
       number = {47},
          eid = {2103},
        pages = {2103},
          doi = {10.21105/joss.02103},
       adsurl = {https://ui.adsabs.harvard.edu/abs/2020JOSS....5.2103M}
}

@ARTICLE{batman,
       author = {{Kreidberg}, Laura},
        title = "{batman: BAsic Transit Model cAlculatioN in Python}",
      journal = {\pasp},
         year = 2015,
        month = nov,
       volume = {127},
       number = {957},
        pages = {1161},
          doi = {10.1086/683602},
archivePrefix = {arXiv},
       eprint = {1507.08285},
 primaryClass = {astro-ph.EP},
       adsurl = {https://ui.adsabs.harvard.edu/abs/2015PASP..127.1161K}
}

@article{emcee,
   title={<tt>emcee</tt>: The MCMC Hammer},
   volume={125},
   ISSN={1538-3873},
   url={http://dx.doi.org/10.1086/670067},
   DOI={10.1086/670067},
   number={925},
   journal={Publications of the Astronomical Society of the Pacific},
   publisher={IOP Publishing},
   author={Foreman-Mackey, Daniel and Hogg, David W. and Lang, Dustin and Goodman, Jonathan},
   year={2013},
   month=mar, pages={306–312} }

@ARTICLE{murray2025_SCEeffects,
       author = {{Murray}, C.~A. and {Berta-Thompson}, Z.},
        title = "{Quantifying the Impact of Starspot-Crossing Events on Retrieved Parameters from Transit Lightcurves}",
      journal = {arXiv e-prints},
         year = 2025,
        month = nov,
          eid = {arXiv:2511.03045},
        pages = {arXiv:2511.03045},
          doi = {10.48550/arXiv.2511.03045},
archivePrefix = {arXiv},
       eprint = {2511.03045},
 primaryClass = {astro-ph.EP},
       adsurl = {https://ui.adsabs.harvard.edu/abs/2025arXiv251103045M}
}

@article{livingston2026_v1298tau,
   title={A young progenitor for the most common planetary systems in the Galaxy},
   volume={649},
   ISSN={1476-4687},
   url={http://dx.doi.org/10.1038/s41586-025-09840-z},
   DOI={10.1038/s41586-025-09840-z},
   number={8096},
   journal={Nature},
   publisher={Springer Science and Business Media LLC},
   author={Livingston, John H. and Petigura, Erik A. and David, Trevor J. and Masuda, Kento and Owen, James and Nesvorný, David and Batygin, Konstantin and de Leon, Jerome and Mori, Mayuko and Ikuta, Kai and Fukui, Akihiko and Watanabe, Noriharu and Orell Miquel, Jaume and Murgas, Felipe and Parviainen, Hannu and Korth, Judith and Libotte, Florence and Abreu García, Néstor and Gallardo, Pedro Pablo Meni and Narita, Norio and Pallé, Enric and Tamura, Motohide and Yonehara, Atsunori and Ridden-Harper, Andrew and Bieryla, Allyson and Trani, Alessandro A. and Mamajek, Eric E. and Ciardi, David R. and Gorjian, Varoujan and Hillenbrand, Lynne A. and Rebull, Luisa M. and Newton, Elisabeth R. and Mann, Andrew W. and Vanderburg, Andrew and Stefánsson, Guðmundur and Mahadevan, Suvrath and Cañas, Caleb and Ninan, Joe and Higuera, Jesus and Todorov, Kamen and Désert, Jean-Michel and Pino, Lorenzo},
   year={2026},
   month=jan, pages={310–314} }

@article{NewEraGrid,
   title={The NewEra model grid},
   volume={698},
   ISSN={1432-0746},
   url={http://dx.doi.org/10.1051/0004-6361/202554171},
   DOI={10.1051/0004-6361/202554171},
   journal={Astronomy \& Astrophysics},
   publisher={EDP Sciences},
   author={Hauschildt, P. H. and Barman, T. and Baron, E. and Aufdenberg, J. P. and Schweitzer, A.},
   year={2025},
   month=may, pages={A47} }

@ARTICLE{CCM89,
       author = {{Cardelli}, Jason A. and {Clayton}, Geoffrey C. and {Mathis}, John S.},
        title = "{The Relationship between Infrared, Optical, and Ultraviolet Extinction}",
      journal = {\apj},
         year = 1989,
        month = oct,
       volume = {345},
        pages = {245},
          doi = {10.1086/167900},
       adsurl = {https://ui.adsabs.harvard.edu/abs/1989ApJ...345..245C}
}

@article{ultranest_algorithm1,
   title={A statistical test for Nested Sampling algorithms},
   volume={26},
   ISSN={1573-1375},
   url={http://dx.doi.org/10.1007/s11222-014-9512-y},
   DOI={10.1007/s11222-014-9512-y},
   number={1–2},
   journal={Statistics and Computing},
   publisher={Springer Science and Business Media LLC},
   author={Buchner, Johannes},
   year={2014},
   month=sep, pages={383–392} }

@article{ultranest_algorithm2,
   title={Collaborative Nested Sampling: Big Data versus Complex Physical Models},
   volume={131},
   ISSN={1538-3873},
   url={http://dx.doi.org/10.1088/1538-3873/aae7fc},
   DOI={10.1088/1538-3873/aae7fc},
   number={1004},
   journal={Publications of the Astronomical Society of the Pacific},
   publisher={IOP Publishing},
   author={Buchner, Johannes},
   year={2019},
   month=aug, pages={108005} }

@ARTICLE{ultranest,
       author = {{Buchner}, Johannes},
        title = "{UltraNest - a robust, general purpose Bayesian inference engine}",
      journal = {The Journal of Open Source Software},
         year = 2021,
        month = apr,
       volume = {6},
       number = {60},
          eid = {3001},
        pages = {3001},
          doi = {10.21105/joss.03001},
archivePrefix = {arXiv},
       eprint = {2101.09604},
 primaryClass = {stat.CO},
       adsurl = {https://ui.adsabs.harvard.edu/abs/2021JOSS....6.3001B}
}

@misc{MURaM_onlinespectrarepo, title={Spot contrasts from 3D MHD MURaM simulations}, url={https://edmond.mpg.de/citation?persistentId=doi:10.17617/3.HS2EE6}, DOI={doi:10.17617/3.HS2EE6}, publisher={Edmond}, author={Narayanamurthy, Smitha H.}, year={2024} }

@article{barat2024_v1298taubc,
       author = {{Barat}, Saugata and {D{\'e}sert}, Jean-Michel and {Goyal}, Jayesh M. and {Vazan}, Allona and {Kawashima}, Yui and {Fortney}, Jonathan J. and {Bean}, Jacob L. and {Line}, Michael R. and {Panwar}, Vatsal and {Jacobs}, Bob and {Shivkumar}, Hinna and {Sikora}, James and {Baeyens}, Robin and {Oklop{\v{c}}i{\'c}}, Antonija and {David}, Trevor J. and {Livingston}, John H.},
        title = "{First comparative exoplanetology within a transiting multi-planet system: Comparing the atmospheres of V1298 Tau b and c}",
      journal = {\aap},
         year = 2024,
        month = dec,
       volume = {692},
          eid = {A198},
        pages = {A198},
          doi = {10.1051/0004-6361/202451127},
archivePrefix = {arXiv},
       eprint = {2407.14995},
 primaryClass = {astro-ph.EP},
       adsurl = {https://ui.adsabs.harvard.edu/abs/2024A&A...692A.198B}
}

@ARTICLE{maunder1922,
       author = {{Maunder}, E.~W.},
        title = "{The sun and sun-spots, 1820-1920}",
      journal = {\mnras},
         year = 1922,
        month = jun,
       volume = {82},
        pages = {534-543},
          doi = {10.1093/mnras/82.9.534},
       adsurl = {https://ui.adsabs.harvard.edu/abs/1922MNRAS..82..534M}
}

@ARTICLE{thao2024,
       author = {{Thao}, Pa Chia and {Mann}, Andrew W. and {Feinstein}, Adina D. and {Gao}, Peter and {Thorngren}, Daniel and {Rotman}, Yoav and {Welbanks}, Luis and {Brown}, Alexander and {Duvvuri}, Girish M. and {France}, Kevin and {Longo}, Isabella and {Sandoval}, Angeli and {Schneider}, P. Christian and {Wilson}, David J. and {Youngblood}, Allison and {Vanderburg}, Andrew and {Barber}, Madyson G. and {Wood}, Mackenna L. and {Batalha}, Natasha E. and {Kraus}, Adam L. and {Murray}, Catriona Anne and {Newton}, Elisabeth R. and {Rizzuto}, Aaron and {Tofflemire}, Benjamin M. and {Tsai}, Shang-Min and {Bean}, Jacob L. and {Berta-Thompson}, Zachory K. and {Evans-Soma}, Thomas M. and {Froning}, Cynthia S. and {Kempton}, Eliza M.-R. and {Miguel}, Yamila and {Pineda}, J. Sebastian},
        title = "{The Featherweight Giant: Unraveling the Atmosphere of a 17 Myr Planet with JWST}",
      journal = {\aj},
         year = 2024,
        month = dec,
       volume = {168},
       number = {6},
          eid = {297},
        pages = {297},
          doi = {10.3847/1538-3881/ad81d7},
archivePrefix = {arXiv},
       eprint = {2409.16355},
 primaryClass = {astro-ph.EP},
       adsurl = {https://ui.adsabs.harvard.edu/abs/2024AJ....168..297T}
}

@misc{Smitha2024_MURaMspotmodels,
      title={First Calculations of Starspot Spectra based on 3D Radiative Magnetohydrodynamics Simulations}, 
      author={H. N. Smitha and Alexander I. Shapiro and Veronika Witzke and Nadiia M. Kostogryz and Yvonne C. Unruh and Tanayveer S. Bhatia and Robert Cameron and Sara Seager and Sami K. Solanki},
      year={2024},
      eprint={2411.14056},
      archivePrefix={arXiv},
      primaryClass={astro-ph.SR},
      url={https://arxiv.org/abs/2411.14056}, 
}

@ARTICLE{Vogler2005_muram,
       author = {{V{\"o}gler}, A. and {Shelyag}, S. and {Sch{\"u}ssler}, M. and {Cattaneo}, F. and {Emonet}, T. and {Linde}, T.},
        title = "{Simulations of magneto-convection in the solar photosphere.  Equations, methods, and results of the MURaM code}",
      journal = {\aap},
         year = 2005,
        month = jan,
       volume = {429},
        pages = {335-351},
          doi = {10.1051/0004-6361:20041507},
       adsurl = {https://ui.adsabs.harvard.edu/abs/2005A&A...429..335V}
}

@ARTICLE{GullySantiago2017_youngspottystar,
       author = {{Gully-Santiago}, Michael A. and {Herczeg}, Gregory J. and {Czekala}, Ian and {Somers}, Garrett and {Grankin}, Konstantin and {Covey}, Kevin R. and {Donati}, J.~F. and {Alencar}, Silvia H.~P. and {Hussain}, Gaitee A.~J. and {Shappee}, Benjamin J. and {Mace}, Gregory N. and {Lee}, Jae-Joon and {Holoien}, T.~W.-S. and {Jose}, Jessy and {Liu}, Chun-Fan},
        title = "{Placing the Spotted T Tauri Star LkCa 4 on an HR Diagram}",
      journal = {\apj},
         year = 2017,
        month = feb,
       volume = {836},
       number = {2},
          eid = {200},
        pages = {200},
          doi = {10.3847/1538-4357/836/2/200},
archivePrefix = {arXiv},
       eprint = {1701.06703},
 primaryClass = {astro-ph.SR},
       adsurl = {https://ui.adsabs.harvard.edu/abs/2017ApJ...836..200G}
}

@ARTICLE{Feinstein2020_youngstarflares,
       author = {{Feinstein}, Adina D. and {Montet}, Benjamin T. and {Ansdell}, Megan and {Nord}, Brian and {Bean}, Jacob L. and {G{\"u}nther}, Maximilian N. and {Gully-Santiago}, Michael A. and {Schlieder}, Joshua E.},
        title = "{Flare Statistics for Young Stars from a Convolutional Neural Network Analysis of TESS Data}",
      journal = {\aj},
         year = 2020,
        month = nov,
       volume = {160},
       number = {5},
          eid = {219},
        pages = {219},
          doi = {10.3847/1538-3881/abac0a},
archivePrefix = {arXiv},
       eprint = {2005.07710},
 primaryClass = {astro-ph.SR},
       adsurl = {https://ui.adsabs.harvard.edu/abs/2020AJ....160..219F}
}

@ARTICLE{Namekata2025_youngstarCMEs,
       author = {{Namekata}, Kosuke and {Maehara}, Hiroyuki and {Notsu}, Yuta and {Honda}, Satoshi and {Ikuta}, Kai and {Nogami}, Daisaku and {Shibata}, Kazunari},
        title = "{Do Young Suns Produce Frequent, Massive CMEs? Results from Five-year Dedicated Optical Observations of EK Draconis and V889 Hercules}",
      journal = {\apj},
         year = 2025,
        month = nov,
       volume = {993},
       number = {1},
          eid = {80},
        pages = {80},
          doi = {10.3847/1538-4357/adfe70},
archivePrefix = {arXiv},
       eprint = {2510.22111},
 primaryClass = {astro-ph.SR},
       adsurl = {https://ui.adsabs.harvard.edu/abs/2025ApJ...993...80N}
}

@ARTICLE{sing2024_wasp107,
       author = {{Sing}, David K. and {Rustamkulov}, Zafar and {Thorngren}, Daniel P. and {Barstow}, Joanna K. and {Tremblin}, Pascal and {Alves de Oliveira}, Catarina and {Beck}, Tracy L. and {Birkmann}, Stephan M. and {Challener}, Ryan C. and {Crouzet}, Nicolas and {Espinoza}, N{\'e}stor and {Ferruit}, Pierre and {Giardino}, Giovanna and {Gressier}, Am{\'e}lie and {Lee}, Elspeth K.~H. and {Lewis}, Nikole K. and {Maiolino}, Roberto and {Manjavacas}, Elena and {Rauscher}, Bernard J. and {Sirianni}, Marco and {Valenti}, Jeff A.},
        title = "{A warm Neptune's methane reveals core mass and vigorous atmospheric mixing}",
      journal = {\nat},
         year = 2024,
        month = jun,
       volume = {630},
       number = {8018},
        pages = {831-835},
          doi = {10.1038/s41586-024-07395-z},
archivePrefix = {arXiv},
       eprint = {2405.11027},
 primaryClass = {astro-ph.EP},
       adsurl = {https://ui.adsabs.harvard.edu/abs/2024Natur.630..831S}
}

@ARTICLE{stauffer2003_spottyyoungKdwarfs,
       author = {{Stauffer}, John R. and {Jones}, Burton F. and {Backman}, Dana and {Hartmann}, Lee W. and {Barrado y Navascu{\'e}s}, David and {Pinsonneault}, Marc H. and {Terndrup}, Donald M. and {Muench}, August A.},
        title = "{Why Are the K Dwarfs in the Pleiades So Blue?}",
      journal = {\aj},
         year = 2003,
        month = aug,
       volume = {126},
       number = {2},
        pages = {833-847},
          doi = {10.1086/376739},
archivePrefix = {arXiv},
       eprint = {astro-ph/0306127},
 primaryClass = {astro-ph},
       adsurl = {https://ui.adsabs.harvard.edu/abs/2003AJ....126..833S}
}

@article{Morris2020_spotsVSage,
   title={A Relationship between Stellar Age and Spot Coverage},
   volume={893},
   ISSN={1538-4357},
   url={http://dx.doi.org/10.3847/1538-4357/ab79a0},
   DOI={10.3847/1538-4357/ab79a0},
   number={1},
   journal={The Astrophysical Journal},
   publisher={American Astronomical Society},
   author={Morris, Brett M.},
   year={2020},
   month=apr, pages={67} }

@article{Roettenbacher2016_spotimaging,
   title={No Sun-like dynamo on the active star ζ Andromedae from starspot asymmetry},
   volume={533},
   ISSN={1476-4687},
   url={http://dx.doi.org/10.1038/nature17444},
   DOI={10.1038/nature17444},
   number={7602},
   journal={Nature},
   publisher={Springer Science and Business Media LLC},
   author={Roettenbacher, R. M. and Monnier, J. D. and Korhonen, H. and Aarnio, A. N. and Baron, F. and Che, X. and Harmon, R. O. and Kővári, Zs. and Kraus, S. and Schaefer, G. H. and Torres, G. and Zhao, M. and ten Brummelaar, T. A. and Sturmann, J. and Sturmann, L.},
   year={2016},
   month=may, pages={217–220} }

@ARTICLE{barat2024_bHST,
       author = {{Barat}, Saugata and {D{\'e}sert}, Jean-Michel and {Vazan}, Allona and {Baeyens}, Robin and {Line}, Michael R. and {Fortney}, Jonathan J. and {David}, Trevor J. and {Livingston}, John H. and {Jacobs}, Bob and {Panwar}, Vatsal and {Shivkumar}, Hinna and {Todorov}, Kamen O. and {Pino}, Lorenzo and {Mraz}, Georgia and {Petigura}, Erik A.},
        title = "{The metal-poor atmosphere of a potential sub-Neptune progenitor}",
      journal = {Nature Astronomy},
         year = 2024,
        month = jul,
       volume = {8},
        pages = {899-908},
          doi = {10.1038/s41550-024-02257-0},
archivePrefix = {arXiv},
       eprint = {2312.16924},
 primaryClass = {astro-ph.EP},
       adsurl = {https://ui.adsabs.harvard.edu/abs/2024NatAs...8..899B}
}

@article{rackham2018_tlseM,
  title={The transit light source effect: false spectral features and incorrect densities for M-dwarf transiting planets},
  author={Rackham, Benjamin V and Apai, D{\'a}niel and Giampapa, Mark S},
  journal={The Astrophysical Journal},
  volume={853},
  number={2},
  pages={122},
  year={2018},
  publisher={IOP Publishing}
}

@article{Murphy2025_wasp107b,
   title={A Panchromatic Characterization of the Evening and Morning Atmosphere of WASP-107 b: Composition and Cloud Variations, and Insight into the Effect of Stellar Contamination},
   volume={170},
   ISSN={1538-3881},
   url={http://dx.doi.org/10.3847/1538-3881/addf38},
   DOI={10.3847/1538-3881/addf38},
   number={1},
   journal={The Astronomical Journal},
   publisher={American Astronomical Society},
   author={Murphy, Matthew M. and Beatty, Thomas G. and Schlawin, Everett and Bell, Taylor J. and Radica, Michael and Kennedy, Thomas D. and Mehta, Nishil and Welbanks, Luis and Line, Michael R. and Parmentier, Vivien and Greene, Thomas P. and Mukherjee, Sagnick and Fortney, Jonathan J. and Ohno, Kazumasa and Wiser, Lindsey and Arnold, Kenneth and Rauscher, Emily and Edelman, Isaac R. and Rieke, Marcia J.},
   year={2025},
   month=jun, pages={61} }

@ARTICLE{Fu2022,
       author = {{Fu}, Guangwei and {Espinoza}, N{\'e}stor and {Sing}, David K. and {Lothringer}, Joshua D. and {Dos Santos}, Leonardo A. and {Rustamkulov}, Zafar and {Deming}, Drake and {Kempton}, Eliza M.-R. and {Komacek}, Thaddeus D. and {Knutson}, Heather A. and et al.},
        title = "{Water and an Escaping Helium Tail Detected in the Hazy and Methane-depleted Atmosphere of HAT-P-18b from JWST NIRISS/SOSS}",
      journal = {\apjl},
         year = 2022,
        month = dec,
       volume = {940},
       number = {2},
          eid = {L35},
        pages = {L35},
          doi = {10.3847/2041-8213/ac9977},
archivePrefix = {arXiv},
       eprint = {2211.13761},
 primaryClass = {astro-ph.EP},
       adsurl = {https://ui.adsabs.harvard.edu/abs/2022ApJ...940L..35F}
}

@ARTICLE{LibbyRoberts2025,
       author = {{Libby-Roberts}, Jessica E. and {Bello-Arufe}, Aaron and {Berta-Thompson}, Zachory K. and {Ca{\~n}as}, Caleb I. and {Chachan}, Yayaati and {Hu}, Renyu and {Kawashima}, Yui and {Murray}, Catriona and {Ohno}, Kazumasa and {Tokadjian}, Armen and et al.},
        title = "{The James Webb Space Telescope NIRSpec-PRISM Transmission Spectrum of the Super-Puff, Kepler-51d}",
      journal = {arXiv e-prints},
         year = 2025,
        month = may,
          eid = {arXiv:2505.21358},
        pages = {arXiv:2505.21358},
          doi = {10.48550/arXiv.2505.21358},
archivePrefix = {arXiv},
       eprint = {2505.21358},
 primaryClass = {astro-ph.EP},
       adsurl = {https://ui.adsabs.harvard.edu/abs/2025arXiv250521358L}
}

@ARTICLE{FournierTondreau2025,
       author = {{Fournier-Tondreau}, Marylou and {Pan}, Yanbo and {Morel}, Kim and {Lafreni{\`e}re}, David and {MacDonald}, Ryan J. and {Coulombe}, Louis-Philippe and {Allart}, Romain and {Albert}, Lo{\"\i}c and {Radica}, Michael and {Piaulet-Ghorayeb}, Caroline and et al.},
        title = "{Transmission spectroscopy of WASP-52 b with JWST NIRISS: water and helium atmospheric absorption, alongside prominent star-spot crossings}",
      journal = {\mnras},
         year = 2025,
        month = may,
       volume = {539},
       number = {1},
        pages = {422-438},
          doi = {10.1093/mnras/staf489},
archivePrefix = {arXiv},
       eprint = {2412.17072},
 primaryClass = {astro-ph.EP},
       adsurl = {https://ui.adsabs.harvard.edu/abs/2025MNRAS.539..422F}
}

@ARTICLE{Roy2025,
       author = {{Roy}, Pierre-Alexis and {Benneke}, Bj{\"o}rn and {Fournier-Tondreau}, Marylou and {Coulombe}, Louis-Philippe and {Piaulet-Ghorayeb}, Caroline and {Lafreni{\`e}re}, David and {Allart}, Romain and {Cowan}, Nicolas B. and {Dang}, Lisa and {Johnstone}, Doug and et al.},
        title = "{Diversity in the haziness and chemistry of temperate sub-Neptunes}",
      journal = {Nature Astronomy},
         year = 2025,
        month = dec,
          doi = {10.1038/s41550-025-02723-3},
archivePrefix = {arXiv},
       eprint = {2512.10876},
 primaryClass = {astro-ph.EP},
       adsurl = {https://ui.adsabs.harvard.edu/abs/2025NatAs.tmp..256R}
}

@ARTICLE{Zaleski2025,
       author = {{Zaleski}, S.~M. and {Valio}, A. and {Carter}, B.},
        title = "{Extracting starspot structures from exoplanet transit photometry}",
      journal = {\aap},
         year = 2025,
        month = oct,
       volume = {702},
          eid = {A227},
        pages = {A227},
          doi = {10.1051/0004-6361/202452779},
       adsurl = {https://ui.adsabs.harvard.edu/abs/2025A&A...702A.227Z}
}

@ARTICLE{Jarvinen2018,
       author = {{J{\"a}rvinen}, S.~P. and {Strassmeier}, K.~G. and {Carroll}, T.~A. and {Ilyin}, I. and {Weber}, M.},
        title = "{Mapping EK Draconis with PEPSI. Possible evidence for starspot penumbrae}",
      journal = {\aap},
         year = 2018,
        month = dec,
       volume = {620},
          eid = {A162},
        pages = {A162},
          doi = {10.1051/0004-6361/201833496},
archivePrefix = {arXiv},
       eprint = {1812.03675},
 primaryClass = {astro-ph.SR},
       adsurl = {https://ui.adsabs.harvard.edu/abs/2018A&A...620A.162J}
}

@ARTICLE{Yamashita2025,
       author = {{Yamashita}, Mai and {Itoh}, Yoichi and {Toriumi}, Shin},
        title = "{Variations in the Magnetic Field Strength of Pre-main-sequence Stars, Solar-type Main-sequence Stars, and the Sun}",
      journal = {\apj},
         year = 2025,
        month = may,
       volume = {985},
       number = {1},
          eid = {46},
        pages = {46},
          doi = {10.3847/1538-4357/adc816},
archivePrefix = {arXiv},
       eprint = {2504.04684},
 primaryClass = {astro-ph.SR},
       adsurl = {https://ui.adsabs.harvard.edu/abs/2025ApJ...985...46Y}
}

@software{Rackham2023,
       author = {{Rackham}, Benjamin V.},
        title = "{speclib}",
         year = 2023,
        month = apr,
          eid = {10.5281/zenodo.7868050},
          doi = {10.5281/zenodo.7868050},
      version = {0.0-beta.0},
    publisher = {Zenodo},
       adsurl = {https://ui.adsabs.harvard.edu/abs/2023zndo...7868050R}
}

@ARTICLE{Rackham2024,
       author = {{Rackham}, Benjamin V. and {de Wit}, Julien},
        title = "{Toward Robust Corrections for Stellar Contamination in JWST Exoplanet Transmission Spectra}",
      journal = {\aj},
         year = 2024,
        month = aug,
       volume = {168},
       number = {2},
          eid = {82},
        pages = {82},
          doi = {10.3847/1538-3881/ad5833},
archivePrefix = {arXiv},
       eprint = {2303.15418},
 primaryClass = {astro-ph.EP},
       adsurl = {https://ui.adsabs.harvard.edu/abs/2024AJ....168...82R}
}

@ARTICLE{feinstein_early_2023,
       author = {{Feinstein}, Adina D. and {Radica}, Michael and {Welbanks}, Luis and {Murray}, Catriona Anne and {Ohno}, Kazumasa and {Coulombe}, Louis-Philippe and {Espinoza}, N{\'e}stor and {Bean}, Jacob L. and {Teske}, Johanna K. and {Benneke}, Bj{\"o}rn and {Line}, Michael R. and {Rustamkulov}, Zafar and {Saba}, Arianna and {Tsiaras}, Angelos and {Barstow}, Joanna K. and {Fortney}, Jonathan J. and {Gao}, Peter and {Knutson}, Heather A. and {MacDonald}, Ryan J. and {Mikal-Evans}, Thomas and {Rackham}, Benjamin V. and {Taylor}, Jake and {Parmentier}, Vivien and {Batalha}, Natalie M. and {Berta-Thompson}, Zachory K. and {Carter}, Aarynn L. and {Changeat}, Quentin and {dos Santos}, Leonardo A. and {Gibson}, Neale P. and {Goyal}, Jayesh M. and {Kreidberg}, Laura and {L{\'o}pez-Morales}, Mercedes and {Lothringer}, Joshua D. and {Miguel}, Yamila and {Molaverdikhani}, Karan and {Moran}, Sarah E. and {Morello}, Giuseppe and {Mukherjee}, Sagnick and {Sing}, David K. and {Stevenson}, Kevin B. and {Wakeford}, Hannah R. and {Ahrer}, Eva-Maria and {Alam}, Munazza K. and {Alderson}, Lili and {Allen}, Natalie H. and {Batalha}, Natasha E. and {Bell}, Taylor J. and {Blecic}, Jasmina and {Brande}, Jonathan and {Caceres}, Claudio and {Casewell}, S.~L. and {Chubb}, Katy L. and {Crossfield}, Ian J.~M. and {Crouzet}, Nicolas and {Cubillos}, Patricio E. and {Decin}, Leen and {D{\'e}sert}, Jean-Michel and {Harrington}, Joseph and {Heng}, Kevin and {Henning}, Thomas and {Iro}, Nicolas and {Kempton}, Eliza M.-R. and {Kendrew}, Sarah and {Kirk}, James and {Krick}, Jessica and {Lagage}, Pierre-Olivier and {Lendl}, Monika and {Mancini}, Luigi and {Mansfield}, Megan and {May}, E.~M. and {Mayne}, N.~J. and {Nikolov}, Nikolay K. and {Palle}, Enric and {Petit dit de la Roche}, Dominique J.~M. and {Piaulet}, Caroline and {Powell}, Diana and {Redfield}, Seth and {Rogers}, Laura K. and {Roman}, Michael T. and {Roy}, Pierre-Alexis and {Nixon}, Matthew C. and {Schlawin}, Everett and {Tan}, Xianyu and {Tremblin}, P. and {Turner}, Jake D. and {Venot}, Olivia and {Waalkes}, William C. and {Wheatley}, Peter J. and {Zhang}, Xi},
        title = "{Early Release Science of the exoplanet WASP-39b with JWST NIRISS}",
      journal = {\nat},
         year = 2023,
        month = feb,
       volume = {614},
       number = {7949},
        pages = {670-675},
          doi = {10.1038/s41586-022-05674-1},
archivePrefix = {arXiv},
       eprint = {2211.10493},
 primaryClass = {astro-ph.EP},
       adsurl = {https://ui.adsabs.harvard.edu/abs/2023Natur.614..670F}
}
\bibliographystyle{aasjournalv7}



\end{document}